\documentclass[manuscript, screen,nonacm]{acmart}

\usepackage{booktabs}
\usepackage{amsfonts}
\usepackage{longtable}
\usepackage{graphicx}
\usepackage{bbding}
\usepackage{multirow}
\usepackage{inputenc}
\usepackage{amsmath}
\usepackage{float}
\usepackage{tabularx}
\usepackage{booktabs}
\usepackage[normalem]{ulem}
\useunder{\uline}{\ul}{}

\AtBeginDocument{%
  \providecommand\BibTeX{{%
    \normalfont B\kern-0.5em{\scshape i\kern-0.25em b}\kern-0.8em\TeX}}}

\setcopyright{none}
\renewcommand\footnotetextcopyrightpermission[1]{}
\begin{document}

\title[Avatars and Environments for Meetings in Social VR]{Avatars and Environments for Meetings in Social VR: What Styles and Choices Matter to People in Group Creativity Tasks?}

\author{Anya Osborne}
\email{anyaosborne@ucsc.edu}
\orcid{0000-0002-5506-623X}
\affiliation{%
  \institution{University of California Santa Cruz}
  \city{Santa Cruz}
  \state{CA}
  \country{USA}
}

\author{Sabrina Fielder}
\affiliation{%
  \institution{University of California Santa Cruz}
  \city{Santa Cruz}
  \state{CA}
  \country{USA}}
\email{ssfielde@ucsc.edu}

\author{Lee Taber}
\orcid{0000-0001-5220-1291}
\affiliation{%
  \institution{University of California Santa Cruz}
  \city{Santa Cruz}
  \country{USA}}
\email{ltaber@ucsc.edu}

\author{Tara Lamb}
\affiliation{%
  \institution{University of California Santa Cruz}
  \city{Santa Cruz}
  \country{USA}}
\email{tnlamb@ucsc.edu}

\author{Joshua McVeigh-Schultz}
\orcid{0000-0003-3419-757X}
\affiliation{%
  \institution{San Francisco State University}
  \city{San Francisco}
    \state{CA}
  \country{USA}}
\email{jmcvs@sfsu.edu}

\author{Katherine Isbister}
\orcid{0000-0003-2459-4045}
\affiliation{%
  \institution{University of California Santa Cruz}
  \city{Santa Cruz}
  \country{USA}}
\email{kisbiste@ucsc.edu}

\renewcommand{\shortauthors}{Osborne,~et~al.}

\begin{abstract}
Due to the COVID-19 pandemic, many professional entities shifted toward remote collaboration and video conferencing (VC) tools. Social virtual reality (VR) platforms present an alternative to VC for meetings and collaborative activities. Well-crafted social VR environments could enhance feelings of co-presence and togetherness at meetings, helping reduce the need for carbon-intensive travel to face-to-face meetings. This research contributes to creating meeting tools in VR by exploring the effects of avatar styles and virtual environments on groups’ creative performance using the Mozilla Hubs platform. We present the results of two sequential studies. Study One surveys avatar and environment preferences in various VR meeting contexts (N=87). Study Two applies these findings to the design of a between-subjects and within-subjects research where participants (N=40) perform creativity tasks in pairs as embodied avatars in different virtual settings using VR headsets. We discuss the design implications of avatar appearances and meeting settings on teamwork.
\end{abstract}

\ccsdesc[500]{Human-centered computing~Virtual reality}
\ccsdesc[500]{Human-centered computing~Collaborative interaction}
\ccsdesc[300]{Human-centered computing~Mixed / augmented reality}
\ccsdesc[100]{Human-centered computing~Scenario-based design}
\ccsdesc[300]{Human-centered computing~User studies}

\keywords{Virtual Reality, Social VR, Creativity Support, Avatars, Environments, VR meetings, Social Augmentation, Mozilla Hubs}



\maketitle

\section{Introduction}

As technology progresses, so does the way we work, communicate, and solve problems. As a consequence of the COVID-19 pandemic, the shift to video conferencing (VC) technologies significantly contributed to the viability of virtual work at a large scale~\cite{Brucks2022}. Social Virtual Reality (VR) has shown a tremendous increase in the commercial market as an emerging technology for rich embodied interactions, relationship building, leisure, and work activities~\cite{Handley2022}. Unlike VC software, the commercial increase of social VR has been driven by the capacity of embodied awareness~\cite{Smith2018}, including a heightened experience of social presence~\cite{Bailenson2004, Yoon2019, Steinicke2020} or the feeling of “being there”~\cite{Heeter1992, Lombard1997} with other people in a shared virtual environment. Social VR can even enable new forms of social augmentation~\cite{Freeman2021, Maloney2020c, Moustafa2018, Roth2019} that exceed what is possible in face-to-face contexts~\cite{Roth2018, Roth2019, McVeigh-Schultz2021, McVeigh-Schultz2021a, McVeigh-Schultz2021b, Isbister2022}. In our larger research agenda, we propose creating novel design interventions in social VR that can unleash new collective human capacities and establish new grounds for effective collaboration and social connection in VR meetings.

This research focuses on the specific workplace task of creative ideation as a crucial workplace activity in social VR. We concentrate on creative ideation as it is fundamental to scientific and commercial innovation at workplace meetings~\cite{Handley2022, Salter2003}. Recent studies on virtual teams found that VC groups generate less creative ideas than in-person groups because of the ‘narrowed visual focus’~\cite{Brucks2022} and that immersive virtual environments where people are presented as avatars contribute to enhancing creative collaboration among distributed teams~\cite{Le2017, Guegan2016, Buisine2016, Bourgeois-Bougrine2020a, Toumi2021, Kohler2009, Riordan2011, Guegan2017}. Prior research empirically and neurologically relates narrowed visual and cognitive focus induced by VC to a constrained associative process inherent to idea generation~\cite{Friedman2003, Posner2011, Rowe2007}, where thoughts diverge into many to be then assimilated and transformed into new creative ideas~\cite{Mednick1962, Nijstad2006, Jung2013}. Experimental studies in 3D virtual worlds like Second Life demonstrated that the embodiment of avatars, perceived as creative, enhances brainstorming performance~\cite{Guegan2017a, Buisine2016, Buisine2020, Bhagwatwar2013, Peña2013, Vosinakis2013}, particularly in the fluency and novelty of generated ideas~\cite{Guegan2016}. Furthermore, deliberate choices of visual and contextual cues to prime creative thinking in these worlds were shown to further augment team performance, outperforming simulations of traditional office settings~\cite{Peña2013, Bhagwatwar2013}. These findings were complemented by recent work on remote collaboration in social VR (Spatial.io), underscoring the role of spatial and embodied interaction in fostering creative thinking among coral scientists over a 1-2 month period during work meetings~\cite{Olaosebikan2022}. Cumulatively, these studies advocate for avatar-mediated brainstorming in multi-user VR as a potent tool for creativity support~\cite{Toumi2021} and improved behavioral, emotional, and social engagement, compared to traditional methods~\cite{Yang2018, Guan2021, Lee2019a}.

If social VR ultimately closes the gap between VC and face-to-face interaction, the question arises whether this new medium could productively supplant in-person collaborative ideation. However, realistic avatars often invoke decreased affinity and increased unease~\cite{Shin2019, Seymour2021} due to the uncanny valley effect~\cite{Schwind2018, Mori2012, Waltemate2018}, while cartoon-style avatars raise concerns about professional appropriateness~\cite{Bailenson2006, Inkpen2011, Junuzovic2012}.  In HCI research on VR, few have examined the symbiotic relationship between avatar styles and virtual environments in meeting scenarios, with some notable exceptions~\cite{Williamson2021, Williamson2022, McVeigh-Schultz2019}. Our research aims to fill this gap by examining these aspects within the Mozilla Hubs platform.

We situated our research in Mozilla Hubs due to its widespread adoption in academic and remote research contexts~\cite{Bredikhina2020, ElAli2021, Eriksson2021, Williamson2021, Yoshimura2020, GomesdeSiqueira2021, Ratcliffe2021, Williamson2021, Williamson2022, Maloney2020b, Li2022, Sanchez2024}. Moreover, its device compatibility and shared link invitation model~\cite{Osborne2023, Osborne2024} facilitate rapid, seamless entry into virtual spaces~\cite{McVeigh-Schultz2019, Osborne2019}, serving requirements for our research objectives.

This paper undertakes a two-fold research approach within the Mozilla Hubs platform to explore avatar and environment impacts on creative VR collaboration. Study One, a survey, identifies preferred avatars and settings in Mozilla Hubs for varied meeting contexts. Study Two applies these preferences in a between-and-within-subjects investigation where participants undertake creative brainstorming tasks in distinct virtual settings embodied in various avatars. The results offer insights into the design implications of avatar and environmental styles to support social interactions in VR meetings, serving as a timely resource for HCI researchers and design practitioners.

\section{Background and Related Work}
\label{background}

Here, we review existing research that tackles two core design elements for technology-mediated collaboration: the characteristics of the environment and the visual representation of users within it, known as avatars~\cite{Benford2000, Bhagwatwar2013}. There is a substantial body of work that discusses how the appearance of avatars and environment design could enhance creative collaboration among virtual teams. We narrowed our focus to studies in HCI, CSCW, and creativity research that informed our understanding of how design choices in avatar and environment styles can either promote or hinder participants’ sense of presence in virtual collaborative contexts.

\subsection{Avatar Appearance and Styles}
\label{back:avatars}

The impacts of avatar perceptions on the quality of social interactions have been studied in multiple domains, including video games~\cite{Isbister2006, Kafai2010, Schroeder2001, Praetorius2020, Wauck2018} and Collaborative Virtual Environments (CVEs)~\cite{Guegan2016, Guegan2017a, Buisine2020, Bhagwatwar2013, Ichino2022, Rooij2017}. Much of this research demonstrated mixed findings about how realistic versus non-realistic avatars should be in social contexts~(e.g.,~\cite{Steed2016, Casaneuva2001, Latoschik2017, Seymour2021, Shami2010}). For example, Steed~et~al. noted a positive influence of avatars on participants' sense of presence in VR~\cite{Steed2016}, as opposed to a no-avatar condition. Research in 3D virtual worlds indicated that creative ideation was positively impacted by self-similar avatars but negatively influenced by `non-creative' ones (represented by a stereotypical office worker)~\cite{Rooij2017}. In contrast, Shami~et~al.'s evaluation of the Olympus e-meeting tool revealed that `lightweight,' or cartoon-style avatars mitigated interaction barriers and enhanced meeting enjoyment through `phatic' communication~\cite{Shami2010}.

In the context of VC, humanoid cartoon avatars were generally deemed inappropriate for professional settings~\cite{Junuzovic2012, Inkpen2011}, except when team members were already acquainted~\cite{Inkpen2011}. In exploring avatar styles for different contexts, Praetorius~et~al. found preferences for low-poly avatars in work contexts (perceiving an avatar as a tool), realistic ones for meetings with friends as an accurate self-presentation, and non-realistic cartoon-style avatars for gaming (perceiving an avatar as a friend)~\cite{Praetorius2021}. On the other hand, Jo~et~al. discovered that cartoon-style avatars resembling participants' attire demonstrated the highest sense of presence in immersive VR~\cite{Jo2016}, though these findings haven't been empirically tested in the context of actual interactions in VR meetings.

HCI research showed that full-body avatars enhanced social presence compared to partial ones~\cite{Heidicker2017, Smith2018, Yassien2020}. Yoon~et~al. found that full-body realistic avatars were optimal for professional settings, while upper-body cartoon-style ones were more suitable for informal contexts in an AR-mediated system for remote collaborative work~\cite{Yoon2019}. Dobre~et~al. studied the longitudinal communicative effect of full-body avatars in Mixed Reality (MR) work meetings (using Microsoft HoloLens 2 devices) and revealed a preference for realistic facial features. However, they also found that recognition among known colleagues occurred regardless of the non-realistic cartoon appearance of participants' avatars~\cite{Dobre2022}. Unlike in computer-mediated systems, users' self-presentations in social VR are notably distinct due to multifaceted factors that affect embodied experiences~\cite{Osborne2019} and social practices~\cite{McVeigh-Schultz2019}. In-depth interviews with users across major social VR platforms~(e.g.,~AltspaceVR, Rec~Room, VRChat, Mozilla~Hubs) revealed a desire for avatars that either closely resemble their physical selves or fit well within the platform's social norms~\cite{Freeman2021}.

Inconsistent findings were observed concerning the impact of avatar appearance on teamwork, largely due to the highly context-dependent nature of such perceptions~\cite{Guegan2016, Guegan2017a}. For instance, in cooperative settings, realistic avatars were found to increase co-presence~\cite{Yoon2019, Dobre2022}, whereas no such effect was found in competitive scenarios~\cite{Freiwald2021}. While prior research mainly explored the modulation of humanoid realistic avatars~\cite{Yee2007, Guegan2017a, Buisine2016, Buisine2020, Ichino2022}, there is a lack of studies on non-realistic avatar styles in collaborative contexts. Recent VR studies, however, showed that creative, non-human avatar embodiments, such as animals~\cite{Ahn2016}, could enhance task performance. Benefits were observed through radical avatar's body transformations, including additional arms~\cite{Won2015}, elongated limbs~\cite{Kilteni2012}, tails~\cite{Steptoe2013}, and extra limbs~\cite{Won2015a}. Our research extends this body of work by examining various design elements of avatar aesthetics~(i.e.,~appearance, styles) in Mozilla Hubs, focusing on users' choices between realistic and non-realistic embodiment tailored to specific meeting contexts in VR.

\subsection{Environmental Contexts in Creative Collaboration}
\label{back:environments}

While contemporary social VR research draws upon studies of traditional CVEs~\cite{Benford1995, Benford2001, Sra2018, Jonas2019, Olin2020, Bleakley2020}, recent research suggests that social VR supports a variety of nuanced activities, play, and entertainment that provide unique experiences compared to traditional CVEs~\cite{Maloney2020b, Olaosebikan2022, Osborne2023}. In addition, the construction of virtual environments could affect the emotional state of participants~\cite{Riva2007, Dey2017}, trigger specific emotional responses~\cite{Felnhofer2015}, and activate particular kinds of perceived contextual norms in stimuli-rich virtual settings~\cite{Peña2013}. However, the design elements of virtual environments and how they may impact social interaction are often neglected when investigating the perception of avatars in VR meetings.

In HCI research, a handful of studies investigated environmental design in social VR platforms along with the perception of avatar styles~\cite{Williamson2021,  Williamson2022, McVeigh-Schultz2019, Osborne2023}. For example, the research on spatial proxemics in Mozilla Hubs discussed how the design configurations of avatars and meeting environments affected the space occupancy, group formation, participants’ perception, and participants' use of the meeting space~\cite{Williamson2021}. Besides HCI, editorials~\cite{GomesdeSiqueira2021, Zibrek2021} and creativity research~\cite{Olaosebikan2022} explored this topic. For example, Gomes~de~Siqueira~et~al. found that Mozilla Hubs users preferred virtual spaces mimicking real-world settings like museums, facilitating team formation~\cite{GomesdeSiqueira2021}. Replicating physical world metaphors in social VR, however, poses design challenges due to the lower fidelity of social and environmental cues~\cite{Williamson2022} and inefficient support of tools for remote collaboration~\cite{Olaosebikan2022, Osborne2023}. As users grow accustomed to avatar-based social VR platforms, their design preferences for VR environments may also evolve~\cite{GomesdeSiqueira2021}, necessitating further research to understand these shifts.

Research on organizational behavior investigated how external environments facilitated or hindered creativity among co-located teams~\cite{West2002, Hunter2007, Ceylan2008, Stone1998}. \textit{‘Material priming’} objects, such as briefcases and desks, led to less cooperative behavior compared to more informal, stimuli-rich settings~\cite{Kay2004}. Environmental contexts may also support creative activities in virtual worlds among distributed teams~\cite{Greeno1994, Sivan2014, Bhagwatwar2013, Peña2013, Vosinakis2013, Riordan2011}, including various aspects of social climate like goal clarity and intellectual stimulation~\cite{West2002, Sivan2014}. Bhagwatwar~et~al. compared brainstorming performance in two Second Life 3D virtual worlds~\cite{Bhagwatwar2013}. One world was designed as a creative-priming environment, while the other mimicked a standard office. Teams in the creative-priming environment demonstrated higher performance, attributed to the virtual world's \textit{‘simulation capabilities’}~\cite{Riordan2011, Sivan2014}. These capabilities enabled creative expression through not just avatars but a range of contextual cues~\cite{Nelson2019}. These cues, termed \textit{‘creativity priming elements,’} included objects of varying shapes, sizes, and colors, designed to induce a mood conducive to creative behavior~\cite{Bhagwatwar2013}.

In industrial design, VR simulations were often employed to explore user preferences for various settings like hotel rooms~\cite{Bogicevic2018}, office lighting~\cite{Heydarian2015}, and study environments~\cite{Applegate2009}, as well as to aid in the creative process of designers~\cite{Coburn2017, Rieuf2015, Toumi2021}. This research often draws from interdisciplinary studies examining the role of physical artifacts—such as windows~\cite{Stone1998}, natural elements~\cite{Shibata2002, Shibata2004}, lighting~\cite{Knez1995}, and color~\cite{Stone2003}—in fostering creativity and mood~\cite{Friedman2003, Rattner2019, Shibata2004}.

Despite discussions of VR's potential to replicate real-world creative settings~\cite{Sivan2014, Lee2019a, Graessler2019, Bourgeois-Bougrine2020a}, there's a gap in examining these in the context of social VR brainstorming activities. Existing work highlighted the benefits of unconventional virtual environments for creativity, as seen in Second Life studies~\cite{Guegan2017, Guegan2021}, indicating the need “to design more imaginary environments, very different from classical workspaces (e.g., working under the sea or in a fairytale forest)”~(\cite{Guegan2017},~p.~205). The deliberate manipulation of virtual design elements~\cite{Williamson2022, Toumi2021} and a wide selection of unfamiliar, non-realistic meeting spaces in social VR offers a new avenue in HCI research to understand how different aesthetics—ranging from skeuomorphic to highly experimental~\cite{Osborne2023}—impact creative collaboration in VR meetings.

\section{Study One: Targeted Survey on Avatars and Environments in VR Meetings}

To investigate the effects of avatar and environmental styles on groups’ creative collaboration in a lab-based setting, we first needed to identify what types of avatars and environments should be included in that study as stimuli based on people’s preferences within different meeting contexts in VR. In this regard, Study One addressed two key questions: (1)~What kinds of avatar attributes (variables) matter to people in completing either business or leisure activities, and why?; (2)~What kinds of environments do people perceive as casual (creative) versus more businesslike, and why?

\subsection{Methods and Materials}

To understand what avatars and VR spaces matter to people in completing different tasks, we conducted an online survey on Qualtrics. The survey encompassed demographics, previous experience with remote meeting tools like VC software and social VR/XR apps, avatar selection in distinct VR meeting contexts, and questions about virtual environments. Guided by previous research~\cite{Inkpen2011, Yoon2019, Dobre2022, Osborne2019, Osborne2023}, the survey scenarios were crafted based on whether participants were meeting with known or unknown individuals. Acknowledging that avatar choices are often influenced by the platform's ambiance and meeting context~\cite{Freeman2021, Williamson2021}, four distinct meeting scenarios were developed~(see~Table~\ref{tab:MeetingScenarios}), detailing the social context and participants' relationships within it.

\begin{table}[ht]
\centering
\caption{Meeting scenarios in VR (I-IV) included in the survey design.}
\label{tab:MeetingScenarios}
\resizebox{\textwidth}{!}{%
\begin{tabular}{lccc}
\hline
\textit{} & 
  With people you do not know: &
  \multicolumn{2}{c}{With people you know:} \\ 
\textit{Avatars for:} &
  Unfamiliars &
  Colleagues &
  Friends \\ 
  \hline
  \addlinespace[0.5ex]
\textit{\begin{tabular}[c]{@{}l@{}} Creative Tasks\end{tabular}} &
  \begin{tabular}[l]{@{} p{0.34\textwidth} l@{}} I. Imagine you are meeting a group of people that you have not met before in VR to work together on a creative project. Your goal is to produce effective results within a short time frame as a group.\end{tabular} &
  \begin{tabular}[l]{@{} p{0.23\textwidth} l@{}} II. Imagine you are meeting with a group of colleagues in VR  to discuss plans for a surprise birthday party for your manager.\end{tabular} &
  \begin{tabular}[l]{@{} p{0.23\textwidth} l@{}} III. Imagine you are meeting a group of friends in VR to discuss plans for a surprise birthday party for a common friend.\end{tabular} \\
  \addlinespace[0.5ex]
  \hline
  \addlinespace[0.5ex]
\textit{\begin{tabular}[c]{@{}l@{}}Business Purposes\end{tabular}} &
  \multicolumn{3}{l}{\begin{tabular}[l] {@{} p{0.8\textwidth} c@{}} 
  IV. Imagine you are applying for a job interview. The position you are applying for is very important to you. Instead of a video call on Zoom, the interview will take place in VR with a few other interviewers.\end{tabular}} \\
  \hline
\end{tabular}%
}
\end{table}

As behaviors and comfort levels often vary in the presence of colleagues versus friends or family, the study differentiated meeting scenarios for these two sub-groups, as outlined in Table~\ref{tab:MeetingScenarios}. Each scenario was accompanied by follow-up questions concerning avatar choices:

\begin{enumerate}
\item[1.] What avatar would you choose (from the given set)?
\item[2.] Please explain why you think it would be a good fit for this meeting scenario.
\item[3.] If you were to create your own full-body avatar for this scenario, what would this avatar look like?
\end{enumerate}

The survey's final section focused on environmental design. Participants chose spaces they deemed most creative and most suitable for business. They then rated the significance of various environmental features, such as movable furniture, availability of productivity or creativity tools, and the quality of lighting and furnishings. Sample questions included the following:

\begin{enumerate}
\item[1.] Select the most business-like space to meet with others to do work in.
\item[2.] Please explain what makes this space seem like a great business meeting space.
\item[3.] Please rate from 1 to 5 the following characteristics of environments that you think are the most important in business settings, where 1 is the most important, and 5 -- the least important.
\end{enumerate}

In the survey, participants chose from 45 images of diverse 3D avatars from Mozilla Hubs for different meeting contexts~(Fig.~\ref{fig:AvatarStimuli}). For meeting spaces, 21 images of Mozilla Hubs environments were provided, each featuring an in-world avatar for context. These spaces varied in size, layout, and theme. Participants could enlarge each image for a detailed view and select multiple preferred options for each scenario.

\subsubsection{Selection of Avatar Stimuli}

Based on previous avatar research, we identified variables such as humanoid vs. non-humanoid and realistic vs. fanciful (Table~\ref{tab:AvatarCodes}). Full-body avatars were excluded due to Mozilla Hubs' platform constraints. The variables were categorized into four groups: (1) realistic self-presentations; (2) non-realistic humanoids; (3) realistic non-humanoids like animals or food; (4) fictional characters such as ghosts or game characters. These categories served to label survey responses for further analysis.

\begin{table}[h]
\caption{Avatar variables.}
\label{tab:AvatarCodes}
\begin{tabular}{lllll}
\hline
\multirow{2}{*}{\textit{Variables}} & \multicolumn{2}{l}{\textbf{Humanoid (H)}} & \multicolumn{2}{l}{\textbf{Non-humanoid (N)}} \\
 & \textit{Label} & \textit{Description} & \textit{Label} & \textit{Description} \\ \hline
 \addlinespace[0.5ex]
\textbf{Realistic (R)} & HR\# & Realistic self & NR\# & Non-humanoid realistic (e.g., food, animal) \\
\textbf{Fanciful (F)} & HF\# & Non-realistic self & NF\# & Non-existing character (e.g., game character) \\ \hline
\addlinespace[0.5ex]
\end{tabular}
\end{table}

\begin{figure}[ht]
\centering
  \includegraphics[width=\linewidth]{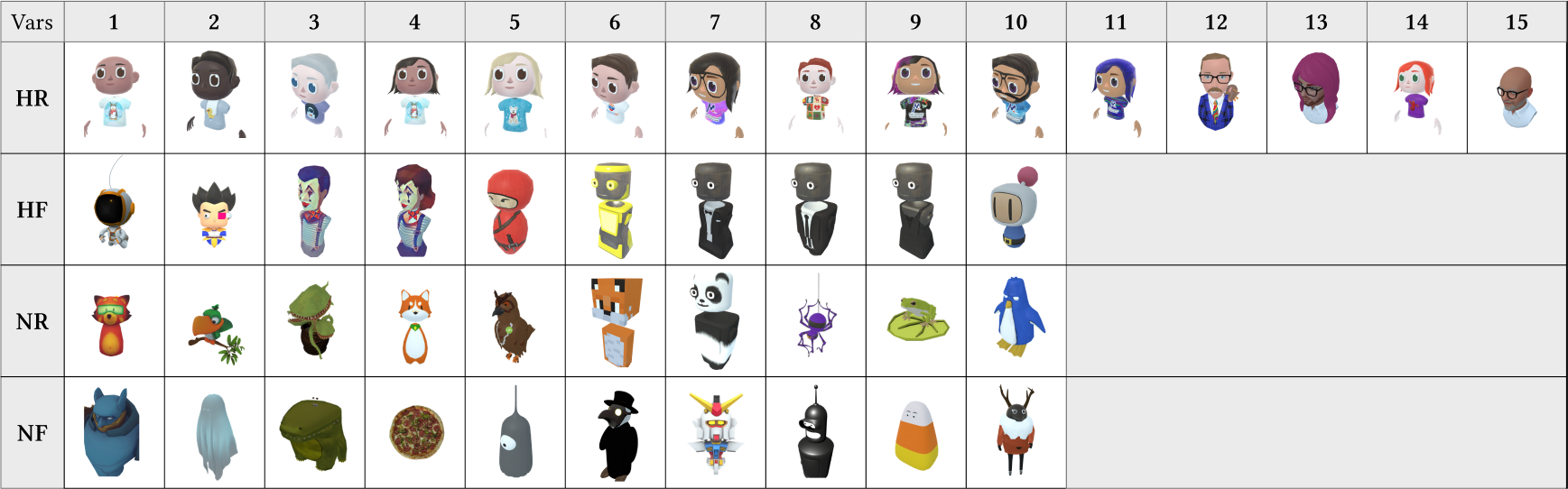}
  \caption {Avatar Stimuli in Study One (\textit{N}=45): Humanoid Realistic (HR), Humanoid Fanciful (HF), Non-humanoid Realistic (NF), and Non-humanoid Fanciful (NF).}
  \label{fig:AvatarStimuli}
  \Description{ Illustration of 45 public avatars from Mozilla Hubs, which we selected as stimuli in Study One. The images depict partial body avatars across four style groups. The first row presents 15 humanoid realistic styles of avatars, followed by humanoid fanciful (10), non-humanoid realistic (10), and non-humanoid fanciful (10).}
\end{figure}

We selected a total of 45 avatars in Mozilla Hubs, divided into four groups (Fig.~\ref{fig:AvatarStimuli}), each containing at least 10 avatars. The humanoid realistic (HR) group had 15 avatars to offer diverse gender and ethnic options. Categorization was challenging due to ambiguities; for instance, a Halloween candy avatar could fit both non-humanoid realistic (NR) and fanciful categories. To address this, collective decisions on categorization were made by a team of three researchers experienced in avatar design.

\subsubsection{Selection of Environment Stimuli}

\begin{figure}[ht]
\centering
  \includegraphics[width=\linewidth]{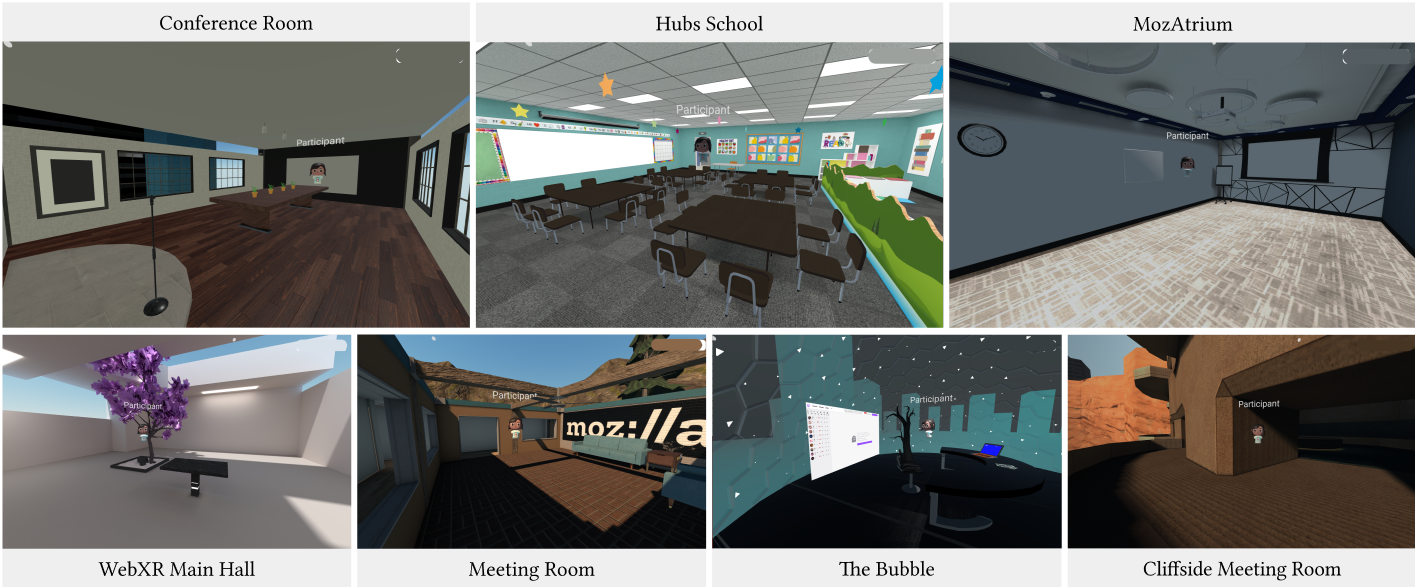}
  \caption {Environmental Stimuli in Study One~(Group~1,~\textit{N}=7): Traditional Style Spaces for Work and Business Meetings. These in-world screenshots of public environments include: \textit{Conference Room, Hub’s School, MozAtrium, WebXR Main Hall, Meeting Room, The Bubble,} and \textit{Cliffside Meeting Room}.}
  \label{fig:EnvStimuliOne_1}
  \Description{Illustration of 7 public environments from the Mozilla Hubs platform selected for the first group of stimuli in Study One. These images are represented by screenshots of environment settings with an in-world avatar. It includes Conference Room, Hub’s School, MozAtrium, WebXR Main Hall, Meeting Room, The Bubble, and Cliffside Meeting Room.}
\end{figure}

\begin{figure}[ht]
\centering
  \includegraphics[width=\linewidth]{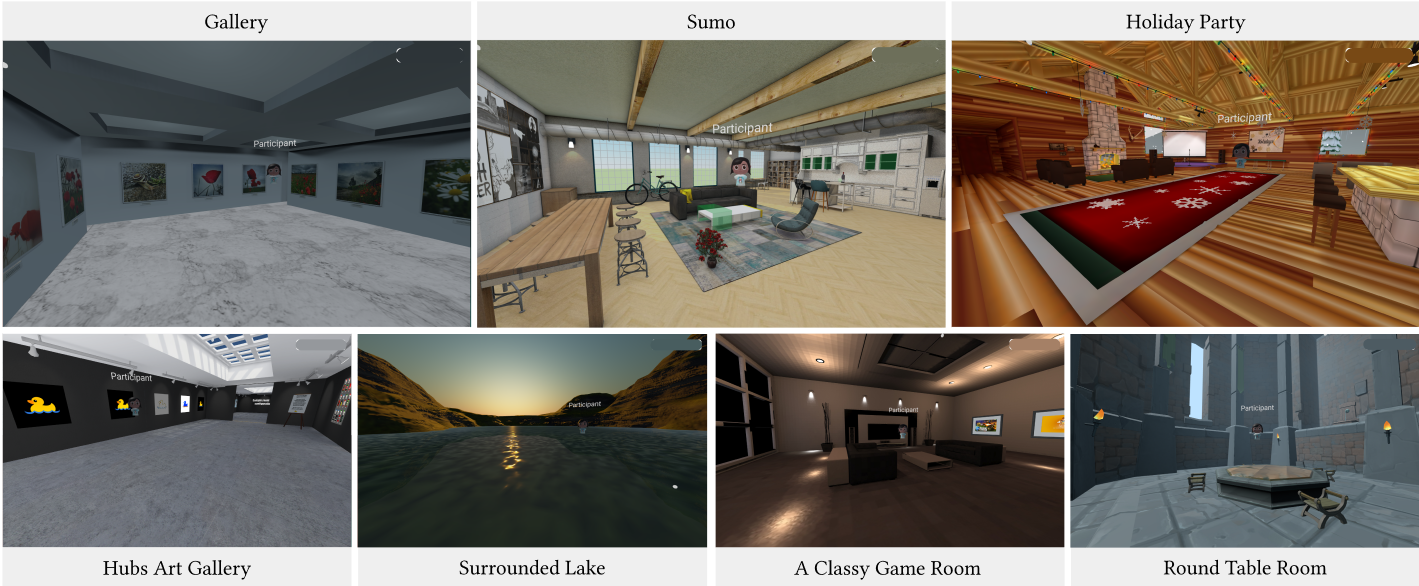}
  \caption {Environmental Stimuli in Study One~(Group~2,~\textit{N}=7): Realistic Styles for Creative Brainstorming. These in-world screenshots of public environments include: \textit{Gallery, Sumo, Holiday Party, Hubs Art Gallery, Surrounded Lake, A Classy Game Room,} and \textit{Round Table Room}.}
  \label{fig:EnvStimuliOne_2}
  \Description{Illustration of 7 public environments from the Mozilla Hubs platform selected for the second group of stimuli in Study One. These images are represented by screenshots of environment settings with an in-world avatar. It includes Gallery, Sumo, Holiday Party, Hubs Art Gallery, Surrounded Lake, A Classy Game Room, and Round Table Room.}
\end{figure}

\begin{figure}[ht]
\centering
  \includegraphics[width=\linewidth]{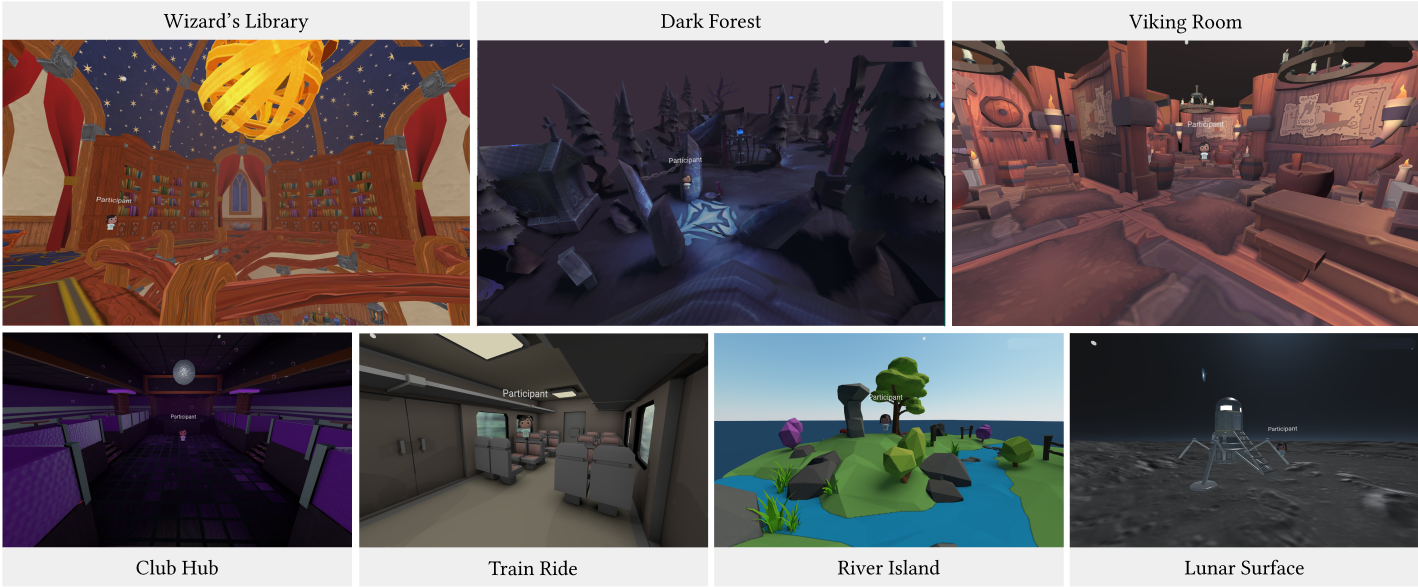}
  \caption {Environmental Stimuli in Study One (Group~3,~\textit{N}=7): Experimental and Fantastical Styles for Business and Creative meetings.” These in-world screenshots of public environments include: \textit{Wizard’s Library, Dark Forest, Viking Room, Club Hub, Train Ride, River Island,} and \textit{Lunar Surface}.}
  \label{fig:EnvStimuliOne_3}
  \Description{Illustration of 7 public environments from the Mozilla Hubs platform selected for the third group of stimuli in Study One. These images are represented by screenshots of environment settings with an in-world avatar. It includes Wizard’s Library, Dark Forest, Viking Room, Club Hub, Train Ride, River Island, and Lunar Surface.}
\end{figure}

Given the limited HCI research on the design details of virtual spaces and their impact on VR meetings~(Section~\ref{back:environments}), we selected a broad array of public environments in Mozilla Hubs. These included traditional business settings like conference rooms (Group~1, Fig.\ref{fig:EnvStimuliOne_1}), spaces suitable for creative brainstorming like art galleries (Group~2, Fig.\ref{fig:EnvStimuliOne_2}), and more experimental VR environments like lunar surfaces (Group~3, Fig.~\ref{fig:EnvStimuliOne_3}). The survey featured 21 such spaces, each represented by a screenshot containing an avatar to offer context on lighting and scale.


\subsubsection{Analysis}

For data analysis, we employed a dual-method approach: quantitative evaluation of closed-ended survey questions about avatar and environment choices, and qualitative scrutiny of open-ended responses explaining those choices. The analysis was conducted by a multidisciplinary team of four, skilled in areas ranging from design research to social VR design. Data reports, generated via Qualtrics, included both graphical and text-based presentations of survey responses. These responses were analyzed against predefined variables and descriptive groups (i.e.,~avatar and environmental stimuli), as informed by existing literature on avatars and environments (Section~\ref{background}).

\subsection{Participants}
\label{participants}

Participants were recruited through various online platforms, including social VR app Discord channels, social media groups, and internal channels at Meta, resulting in 87 voluntary respondents. The recruitment post included brief details about the research and the requirement of being familiar with social VR/XR platforms before participants proceeded with the link to the consent form and survey questions. The age distribution ranged between 18-25 years old (53\%), 26-35 years old~(30\%), 36-45 years old~(11\%), 46-55 years old~(3\%), and 56-60 years old (2\%). The majority identified as male~(63\%), followed by female~(25\%), non-binary~(8\%), and some declined to respond~(3\%). 

Participants' remote collaboration habits revealed frequent use of video conferencing (VC) software for business meetings, with 31\% using it daily. Casual VC meetings with friends/family (e.g.,~Zoom, Skype, Facetime) were less common, mostly occurring 1-2 times a week (20\%) or not at all (17\%). In terms of VR usage, 76\% reported using XR/VR platforms at least annually. VRChat was the most popular for casual interactions, with 42\% using it weekly. Other platforms like Rec Room, AltspaceVR, and Mozilla Hubs were also used, albeit less frequently, for both business and casual meetings with colleagues, friends, and unfamiliars.

\subsection{Results}
\label{One:Results}

We investigated avatar preferences across four meeting contexts in VR, depending on who participants were meeting with and what activities they could potentially engage in together in these meetings. We found that in creative and casual meetings, participants preferred more playful, non-realistic~(fanciful, fantastical) avatar styles whereas in formal, business-focused VR meetings – mostly realistic, humanoid avatars. We also examined what VR environments in Mozilla Hubs were generally viewed as more supportive of working on creative tasks versus productivity-focused tasks using familiar business spaces (e.g.,~conference and meeting rooms) and more unfamiliar non-realistic styles of environments (e.g.,~meeting on a Moon, on a moving train ride, ect.). We found that business environments like \textit{Conference Room}~(Fig.~\ref{fig:EnvStimuliOne_1}) were generally described as functional and conducive to productive meetings and business interactions. The unfamiliar non-realistic environments like \textit{Wizard’s library}~(Fig.~\ref{fig:EnvStimuliOne_3}) were mostly perceived to inspire creativity and spark new ideas due to a range of unique and interesting design features with respect to architecture, layout, vista view, and detailed decorations. We present these results in more detail below.

\subsubsection{Avatar Preferences in Examined Meeting Contexts}
\label{One:ResultsAvatars}

The survey results showed a similar avatar preference for three cases: creative projects with unfamiliars, creative projects with colleagues, and casual meetings with friends. In each of these cases, most participants preferred humanoid fanciful avatar styles~(Table~\ref{tab:AvatarPreferences}). For business meetings (e.g., job interview), participants predominantly preferred using humanoid realistic avatars to represent themselves. We also noted an interesting quantitative difference between the participants' ratings for specific groups of avatar styles among two contexts: casual VR meetings with friends and business meetings for a job interview in VR.

For casual meetings with friends, for example, there was only a slight percentage difference between ratings of humanoid fanciful (31\%) and non-humanoid fanciful (30\%) avatar groups among 70 participants~(Table~\ref{tab:AvatarPreferences}). On the other hand, for formal business contexts like a job interview, participants showed the highest preference for humanoid realistic avatars (49\%), compared to non-humanoid avatars (7-8\%) among 87 participants~(Table~\ref{tab:AvatarPreferences}). The latter confirms prior research findings on the appropriateness of using humanoid realistic avatar styles in related work contexts among VC teams (e.g.,~\cite{Junuzovic2012, Inkpen2011}).

\begin{table}[ht]
\centering
\caption{Participant's avatar style preferences across given meeting scenarios in VR, \textit{N}=87.}
\label{tab:AvatarPreferences}
\resizebox{0.98\textwidth}{!}{%
\begin{tabular}{lcccc}
\hline
\multirow{2}{*}{\textit{\begin{tabular}[c]{@{}l@{}} \addlinespace[1.5ex] Avatar Selections (within\\pre-determined style groups), \%\end{tabular}}} & I. & II. & III. & IV. \\
 & \begin{tabular}[c]{@{}c@{}}Creative project \\ w/people you do not know\end{tabular} & \begin{tabular}[c]{@{}c@{}}Creative project \\ w/colleagues\end{tabular} & \begin{tabular}[c]{@{}c@{}}Casual meeting \\ w/friends\end{tabular} & \begin{tabular}[c]{@{}c@{}}Business meeting \\ for a job interview\end{tabular} \\ \hline \addlinespace[0.5ex]
Humanoid Realistic (HR) & 30\% & 29\% & 23\% & \textbf{49\%} \\
Humanoid Fanciful (HF) & \textbf{39\%} & \textbf{41\%} & \textbf{31\%} & 36\% \\
Non-humanoid Realistic (NR) & 13\% & 13\% & 16\% & 7\% \\
Non-humanoid Fanciful (NF) & 17\% & 17\% & 30\% & 8\% \\ \addlinespace[1ex]
Total \textit{N} of responses & 76 & 70 & 70 & 87 \\ \hline
\end{tabular}%
}
\end{table}

\begin{figure}[ht]
\centering
  \includegraphics[width=\linewidth]{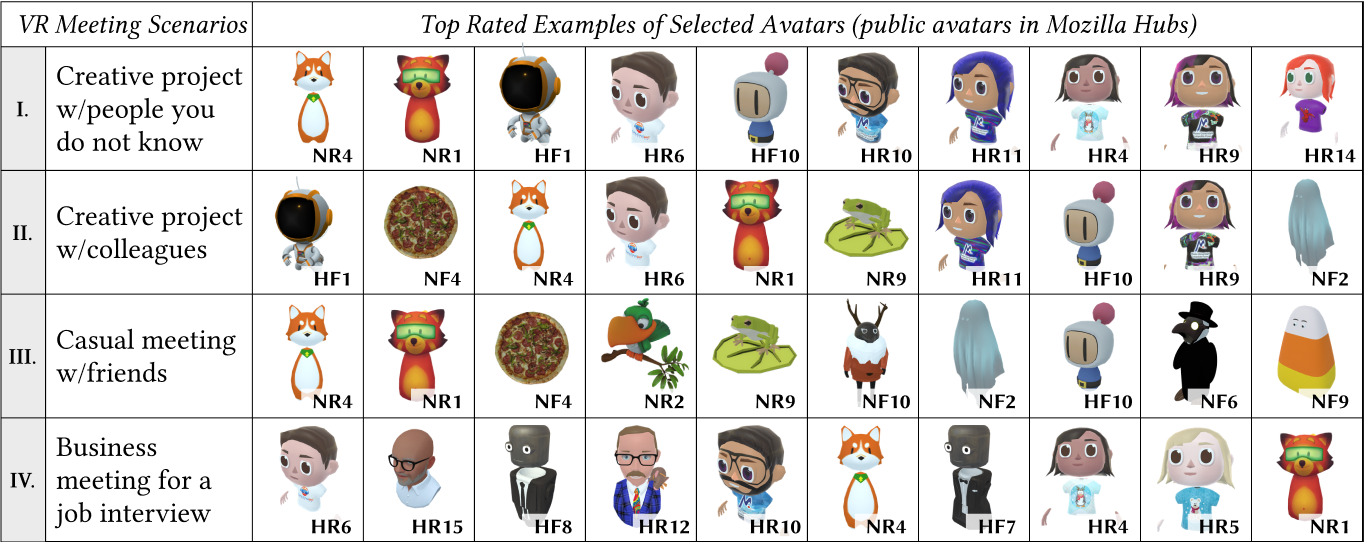}
  \caption {Top 10 examples of avatars participants selected as the most appropriate for each meeting scenario in VR,~\textit{N}=87 (public avatars in the Mozilla Hubs platform).}
  \label{fig:AvatarResults}
  \Description{Illustration of avatar images with avatar style codes that received the highest participants' ratings (top ten) for each meeting scenario in Study One. The first raw presents avatar results for a creative project with people you do not know (meeting scenario I), followed by the creative project with colleagues (meeting scenario II), casual meeting with friends (meeting scenario III), and a business meeting for a job interview (meeting scenario IV).}
\end{figure}

To interpret findings from Table~\ref{tab:AvatarPreferences}, we demonstrate examples of the top ten avatars of different style groups that received the highest ratings~(Fig.~\ref{fig:AvatarResults}) across four potential meeting scenarios in VR. Within the first two meeting contexts (Scenarios~I,~II), participants selected 7~instances of the precise same avatars. These instances ranged from non-humanoid realistic avatars representing animals (e.g.,~foxes - NR4, NR1) and food (e.g.,~pizza - NF4) to humanoid non-realistic styles, such as a Bomberman game character~(HF10) and a floating space astronaut~(HF1), with a couple of gender-specific ones~(e.g.,~HR6 and HR11). This indicates that participants did not perceive a big difference between the contexts of creative meetings with unfamiliars and colleagues in terms of their avatar representations in VR, compared to casual meetings with friends or having a business meeting in VR, like a job interview.

In creative meetings with unknown individuals (Scenario~I), participants leaned towards “casual,” “abstract,” and “fun” humanoid avatars. These avatars were considered beneficial both as icebreakers and for their interactive affordances. For example, P60 highlighted the utility of being able to see avatar interactions with the environment, stating, “I think being able to see what the avatars grab/pick up is beneficial.” In creative meetings with colleagues (Scenario~II), there was more latitude in avatar selection, ranging from realistic to fanciful types. P12 encapsulated this by noting, “this is a more creative-like setting, so I’m willing to either look like myself or pick a creative/weird avatar.” Some participants felt that such meetings could be “a side meeting, and not too formal,” allowing for more playful avatar choices. 

For casual meetings with friends (Scenario~III,~Fig.~\ref{fig:AvatarResults}), participants selected a much larger number of unique avatar instances that were not repeated elsewhere but in this particular context. More specifically, among the first ten top-rated avatars, there were no gender-specific avatars of humanoid realistic style, rather 4~unique avatar instances of non-humanoid fanciful styles~(NF10, NF2, NF6, NF in~Fig.~\ref{fig:AvatarResults}); and 6~-- humanoid realistic avatars that were selected in the other two creative work contexts (e.g.,~NR4, NR1, NF4, NR9 in~Fig.~\ref{fig:AvatarResults}). participants predominantly chose avatars that were “goofy," “entertaining,” and "silly-looking,” often switching avatars during the meeting. P63 elucidated this by saying, “I'd probably have something a little goofy just to mess with friends, like NF2, HF6, HF7, but then switch to one that is closer to me when more planning was actually going on.”

For job interviews in VR (Scenario~IV in~Fig.~\ref{fig:AvatarResults}), participants selected mostly humanoid avatars, which were represented by 5 unique avatar instances within the top ten selections. Among them, the predominant style group included humanoid realistic avatars (e.g., HR15, HR12, HR5 in~Fig.~\ref{fig:AvatarResults}). P24 indicated that this choice was motivated by a desire “(...) to be taken seriously,” and P10 further elaborated that photorealistic avatars were chosen to “convey a feeling of remote professionalism.” Outfit choices were considered crucial, varying based on the type of job interview. P87 summed this up by stating, “Depending on what the interview is for, it would probably have varying levels of dress—from business casual to full-on suit and tie, groomed for success.” Participants were very considerate about the social nuances and expectations of a business setting, aiming to fit into these rather than disrupt them.

\subsubsection{Environment Preferences for Business and Creative meetings}
\label{One:ResultsEnvironments}

We asked participants to select environments that they perceived as business-like to do work in, and spaces they thought would be best for a creative brainstorming with others. Unlike in avatar questions, these meeting contexts did not specify prior relationships with people because we wanted participants to focus on the common design characteristics of environments rather than their personal preferences. To address participants' personal preferences, we asked them to describe why they preferred selected spaces for each of these two meeting contexts in VR.

As shown in Table~\ref{tab:EnvResults}, the majority of participants characterized realistic work environments from group~1 (Fig.~\ref{fig:EnvStimuliOne_1}) as the most business-like spaces to hold meetings with others (70\%, \textit{N}=87). In contrast, experimental (fantastical) environments from group~3~(Fig.~\ref{fig:EnvStimuliOne_3}) were characterized as the best spaces for creative brainstorming (66\%, \textit{N}=68). Realistic environments for creative brainstorming from group~2~(Fig.~\ref{fig:EnvStimuliOne_2}), which represented the analogues from the real-life settings, were perceived as relatively appropriate for both work meetings (24\%, \textit{N}=87) and creative brainstorming tasks (28\%, \textit{N}=68).

\begin{table}[ht]
\centering
\caption{Participants’ perceptions of the appropriateness of different environmental settings in VR for conducting business/work meetings (meeting context I) and/or creative brainstorming with others (meeting context II), \textit{N}=87. Participants were asked to select at least one screenshot among 21 environments (clustered into three stimuli groups) that they thought would work best for each meeting context.}
\label{tab:EnvResults}
\resizebox{0.98\textwidth}{!}{%
\begin{tabular}{lcc}
\hline
\textit{\begin{tabular}[c]{@{}l@{}}Environmental Stimuli Selections, \%\\(within pre-determined stimuli style groups)\end{tabular}} & \begin{tabular}[l]{@{}c@{}} I.~“The most business-like spaces\\to meet with others to do work in.”\end{tabular} & \begin{tabular}[l]{@{}c@{}} II.~“The best spaces for a creative\\brainstorming with others.”\end{tabular} \\ \hline \addlinespace[0.5ex]
Group 1: Traditional conference room styles (Fig.~\ref{fig:EnvStimuliOne_1}) & \textbf{70\%} & 19\% \\
Group 2: Realistic room styles for brainstorming (Fig.~\ref{fig:EnvStimuliOne_2}) & 24\% & 28\% \\
Group 3: Experimental, fantastical styles (Fig.~\ref{fig:EnvStimuliOne_3}) & 6\% & \textbf{66\%} \\ \addlinespace[1ex]
Total \textit{N} of responses & 87 & 68 \\ \hline
\end{tabular}%
}
\end{table}

\begin{figure}[ht]
\centering
  \includegraphics[width=\linewidth]{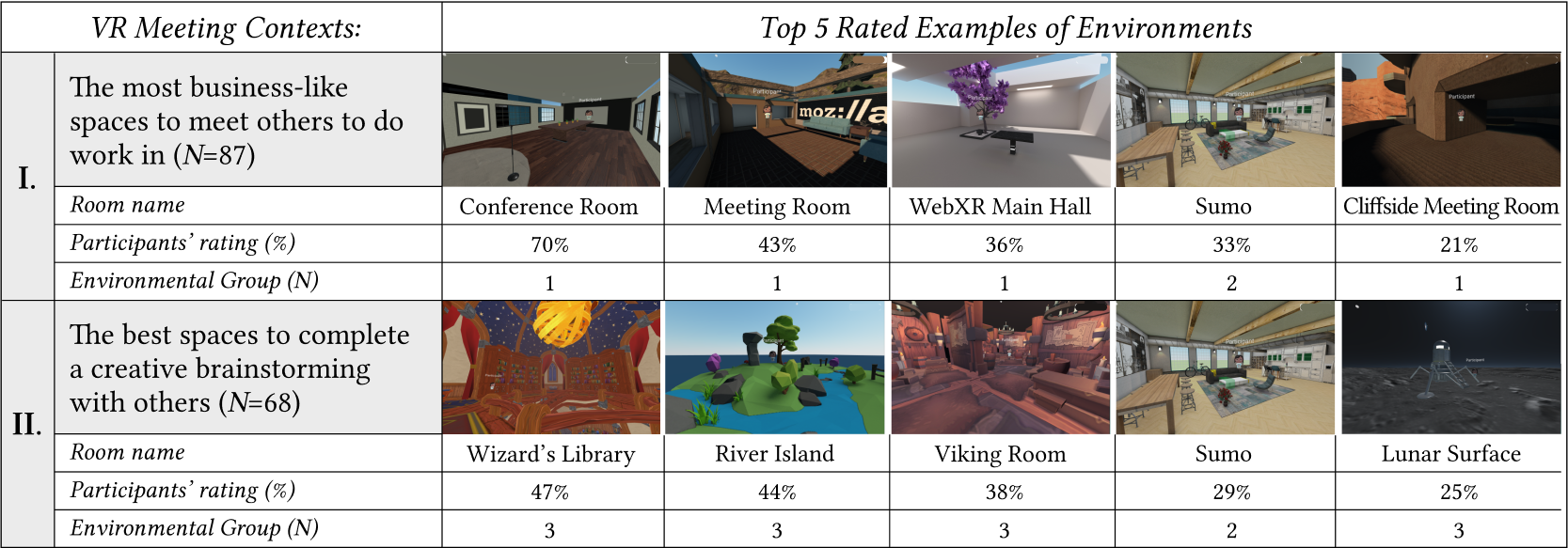}
  \caption {Top 5 examples of environments participants selected as the most appropriate for each VR meeting context, \textit{N}=87 (public environments in the Mozilla Hubs platform).}
  \label{fig:Top5Env}
  \Description{A table with ten screenshots of the Mozilla Hubs environments that received the highest ratings across two VR meeting contexts. Each screenshot is followed by a description below that specifies the room name, participants’ rating in percentages, and the number of the environmental group presented earlier in the stimuli description. The most business-like spaces to meet others to do work in (\textit{N}=87) are represented by Conference Room (70\%, Group 1), Meeting Room (43\%, Group 1), WebXR Main Hall (36\%, Group 1), Sumo (33\%, Group 2), and Cliffside Meeting Room (21\%, Group 1). The best spaces to complete a creative brainstorming with others (\textit{N}=68) are presented with screenshots of Wizard’s Library (47\%, Group 3), River Island (44\%, Group 3), Viking Room (38\%, Group 3), Sumo (29\%, Group 2), and Lunar Surface (25\%, Group 3).}
\end{figure}

Among the top 5 environments selected by participants (\textit{N}=87) for a business meeting~(Fig.~\ref{fig:Top5Env}), the Conference Room received the majority of votes~(70\%), followed by the Meeting Room~(43\%) and WebXR Main Hall~(36\%). These environments were mostly represented by realistic indoor spaces to do work in, which included contextual artifacts like tables, chairs, whiteboards, windows, and wall decorations. Among the best spaces for a creative brainstorming, the majority of participants’ votes~(\textit{N}=68) were given to the Wizard’s Library~(47\%), River Island~(44\%), and Viking Room~(38\%). In contrast to business-like spaces, this set of environments represented more experimental, unfamiliar settings of very divergent styles and contextual cues, which are rather unlikely to encounter or experienced in the real world. Interestingly, the Sumo environment received a relatively similar percentage of participants’ votes as an appropriate space to conduct both business (33\%, \textit{N}=87) and creative (29\%, \textit{N}=68) meeting contexts~(Fig.~\ref{fig:Top5Env}).

In business-focused virtual meetings (Context I in Fig.~\ref{fig:Top5Env}), participants highly favored minimalistic environments with elements focused on business productivity. Spaces like the Conference Room and Meeting Room were lauded for their “no distractions” design. P81 said the Conference Room was “Realistic, not distracting, nice that there’s a table in the center,” and P78 described the Meeting Room as having a “clean design, not much distraction.” P66 mentioned that smaller, more intimate spaces like the Conference Room and Meeting Room were “better due to natural light and smaller space in general.” Contrastingly, creative brainstorming sessions (Context II in Fig.~\ref{fig:Top5Env}) were dominated by stimuli-rich, colorful, and ‘dreamlike’ spaces~\cite{Guegan2021} such as the Wizard's Library and Lunar Surface. P87 mentioned that the Wizard's Library helps to “get the creative juices flowing,” and P62 described the Lunar Surface as an “exotic and conductive [space] to creativity.” Participants also emphasized the importance of brighter colors and outdoor environments for creative thinking.

As demonstrated in Table~\ref{tab:EnvRatings}, participants rated appropriate lighting and ‘material priming’ objects~\cite{Kay2004} represented by productivity tools as “the most important” in business-like spaces (53\%, \textit{N}=68). Conversely, ‘creativity priming’ tools~\cite{Bhagwatwar2013} were highly valued in creative brainstorming contexts (55\%, \textit{N}=66). The Sumo room emerged as a balanced space for both contexts. P75 described it as a “laid-back environment without coming off as unprofessional,” and P85 highlighted its “artistic vibes mixed with the formal setup.” These qualitative findings were supported by quantitative ratings, with the ability to manipulate objects being “not important” in business contexts (26\%, \textit{N}=68) but “the most important” in creative contexts (29\%, \textit{N}=66).

\begin{table}[ht]
\centering
\caption{Participants' ratings of environmental characteristics across two given meeting contexts in VR,~\textit{N}=87.}
\label{tab:EnvRatings}
\resizebox{\textwidth}{!}{%
\begin{tabular}{llccccc}
\hline
\multicolumn{2}{l}{\multirow{2}{*}{\textit{\begin{tabular}[c]{@{}l@{}} \addlinespace[1ex] Environmental characteristics per each \\ VR meeting context\end{tabular}}}} & \multicolumn{5}{c}{\textit{Participants' ratings, \%}} \\ \cline{3-7} 
\multicolumn{2}{l}{} & \begin{tabular}[c]{@{}c@{}}Most \\ Important\end{tabular} & Important & \begin{tabular}[c]{@{}c@{}}Somewhat\\ Important\end{tabular} & \begin{tabular}[c]{@{}c@{}}Not \\ Important\end{tabular} & \begin{tabular}[c]{@{}c@{}}Least \\ Important\end{tabular} \\ \hline \addlinespace[0.5ex]
\textbf{I.} & \multicolumn{6}{l}{\textbf{``The most business-like spaces to meet others to do work in,'' \textit{N}=68}} \\
\multicolumn{1}{c}{\textbf{}} & Appropriate furnishings (table, chairs) & 18\% & \textbf{29\%} & \textbf{29\%} & 12\% & 12\% \\
 & Appropriate lighting & \textbf{53\%} & 18\% & 16\% & 7\% & 6\% \\
 & Appropriate decorations/decor & 13\% & \textbf{35\%} & 21\% & 16\% & 15\% \\
 & Productivity tools (whiteboards, pens, markers) & \textbf{53\%} & 26\% & 7\% & 7\% & 6\% \\
 & \begin{tabular}[c]{@{}l@{}}I can manipulate objects (furniture) in \\ the environment (e.g., move chairs)\end{tabular} & 12\% & 21\% & 21\% & \textbf{26\%} & 21\% \\ \hline \addlinespace[0.5ex]
\textbf{II.} & \multicolumn{6}{l}{\textbf{``The best spaces to complete a creative brainstorming with others,'' \textit{N}=66}} \\
\multicolumn{1}{c}{\textbf{}} & Appropriate furnishings (sofas, lounge areas, arcades) & 15\% & \textbf{29\%} & 20\% & 27\% & 9\% \\
 & Appropriate lighting & 29\% & \textbf{35\%} & 27\% & 3\% & 6\% \\
 & Appropriate decorations/decor & 26\% & \textbf{30\%} & 20\% & 11\% & 14\% \\
 & Creativity tools (whiteboards, pens, markers) & \textbf{55\%} & 15\% & 14\% & 8\% & 9\% \\
 & \begin{tabular}[c]{@{}l@{}}I can manipulate objects (furniture) in \\ the environment (e.g., move chairs)\end{tabular} & \textbf{29\%} & 17\% & 26\% & 11 \% & 18\% \\ \hline
\end{tabular}%
}
\end{table}

\subsection{Limitations}
\label{One:limitations}

Study One faced several constraints that need to be considered when interpreting the findings. Firstly, the sample had a disproportionate gender representation with 63\% identifying as male, 25\% as female, 8\% as non-binary, and 3\% declining to respond. This likely led to higher preferences for male avatars, particularly in the humanoid realistic category. We attempted to mitigate this limitation in Study Two by incorporating a more diverse array of humanoid avatars.


Lastly, there was an attrition issue. Approximately 25\% of participants dropped out of the study before completing all survey questions. Of the original 87 participants, only 66 fully completed the survey, which was voluntary and offered no monetary incentives. Given this dropout rate, the decision was made to analyze all the responses received, regardless of whether the participants completed the survey in its entirety.

\subsection{Study One: Discussion}
\subsubsection{Avatar Preferences.}

The results showed that participants did not perceive a big difference between creative meetings with unfamiliars and colleagues in terms of their avatar presentations in VR. Among the first 10 top-rated avatars for these two meeting contexts, participants chose the same avatar instances ranging from non-humanoid realistic (e.g., animals or food) to humanoid non-realistic styles (e.g., game characters or a floating astronaut) (Fig.~\ref{fig:AvatarResults}).  For creative meetings with unfamiliars and colleagues, participants generally preferred avatars that were fun, casual, and acted as an icebreaker. These findings confirm the results of prior research on avatar systems in social VR apps showing that avatars often become “a conversation starter” in social VR communities~\cite{Osborne2019}. In this respect, participants mentioned that an avatar doesn't necessarily need to look like them and could be more abstract or fun, and casual. This also corresponds to the results in earlier work on VC teams, where the use of cartoon-style avatars showed to reduce interaction barriers in meetings and made the collaborative experience more enjoyable through `phatic' communication ~\cite{Shami2010}. Furthermore, participants noted the importance of the avatar being not too distracting and also having visible hands and arms in order to help pass on information in a creative brainstorming task, which greatly informed our approach for the avatar stimuli selection in Study Two (see Section~\ref{Two:Stimuli}).

For a casual meeting with friends, participants preferred avatar styles that were funny, silly, unique, or felt right in the moment. They often mentioned the importance of the avatar being informal, matching the nature of the meeting, and representing them personally besides their physical appearance. Additionally, they noted that friends usually do not care about professionalism and funny/casual avatars were generally better for exploring the full potential of meeting in VR. Some participants chose to use their own avatars, which are based on anime models or a robot with unique designs.

In formal business contexts in VR, like a job interview, participants showed the highest preference for humanoid realistic avatars compared to other meeting scenarios in VR. While participants generally preferred avatars that were neutral, inoffensive, and professional, with some resemblance to their real-life appearance, they often noted the importance of the avatar looking photorealistic, similar to themselves -- in order to convey a feeling of remote professionalism. This relates to earlier findings on the appropriateness of using humanoid realistic avatar styles in related work contexts among VC teams (e.g.,~\cite{Junuzovic2012, Inkpen2011}).

\subsubsection{Environment Preferences.}

Business environments were generally described as functional and conducive to productive meetings and business interactions. Conference Room, Meeting Room, WebXR Main Hall, and Sumo spaces~(Fig.~\ref{fig:Top5Env}) were perceived with a focus on minimizing distractions and creating a comfortable and professional work environment. Interestingly, some spaces were perceived for specific activities such as presentations or stand-up style meetings, which strongly relates to the concept of \textit{“Rooms are behavior!”} presented from earlier research on shaping pro-social interactions in social VR apps~\cite{McVeigh-Schultz2019}. Indeed, some of these spaces had interactive elements or tools for visualizing ideas or even a vista view that could enhance productivity and focus~\cite{Stone1998} while also promoting social expectations of behavioral patterns induced from similar contextual cues in real-life~\cite{Peña2013, Nelson2019, McVeigh-Schultz2019}. In general, these spaces were all perceived to be functional and conducive to productive meetings and business interactions. 

Conference Room was described as the most formal and business-oriented environment. Participants liked that it had a familiar "big round table" geometry, which allowed them to orient themselves and their attention. This emphasizes the importance of  ‘circular gatherings as conversation anchors’ discussed in prior work on social VR~\cite{McVeigh-Schultz2019}. Participants described Conference Room as minimalist and neutral, with no distractions, clean design, and well-lit rooms, which showed to promote productivity in real-life settings (e.g.,~\cite{Knez1995}). Conference Room was also viewed as a realistic and fairly standard space for business meetings that might have useful work tools, simulating a conference room from~real-life.

The top-rated creative environments included Wizard's Library, River Island, Viking Room, and Sumo (see~Fig.~\ref{fig:Top5Env}). The designs of these spaces were perceived to inspire creativity and spark new ideas. They were generally described as colorful, fun, and creative, with unique and interesting features such as strange architecture, outdoor settings, open areas, relaxing atmospheres and detailed decorations, which were discussed in earlier research as ‘stimuli-rich’ environments~\cite{Kay2004, Bhagwatwar2013}. Some spaces were perceived for specific creative activities such as game design, movie production, and brainstorming around similar themes.

Wizard's Library was one of the top-rated creative environments described with a unique and interesting design. Participants liked that it had a lot of detail and color, which could spark creative thinking and inspiration. They also found it fun and conducive to creativity noting that its unique design could be useful for brainstorming around similar themes. Some participants specifically mentioned that the library's witchy theme seemed unique and enjoyable.

\subsubsection{Insights for Study Two.}

The findings on avatar preferences revealed a nuanced design space that varies by meeting context and participant relationships, a topic underexplored in HCI literature~\cite{Yoon2019, Inkpen2011, Dobre2022, Zamanifard2019}. Study Two aims to examine how these avatar preferences influence group creativity in specific Mozilla Hubs environments during brainstorming tasks.

In light of environmental preferences, we selected the two highest-rated settings—Conference Room and Wizard’s Library—for Study Two. These choices aim to explore how environment and avatar styles collectively affect creative ideation. The role of avatar embodiment in these settings is critical, as it interacts with conversational orientation~\cite{Kendon1990, Marshall2011} and social atmospheres in VR~\cite{Williamson2021}, promoted through contextual cues in stimuli-rich environments~\cite{Kay2004, Bhagwatwar2013, Peña2013, Guegan2017, Nelson2019} may “only operate when they activate shared cultural understandings with others”~\cite[p.~10]{McVeigh-Schultz2019}.

\section{Study Two: Effects of Avatar and Environment Styles on Group Creative Performance}

Study Two aimed to explore the impact of avatar and environmental styles on group creativity in ideation tasks, guided by Research through Design~(RtD) principles~\cite{Gaver2012, Sun2019, Won2014}. Unlike prior studies that emphasized individual creativity~\cite{Sivan2014, Parks1994, Spitzberg1988}, this study focuses on group creativity, recognizing the team processes that catalyze individual creativity into collective output~\cite{Taggar2002, Paulus2000}. The study design drew inspiration primarily from Sun~et~al. and Won~et~al.~\cite{Sun2019, Won2014}, adopting measures that evaluated groups on generating resource conservation strategies~(see Appendix~\ref{appendix: A.1}). Adaptations in study design, measures, and procedures are detailed in the following section.

\subsection{Methods and Materials}
To examine the influence of avatar and environmental styles on group creativity, Study Two employed an experimental approach to test the results of Study One in a controlled setting. Methods and materials were adopted from prior HCI research for investigating creative ideation outcomes among participants pairs~\cite{Sun2019, Won2014}, using four custom-built environments in Mozilla Hubs to rigorously test all conditions. The study incorporated both quantitative and qualitative measures, sourcing data from video and audio recordings, observation notes, and the post-study survey.

\subsubsection{Study Design}
\label{Two:StudyDesign}

In Study Two, we used a 2x2 study design with both within-subjects and between-subjects variables~(Table~\ref{tab:2x2}). The control condition was represented by an assigned avatar for both business-style and creative-style environments, whereas the experiment condition provided participants with an opportunity to make their own avatar choice using the given tool for avatar selection embedded in the environment. The study is divided into two parts. For Part 1 (between-subjects variable), each participant pair was assigned to either a business-style environment or a creative-style environment for the entirety of the experiment. In Part 2 (within-subjects variable), each pair of subjects were asked to select an avatar, using the embedded avatar selection tool.

\begin{table}[h]
\centering
\caption{2x2 Design of Study Two.}
\label{tab:2x2}
\begin{tabular}{@{}lll@{}}
\toprule
\textit{2x2 Study design} & Business env. (Conference Room) & Creative env. (Wizard's Library) \\ \midrule
Preassigned avatar (Part 1) & Assigned business-style avatar & Assigned creative-style avatar \\
Avatar choice (Part 2) & Free choice & Free choice \\ \bottomrule
\end{tabular}
\end{table}

\subsubsection{Stimuli Selection}
\label{Two:Stimuli}

The avatar and environmental choices for the 2x2 study design~(Table~\ref{tab:2x2}) were guided by Study One's results. Avatar selection was a three-step process: First, the top five unique avatars preferred for business and unfamiliar creative meetings were chosen~(Scenario~I,~IV). Second, the top five avatars for creative meetings with colleagues were selected~(Scenario~II). The study excluded avatars preferred for meetings with friends~(Scenario~III), focusing instead on unfamiliar or business-related interactions. Finally, eight additional humanoid realistic avatars were added to offer more demographic representation, as Study One showed a preference for humanoid forms in business and unfamiliar creative contexts.

\begin{figure}[ht]
\centering
  \includegraphics[width=\linewidth]{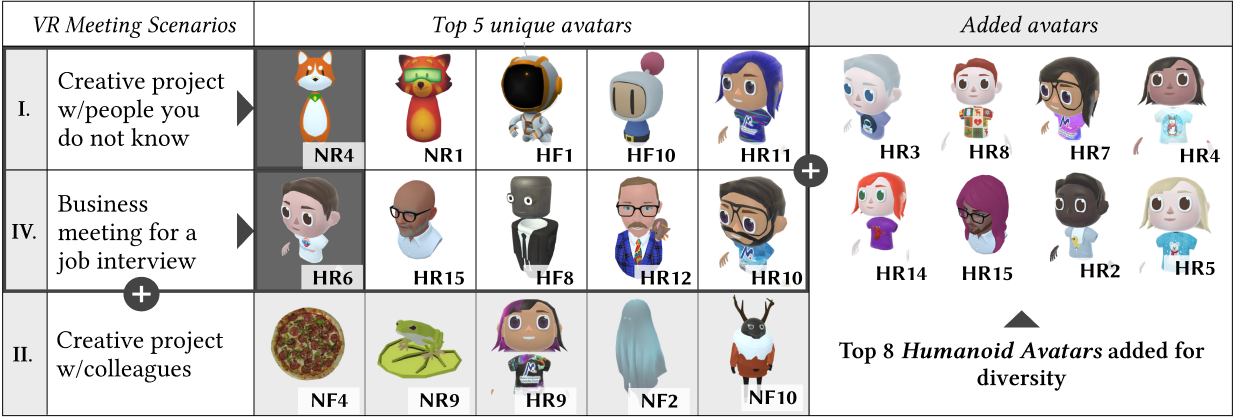}
  \caption {Avatars stimuli selected for Study Two.}
  \label{fig:Study2AvatarStimuli}
  \Description{A table showing avatar images with an avatar style code, which we selected as stimuli for Study Two. These avatars represent top-rated unique avatars discussed earlier in Study One across three VR meeting scenarios, such as a creative project with people you do not know (meeting scenario I), a business meeting for a job interview (meeting scenario IV), and a creative project with colleagues (meeting scenario II). The second column illustrates the top eight unique humanoid realistic avatars, which were added for diversity in terms of gender, hair color, and skin tone.}
\end{figure}

\begin{figure}[ht]
\centering
  \includegraphics[width=\linewidth]{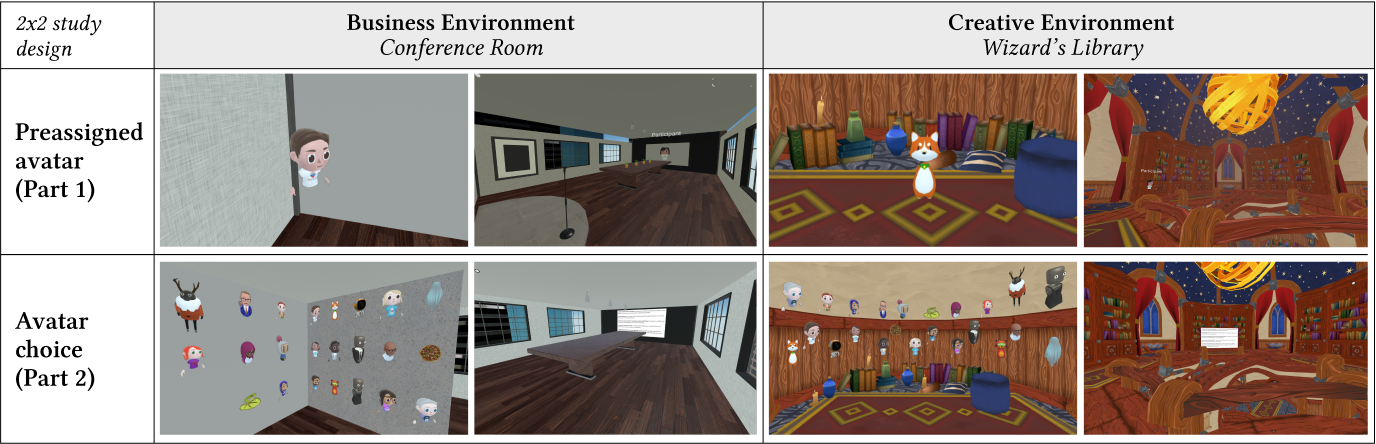}
  \caption {Avatar and environment stimuli used in 2x2 study design to understand their effect on participants’ performance in a creative ideation task.}
  \label{fig:Study2Stumuli}
  \Description{Illustration of 2x2 study design set-up with screenshots of business and creative environments with in-world avatar links in Mozilla Hubs that users could point at to change their avatar instantly. The business environment displays screenshots from the Conference Room setting, and the creative environment includes screenshots of the setting in Wizard’s Library. The first row of images demonstrates a single preassigned avatar for Part 1, and the second row shows screenshots of multiple avatars participants could select in Part 2 of Study Two.}
\end{figure}

For Part 1 (Table~\ref{tab:2x2}), we used the assigned humanoid realistic avatar (HR6) in the business-style environment, and the non-humanoid realistic avatar (NR4) in the creative-style environment. In Part 2, participants were presented with a total of 23 avatars to select from (Fig.~\ref{fig:Study2AvatarStimuli}). Drawing from the Study One results on Mozilla Hubs environments presented in Figure~\ref{fig:Study2AvatarStimuli}, we selected Conference Room as the most business-style environment, and the Wizard’s Library as the most creative one. The avatar stimuli were then situated in each of these environments and ingegrated into the 2x2 study design~(see~Fig.~\ref{fig:Study2Stumuli}).

\subsubsection{Procedure}

To participate in this study, subjects needed a device capable of connecting to Zoom and a VR headset. At the beginning of the experiment, participants met with researchers on Zoom to receive their Participant ID, instructions for entering the environment on their headsets, and the link for the first environment. After giving verbal consent for recording, they entered their assigned Mozilla Hubs lobby, switched to their IDs, and proceeded to the assigned room.

Participants selected their designated avatar by hovering over and selecting the “use avatar” button and then took a selfie using the Mozilla Hubs camera tool. Participants explored the environment before convening near a whiteboard with prompts for the collaborative activity. The activity involved brainstorming new ways to save energy or water (depending on which was randomly selected to be given in Part 1 or Part 2) other than the five examples displayed on the whiteboard~(see~Appendix~\ref{appendix: A.1}). We asked participants to brainstorm novel ideas for the next 10 minutes, with a reminder at the 5-minute mark.

Once the time was up, participants returned to Zoom for a 5-minute break while researchers ended the first recording and prepared for the next. Participants were then sent a new link to the same virtual environment, featuring a different set of avatars and brainstorming prompts.

\subsubsection{Measures}
\label{Two:Measures}
To assess the influence of avatar and environment styles on group creativity in ideation tasks, we implemented the following measures:

\begin{enumerate}
    \item [1.]\textit{Creative Performance.} In two brainstorming tasks focused on energy and water conservation (see Appendix~\ref{appendix: A.1}), participants generated ideas. Scoring ranged from zero to two points per idea beyond the initial 5 given as prompts. Zero points were assigned for redundant or facetious ideas, one point for similar yet distinct ideas, and two points for novel contributions. This scoring system was adopted from previous studies~\cite{Sun2019, Won2014}. Researchers tallied scores in real-time, later confirming them via audio and screen recordings. Task orders were randomized to control for sequence effects.
    \item [2.]\textit{Perceived Group Performance.}  Participants provided open-ended feedback on their perceived creative performance in tasks, specifying in which task they felt more creative and why.
    \item [3.]\textit{Participant Avatar Choice.} Choices of avatars, categorized as humanoid or fanciful, were recorded to explore the environment's influence on avatar selection.
    \item [4.]\textit{Participant Perceptions of Avatar in Relation to Self.} Adopting Ducheneaut et al.'s framework~\cite{Ducheneaut2009}, this qualitative metric evaluated how participants felt their chosen and assigned avatars represented them (e.g.,~Why did you choose this avatar? What was it like to be the avatar you were assigned? What was it like to be the avatar you chose?).
    \item [5.]\textit{Participant Perception of Avatar in Relation to Others.} Building on prior studies~\cite{Isbister2006, Kafai2010, Trepte2010}, we queried participants on their experiences brainstorming with others based on avatar appearances (e.g.,~What was it like to brainstorm with the other person? How do you think the other person felt brainstorming with your avatar in Part 1 and 2?).
    \item [6.]\textit{Environment Perceptions.} Participants reflected on how the virtual setting influenced their ideation performance, noting favorable or unfavorable elements (e.g.,~How was it to brainstorm in the environment you were in? Were there aspects of it that you liked/disliked?).
    \item [7.]\textit{Participants' Prior VR Experience and Demographics.} To capture participants' prior VR experiences and demographic information, questions were included about their frequency of social VR usage, avatar preferences, and prior relationships with other study participants (i.e.,~Stranger (Unfamiliars), Acquaintance (Colleague), or Friend).
    
\end{enumerate}

\subsubsection{Participants}

We recruited 20 pairs of participants (\textit{N}=40) who qualified to participate in the study via a screening survey. The recruitment process involved sharing physical and digital flyers via three Northern California college campuses and social media platforms such as Instagram and Reddit. There were 384 responses to the screening survey which resulted in 40 viable participants. The survey screening questions qualified participants who were 18 years of age or older, lived in the United States, owned a VR headset, and had access to Zoom. Participants were compensated with a \$5 Amazon gift card and entry to a raffle for an Oculus Quest 2. The demographics of the participants were as follows: 25-34 years of age (63\%), 18-24 years of age (28\%), 35-44 years of age (8\%), and 45-54 years of age (3\%). The race was reported as White (48\%), Asian or Pacific Islander (20\%), Hispanic (20\%), mixed race (10\%), and Black (3\%). The devices participants used in the study were the Oculus Quest 1 and 2, Valve Index, and Oculus Rift. There were technical difficulties involved with participants entering the VR environment when using a tethered headset, so some participants had to be rescheduled for new sessions. 

\subsection{Results}

We examined how factors such as avatar selection, task sequence, environmental context, and pre-existing relationships influenced paired creativity scores across tasks. While no statistical significance was observed, we did note trends suggesting a favorability for realistic avatars in professional contexts and non-humanoid or fantastical avatars in creative scenarios. Additional qualitative insights from participants indicated that avatar choices contributed to a sense of connection with their partner, enhanced creativity, and increased identification with their own avatar.

\subsubsection{Avatar Choice.}

\begin{figure}[ht]
\centering
  \includegraphics[width=\linewidth]{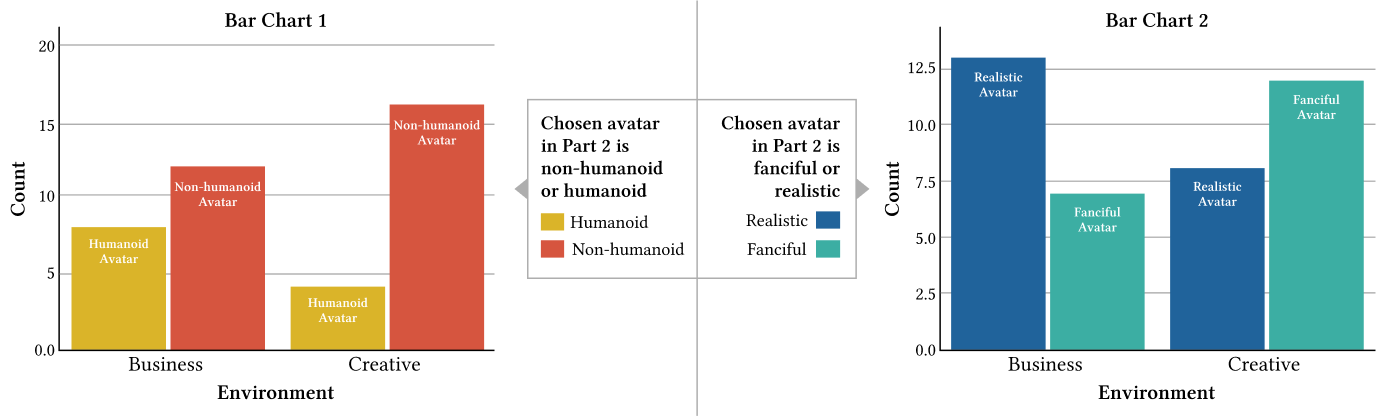}
  \caption {Avatar choice results on a Humanoid dimension (Bar Chart 1), and on the Realism dimension (Bar Chart 2).}
  \label{fig:BarCharts}
  \Description{Illustration of two Bar Charts for chosen avatar in Part 2 of Study Two resulting from performing the Chi-Square tests of the Association for business and creative environments. On the left-hand side, Bar Chart 1 displays the results on the humanoid (yellow bar) and non-humanoid (red bar) avatar styles, showing a trend in favor of non-humanoid avatars. On the right-hand side, Bar Chart 2 illustrates the findings on the realistic (blue bar) and fanciful (green bar) dimensions, showing a trend for more realistic avatars in a business environment, and fanciful avatar styles for creative spaces.}
\end{figure}

Participants had the option to select from 23 different avatars, categorized along two dimensions: humanoid vs. non-humanoid and realistic vs. fanciful. These choices were made within two distinct environmental styles: business and creative. We conducted two Chi-Square tests to examine if environmental context influenced avatar selection along these dimensions. For the humanoid dimension, our results showed no statistically significant relationship between environment and avatar type, $\chi^2$(1)=~1.91, p=~.17, $\varphi$=~.22. While no significant relationship was found, a trend suggested a preference for non-humanoid avatars, especially in creative settings (Bar Chart~1~in~Fig.~\ref{fig:BarCharts}). Similarly, for the realism dimension, no significant relationship was found, $\chi^2$(1)=~2.51, p=~.11, $\varphi$=~.25. However, a trend indicated a preference for realistic avatars in business environments and fanciful avatars in creative settings (Bar Chart~2~in~Fig.~\ref{fig:BarCharts}).


\subsubsection{Effects of Task, Avatars, Environment, and Prior Relationship on Creativity Scores.}

As we had a low \textit{N} of participants, we decided to analyze each variable by itself as it was likely that interactions between variables would be difficult to find. We report the results for the following between subjects variables: participant familiarity within pair, environment, and group avatar choice on the dependent variable of the average creativity task scores.

\subsubsection{Effect of Familiarity on Creativity Score.}

Participants were asked to identify their relationship with their paired participant as either a Stranger, Acquaintance, or Friend. Out of the 40 participants, 24 reported their partner as a Stranger, 8 as an Acquaintance, and 8 as a Friend. Only two pairs had mismatched relationship perceptions. We categorized the data into two groups: those who knew each other (reported as Acquaintance or Friend) and those who did not (reported as Stranger). Statistical analysis revealed no significant relationship between pre-existing familiarity and creativity score, t(18)=~.96, p~>~.05, \textit{d}=~.4. Thus, prior acquaintance between paired participants did not influence their performance on creativity tasks.


\subsubsection{Effect of Environment on Creativity Score.}

We found, when looking at the effect of environment on creativity scores, that there were no significant differences in creativity scores, t(18)=~.31, p~>~.05, \textit{d}=~.14. This result indicates that there was no effect of the environment on creativity score.

\subsubsection{Effect of Group-Level Avatar Choice on Creativity Score.}

We examined the impact of group-level avatar choice on creativity scores, focusing on two variables: the number of fanciful avatars and the number of humanoid avatars chosen within each pair. Statistical analysis revealed that neither variable significantly influenced creativity scores. For fanciful avatars, the result was F(2,17)=~.02, p~>~.05, $\eta^2$=~.002,  indicating no significant effect on creativity scores. Similarly, for humanoid avatars, the result was  F(2,17)=~1.11, p~>~.05, $\eta^2$=~.12, also showing no significant impact on creativity scores. Thus, the avatar style, whether fanciful or humanoid, did not significantly affect participants' performance on creativity tasks.




\subsubsection{Creative Performance Perceptions.}

Participants were queried about the task in which they felt more creative, either the first or the second. A significant majority, 30 out of 40 participants, reported feeling more creatively engaged in the second task, irrespective of the task sequence (be it energy-saving or water-saving). Three main themes emerged from the participants' explanations for this increased sense of creativity in the second task: greater comfort with their chosen avatar, increased familiarity with the study procedures, and the influence of their co-participant’s avatar.

Multiple participants felt that their ability to choose their own avatars led to a heightened sense of comfort and empowerment. For instance, one articulated this sentiment by stating, “I felt more empowered to share ideas since I was able to make a decision on how I would be perceived by the other participant” (P15). The freedom to choose their avatar facilitated a more empowered engagement in collaborative creativity. P17 found that this choice added an element of authenticity to their contributions. Similarly, P26 viewed avatar selection as a form of non-verbal communication that heightened engagement: “Choosing how to present myself made me feel a little bit more engaged with the virtual space.” Moreover, the freedom to choose an avatar appeared to create a more relaxed and creative atmosphere, as noted by P33: “In task 2, it was much more relaxed than task 1...They may not have been correct, but they were fun.”

Participants indicated that the second task allowed them to feel more relaxed and accustomed to the study procedures. One participant explained, “I had time to warm up a bit and get the creative juices flowing. Plus, some of the options I came up with in task 1 were applicable to task 2” (P14). This sense of `warming-up' was further enriched by the avatar choices, as another participant noted, “It was the second task so we already had a bit of practice in the first.  We also got to make a creative choice on something to 'embody' that let me be a little more silly” (P10). This indicates that being able to choose to embody something that lets them be more silly is in itself something of a creative choice.

An interesting insight arose around how the avatars chosen by co-participants also influenced creative performance. While participants acknowledged the importance of their own avatar choices in feeling creative, they also pointed out instances where the avatar chosen by their partner had a significant impact. One participant noted a heightened sense of compatibility and creative synergy when they and their partner accidentally chose the same avatar (P26), underscoring how avatar choices can serve as implicit signals of personal preferences. Another participant emphasized the role of avatar animation in enhancing the sense of presence and engagement, stating, “If I had to guess, it was probably my partner's avatar. I couldn't actually see mine, but seeing his move and animate felt more like an actual conversation” (P5). These observations highlight the nuanced ways in which both one's own and co-participants' avatars can significantly influence the dynamics of collaborative creativity.

\subsubsection{Avatar Perceptions.}

Participants reported a wide range of emotional responses to the avatars they were assigned, especially in contrasting environment styles for business and creativity. For instance, in a creative setting where participants embodied fox avatars, reactions ranged from endearment to strong aversion, such as P6 who stated, “I did not like the avatar, therefore I was not excited to be represented as it.” Others, like P18, expressed a sense of detachment and loss of personal identity, likening the assigned avatar to a “uniform.”

The lack of agency in avatar selection led to a general feeling of disengagement. P26 described the assigned avatar as akin to a mere “poker chip on a map,” serving only as a placeholder for physical presence rather than an embodiment of self. This sentiment was echoed in the business-style environment, where the mismatch between the avatar's and participants' gender, as noted by P27, led to feelings of alienation and sterility. Approximately 76\% of participants in both settings expressed that the avatars did not embody them well, primarily attributing this to the lack of choice.

Conversely, when given the option to choose, participants reported higher levels of engagement and satisfaction. P10 reveled in the freedom to select a whimsical avatar, stating, “My thought was oh hell yes I can be a PIZZA. A very odd and silly character because VR is supposed to be fun.” Others, like P9, felt empowered by the ability to choose an avatar that closely resembled them. About 77\% of participants felt that their chosen avatars in the second part of the study embodied them well. Even when choices were limited, the mere act of choosing enhanced the sense of embodiment and engagement, as noted by P26 who chose a “ghost” avatar for its “unassuming” and adaptable nature.

\subsubsection{Environment Perceptions.}

Participants demonstrated a marked preference for the creative environment over a business-oriented setting within the VR context. While some participants found the creative environment (Wizard's Library) to be lacking in terms of novelty, they were generally more engaged and pleased. P6, for instance, craved a more fantastical experience, stating, “Being in an area confined by the laws of physics we experience in the real world is boring. I want to live in Harry Potter.” P18 found the fantastical elements added personality to the setting, suggesting that users may desire more audacious and imaginative design features in VR spaces.

Conversely, the business environment was often described as uninspiring or even dull. Although P3 found it “standard” and functional, stating, “It got the job done well by having everything you need in the room,” this was a rare positive sentiment. Most participants criticized the Conference Room for its blandness, as exemplified by P35's comment, “It was terribly boring... the environment was almost distractingly bland,” and P40's note that the setup was “too business/cold.” At best, this setting was viewed as functional but uninspiring, with one participant mentioning it was better than Zoom, albeit this felt like faint praise in comparison to the Wizard's Library.

\subsection{Limitations}

The execution of Study Two faced several limitations, most notably in participant recruitment and equipment compatibility. The remote nature of the study, involving participants from different U.S. time zones, led to scheduling issues and absences. In some instances, researchers had to find last-minute replacements through personal networks to maintain the session's viability. Importantly, these substitutes were unique and not repeated across multiple studies. This recruitment challenge resulted in a smaller-than-desired sample size, potentially affecting the robustness of the quantitative results. Despite these limitations, the study revealed a general trend: participants showed a preference for self-selected, fanciful, non-humanoid avatars in creative settings. This preference was not only for the avatars but also seemed to positively influence their creative output in the tasks at hand.


\subsection{Study Two: Discussion}

The lack of statistically significant quantitative outcomes in Study Two did not deter the emergence of valuable insights. Although no decisive impact on creativity scores was observed based on avatar style, environment, or familiarity levels, qualitative data offered a nuanced understanding of user experience and preferences. One key takeaway was the significance of avatar choice, corroborating findings from Study One. Participants reported feeling constrained when assigned a specific avatar, even in creatively designed environments. In contrast, the freedom to choose avatars led to greater self-expression and facilitated better interpersonal interactions, resonating with Osborne~et~al.'s work on avatars as “conversation starters”~\cite{Osborne2019}.

Furthermore, participants conveyed a desire for a balanced virtual environment style—creative yet focused—which aligns with earlier studies on platforms like Second Life~(e.g.,~\cite{Guegan2017}). The data suggests a nuanced approach to VR design could involve interactive elements that enhance, rather than distract from, the primary objective of the virtual space. For instance, a magical whiteboard that follows users could facilitate brainstorming, while overly interactive elements like a shootable bow and arrow could prove disruptive.

The autonomy to choose one's avatar and environment emerged as a pivotal factor for users. It serves as a mechanism for self-presentation and a medium for signaling intentions and personality traits to others. This highlights the untapped potential of VR technology in shaping social interactions and group dynamics.

\section{General Reflections and Future Work}
\label{reflections}

\subsection{Beyond Imitation of Reality}

While imitating reality could be beneficial to some degree in VR meetings~\cite{Bailey2012, Vosinakis2013, Riordan2011}, this research shows that it is not always necessary because of both positive and negative effects on creative collaboration discussed in prior work~\cite{Peña2013, Bhagwatwar2013}. Researchers suggest that the goal of social VR mediums should not be to fully replicate reality, but rather to enable and extend existing communication channels of the physical world~\cite{Li2020}. Our findings point to the potential for pushing boundaries/expectations of what is “normal” for a work context in VR, serving more creative and social bonding purposes in less formal contexts. Guegan~et.~al’s research on creativity similarly showed that virtual workspaces in Second Life that imitated classic real-life workspaces promoted concepts of ‘rule-following’ and had a negative impact on idea generation compared to more fantastical contexts~\cite{Guegan2021}. 

While the results of Study One demonstrated that participants anticipate preferring the use of humanoid photo-realistic avatars in VR environments for business-focused activities (i.e.~interviews), the results of Study Two showed that, in practice, people may enjoy non-realistic, playful avatars in rather fantastical settings, even for business-style brainstorming. This shows how the pursuit of transferring the metaphors from the real world into VR could actually sometimes have a negative effect on productivity not only due to the specificity of the platform’s social affordances~\cite{Osborne2019, Osborne2023, McVeigh-Schultz2019, Olaosebikan2022} but also due to a generally lower fidelity of social and environmental cues that are practical to deliver in current platforms~\cite{Williamson2022}. While social VR cannot fully replace in-person meetings, virtual and hybrid teams continue adapting to new modes of technology-mediated communication, where deliberate choices of avatars and environment design offer new exciting possibilities for people to creatively express themselves, collaborate, and innovate.

\subsection{Avatar and Environment Choice Matters in Supporting Creativity}

Both studies indicated the benefits of choice, especially for avatars, aligning with recent research on avatar-mediated communication in social VR, where participants appreciated the possibility to play with their avatar appearances in order to fit into specific social contexts~\cite{Baker2021}. Although Study Two did not establish any statistical relationship between the avatar choice and the environment style, the qualitative results from both studies suggested that, on the perceptual level, participants valued the ability to control how they presented themselves. This self-presentation was influenced by a complex interplay of social context, virtual environment, and avatar appearance. Participants reported feeling more creatively engaged when they could select their own avatars, particularly in more creative, fantastical settings. Study Two did not explore the impact of environment selection on team performance, pointing to an avenue for future research, ideally involving a larger sample size to tease out the nuanced interactions between these variables.

As Gomes de Siqueira~et.~al’s research on virtual environments in Mozilla Hubs~\cite{GomesdeSiqueira2021} indicates, while teams continue to become familiar with avatar-based communication in social VR platforms, they are as well more likely to develop particular preferences for the design of virtual spaces (along with avatars) in the long term. Future research could explore how avatar affordances of particular styles in social VR platforms (e.g.,~the ability to fly, teleport on elevated surfaces, reach objects from afar, scale objects, instantly change appearances, etc.) could impact teams’ preferences for the design of workspaces to best accommodate their needs, especially for collaborative creative activities, such as brainstorming~\cite{Graessler2019, Yang2018}, embodied 3D sketching~\cite{Segura2016}, bodystorming~\cite{Segura2019, Chilufya2021, McVeigh-Schultz2024} and others.

\subsection{Design Considerations}

In light of the research findings, several design considerations can be outlined to enhance virtual team collaboration in social VR settings. First and foremost, offering a diverse range of avatar styles, from whimsical to professional, provides participants with the agency to tailor their self-presentation, resonating with our findings and previous research~\cite{Kay2004, Bhagwatwar2013, Nelson2019}. Moreover, our data suggest that the environment should be customizable to facilitate the specific goals of the meeting, whether they are creative or business-oriented. Elements such as lighting, color schemes, and interactive objects play pivotal roles in shaping the user experience.


The geometry and layout of virtual spaces are also crucial, serving as an anchor for conversations and group dynamics, a point underscored by existing literature on proxemics in VR~\cite{Williamson2021, Williamson2022}. Future research could aim to develop tools for avatar and environment modulations that are more aligned with real-world team interactions in natural settings, as opposed to controlled experimental setups. For virtual ad-hoc teams~\cite{Teevan2022}, the utility of avatar modulation tools may take a backseat to the design of the virtual environment, particularly when highly productivity-oriented activities are expected to be executed within tight project timelines.

Lastly, Study Two suggests that the avatar selection process can be transformed into a shared social experience, contributing to team building and social connections by using an in-world approach to displaying avatar options to choose from (e.g.,~see Part~2 in~Fig.~\ref{fig:Study2Stumuli}). This approach drew inspiration from the Avatar Hubs worlds in VRChat and could be beneficial for both researchers and designers aiming to leverage social VR for enhanced group cohesion and collaboration.

\section{Conclusion}

This research aimed to explore the design principles for effective social VR meeting tools, focusing on the impact of avatar styles and virtual environments on creative group performance. Utilizing Mozilla Hubs, we conducted two interconnected studies. Study One surveyed preferred avatars and environments for various VR meeting contexts~(\textit{N}=87). Building on these findings, Study Two employed between-subjects and within-subjects research that engaged participants~(\textit{N}=40) in pair-based creativity tasks in different virtual settings. The research unveiled a nuanced relationship between social contexts, avatar styles, and the design of virtual environments. Specifically, it indicated a preference for more intricate and playful self-presentation in creativity-oriented meetings. These insights could serve as valuable guidelines for researchers and designers aiming to advance the field of social VR, particularly as virtual meetings are poised to remain a staple in collaborative work and social interaction.


\begin{acks}
This material is based upon work supported by the National
Science Foundation under Grant No. 2007627 and No. 2007755.
\end{acks}

\bibliographystyle{ACM-Reference-Format}
\bibliography{bib}


\begin{thebibliography}{138}


\ifx \showCODEN    \undefined \def \showCODEN     #1{\unskip}     \fi
\ifx \showDOI      \undefined \def \showDOI       #1{#1}\fi
\ifx \showISBNx    \undefined \def \showISBNx     #1{\unskip}     \fi
\ifx \showISBNxiii \undefined \def \showISBNxiii  #1{\unskip}     \fi
\ifx \showISSN     \undefined \def \showISSN      #1{\unskip}     \fi
\ifx \showLCCN     \undefined \def \showLCCN      #1{\unskip}     \fi
\ifx \shownote     \undefined \def \shownote      #1{#1}          \fi
\ifx \showarticletitle \undefined \def \showarticletitle #1{#1}   \fi
\ifx \showURL      \undefined \def \showURL       {\relax}        \fi
\providecommand\bibfield[2]{#2}
\providecommand\bibinfo[2]{#2}
\providecommand\natexlab[1]{#1}
\providecommand\showeprint[2][]{arXiv:#2}

\bibitem[Ahn et~al\mbox{.}(2016)]%
        {Ahn2016}
\bibfield{author}{\bibinfo{person}{Sun Joo~(Grace) Ahn}, \bibinfo{person}{Joshua Bostick}, \bibinfo{person}{Elise Ogle}, \bibinfo{person}{Kristine~L. Nowak}, \bibinfo{person}{Kara~T. McGillicuddy}, {and} \bibinfo{person}{Jeremy~N. Bailenson}.} \bibinfo{year}{2016}\natexlab{}.
\newblock \showarticletitle{Experiencing {{Nature}}: {{Embodying Animals}} in {{Immersive Virtual Environments Increases Inclusion}} of {{Nature}} in {{Self}} and {{Involvement}} with {{Nature}}}.
\newblock \bibinfo{journal}{\emph{Journal of Computer-Mediated Communication}} \bibinfo{volume}{21}, \bibinfo{number}{6} (\bibinfo{date}{Nov.} \bibinfo{year}{2016}), \bibinfo{pages}{399--419}.
\newblock
\showISSN{1083-6101}
\urldef\tempurl%
\url{https://doi.org/10.1111/jcc4.12173}
\showDOI{\tempurl}


\bibitem[Applegate(2009)]%
        {Applegate2009}
\bibfield{author}{\bibinfo{person}{Rachel Applegate}.} \bibinfo{year}{2009}\natexlab{}.
\newblock \showarticletitle{The {{Library Is}} for {{Studying}}: {{Student Preferences}} for {{Study Space}}}.
\newblock \bibinfo{journal}{\emph{The Journal of Academic Librarianship}} \bibinfo{volume}{35}, \bibinfo{number}{4} (\bibinfo{date}{July} \bibinfo{year}{2009}), \bibinfo{pages}{341--346}.
\newblock
\showISSN{0099-1333}
\urldef\tempurl%
\url{https://doi.org/10.1016/j.acalib.2009.04.004}
\showDOI{\tempurl}


\bibitem[Bailenson and Beall(2006)]%
        {Bailenson2006}
\bibfield{author}{\bibinfo{person}{Jeremy~N. Bailenson} {and} \bibinfo{person}{Andrew~C. Beall}.} \bibinfo{year}{2006}\natexlab{}.
\newblock \showarticletitle{Transformed {{Social Interaction}}: {{Exploring}} the {{Digital Plasticity}} of {{Avatars}}}.
\newblock In \bibinfo{booktitle}{\emph{Avatars at {{Work}} and {{Play}}: {{Collaboration}} and {{Interaction}} in {{Shared Virtual Environments}}}}, \bibfield{editor}{\bibinfo{person}{Ralph Schroeder} {and} \bibinfo{person}{Ann-Sofie Axelsson}} (Eds.). \bibinfo{publisher}{{Springer Netherlands}}, \bibinfo{address}{{Dordrecht}}, \bibinfo{pages}{1--16}.
\newblock
\showISBNx{978-1-4020-3898-3}
\urldef\tempurl%
\url{https://doi.org/10.1007/1-4020-3898-4_1}
\showDOI{\tempurl}


\bibitem[Bailenson et~al\mbox{.}(2004)]%
        {Bailenson2004}
\bibfield{author}{\bibinfo{person}{Jeremy~N. Bailenson}, \bibinfo{person}{Andrew~C. Beall}, \bibinfo{person}{Jack Loomis}, \bibinfo{person}{Jim Blascovich}, {and} \bibinfo{person}{Matthew Turk}.} \bibinfo{year}{2004}\natexlab{}.
\newblock \showarticletitle{Transformed {{Social Interaction}}: {{Decoupling Representation}} from {{Behavior}} and {{Form}} in {{Collaborative Virtual Environments}}}.
\newblock \bibinfo{journal}{\emph{Presence: Teleoper. Virtual Environ.}} \bibinfo{volume}{13}, \bibinfo{number}{4} (\bibinfo{date}{Aug.} \bibinfo{year}{2004}), \bibinfo{pages}{428--441}.
\newblock
\showISSN{1054-7460}
\urldef\tempurl%
\url{https://doi.org/10.1162/1054746041944803}
\showDOI{\tempurl}


\bibitem[Bailey et~al\mbox{.}(2012)]%
        {Bailey2012}
\bibfield{author}{\bibinfo{person}{Diane~E. Bailey}, \bibinfo{person}{Paul~M. Leonardi}, {and} \bibinfo{person}{Stephen~R. Barley}.} \bibinfo{year}{2012}\natexlab{}.
\newblock \showarticletitle{The {{Lure}} of the {{Virtual}}}.
\newblock \bibinfo{journal}{\emph{Organization Science}} \bibinfo{volume}{23}, \bibinfo{number}{5} (\bibinfo{year}{2012}), \bibinfo{pages}{1485--1504}.
\newblock
\showISSN{1047-7039}
\urldef\tempurl%
\url{https://www.jstor.org/stable/23252319}
\showURL{%
\tempurl}


\bibitem[Baker et~al\mbox{.}(2021)]%
        {Baker2021}
\bibfield{author}{\bibinfo{person}{Steven Baker}, \bibinfo{person}{Jenny Waycott}, \bibinfo{person}{Romina Carrasco}, \bibinfo{person}{Ryan~M. Kelly}, \bibinfo{person}{Anthony~John Jones}, \bibinfo{person}{Jack Lilley}, \bibinfo{person}{Briony Dow}, \bibinfo{person}{Frances Batchelor}, \bibinfo{person}{Thuong Hoang}, {and} \bibinfo{person}{Frank Vetere}.} \bibinfo{year}{2021}\natexlab{}.
\newblock \showarticletitle{Avatar-{{Mediated Communication}} in {{Social VR}}: {{An In-depth Exploration}} of {{Older Adult Interaction}} in an {{Emerging Communication Platform}}}. In \bibinfo{booktitle}{\emph{Proceedings of the 2021 {{CHI Conference}} on {{Human Factors}} in {{Computing Systems}}}} \emph{(\bibinfo{series}{{{CHI}} '21})}. \bibinfo{publisher}{{Association for Computing Machinery}}, \bibinfo{address}{{New York, NY, USA}}, \bibinfo{pages}{1--13}.
\newblock
\showISBNx{978-1-4503-8096-6}
\urldef\tempurl%
\url{https://doi.org/10.1145/3411764.3445752}
\showDOI{\tempurl}


\bibitem[Benford et~al\mbox{.}(1995)]%
        {Benford1995}
\bibfield{author}{\bibinfo{person}{Steve Benford}, \bibinfo{person}{John Bowers}, \bibinfo{person}{Lennart~E. Fahlen}, \bibinfo{person}{Chris Greenhalgh}, {and} \bibinfo{person}{Dave Snowdon}.} \bibinfo{year}{1995}\natexlab{}.
\newblock \showarticletitle{User Embodiment in Collaborative Virtual Environments}. In \bibinfo{booktitle}{\emph{Conference on {{Human Factors}} in {{Computing Systems}} - {{Proceedings}}}}, Vol.~\bibinfo{volume}{1}. \bibinfo{publisher}{{ACM}}, \bibinfo{pages}{242--249}.
\newblock
\urldef\tempurl%
\url{https://doi.org/10.1145/223904.223935}
\showDOI{\tempurl}


\bibitem[Benford et~al\mbox{.}(2000)]%
        {Benford2000}
\bibfield{author}{\bibinfo{person}{Steve Benford}, \bibinfo{person}{Paul Dourish}, {and} \bibinfo{person}{Tom Rodden}.} \bibinfo{year}{2000}\natexlab{}.
\newblock \showarticletitle{Introduction to the Special Issue on Human-Computer Interaction and Collaborative Virtual Environments}.
\newblock \bibinfo{journal}{\emph{ACM Transactions on Computer-Human Interaction (TOCHI)}} \bibinfo{volume}{7}, \bibinfo{number}{4} (\bibinfo{date}{Dec.} \bibinfo{year}{2000}), \bibinfo{pages}{439--441}.
\newblock
\showISSN{15577325}
\urldef\tempurl%
\url{https://doi.org/10.1145/365058.365080}
\showDOI{\tempurl}


\bibitem[Benford et~al\mbox{.}(2001)]%
        {Benford2001}
\bibfield{author}{\bibinfo{person}{Steve Benford}, \bibinfo{person}{Chris Greenhalgh}, \bibinfo{person}{Tom Rodden}, {and} \bibinfo{person}{James Pycock}.} \bibinfo{year}{2001}\natexlab{}.
\newblock \showarticletitle{Collaborative {{Virtual Environments}}}.
\newblock \bibinfo{journal}{\emph{Commun. ACM}} \bibinfo{volume}{44}, \bibinfo{number}{7} (\bibinfo{date}{July} \bibinfo{year}{2001}), \bibinfo{pages}{79--85}.
\newblock
\showISSN{00010782}
\urldef\tempurl%
\url{https://doi.org/10.1145/379300.379322}
\showDOI{\tempurl}


\bibitem[Bhagwatwar et~al\mbox{.}(2013)]%
        {Bhagwatwar2013}
\bibfield{author}{\bibinfo{person}{Akshay Bhagwatwar}, \bibinfo{person}{Anne Massey}, {and} \bibinfo{person}{Alan~R. Dennis}.} \bibinfo{year}{2013}\natexlab{}.
\newblock \showarticletitle{Creative {{Virtual Environments}}: {{Effect}} of {{Supraliminal Priming}} on {{Team Brainstorming}}}.
\newblock \bibinfo{journal}{\emph{2013 46th Hawaii International Conference on System Sciences}}  \bibinfo{volume}{1} (\bibinfo{date}{Jan.} \bibinfo{year}{2013}), \bibinfo{pages}{215--224}.
\newblock
\showISBNx{978-1-4673-5933-7}
\showISSN{1530-1605}
\urldef\tempurl%
\url{https://doi.org/10.1109/HICSS.2013.152}
\showDOI{\tempurl}


\bibitem[Bleakley et~al\mbox{.}(2020)]%
        {Bleakley2020}
\bibfield{author}{\bibinfo{person}{Anna Bleakley}, \bibinfo{person}{Vincent Wade}, {and} \bibinfo{person}{Benjamin~R Cowan}.} \bibinfo{year}{2020}\natexlab{}.
\newblock \showarticletitle{Finally a {{Case}} for {{Collaborative VR}}? {{The Need}} to {{Design}} for {{Remote Multi-Party Conversations}}}. In \bibinfo{booktitle}{\emph{Proceedings of the 2nd {{Conference}} on {{Conversational User Interfaces}}}}. \bibinfo{publisher}{{Association for Computing Machinery}}, \bibinfo{pages}{1--3}.
\newblock
\showISBNx{978-1-4503-7544-3}
\urldef\tempurl%
\url{https://doi.org/10.1145/3405755.3406144}
\showDOI{\tempurl}


\bibitem[Bogicevic et~al\mbox{.}(2018)]%
        {Bogicevic2018}
\bibfield{author}{\bibinfo{person}{Vanja Bogicevic}, \bibinfo{person}{Milos Bujisic}, \bibinfo{person}{Cihan Cobanoglu}, {and} \bibinfo{person}{Andrew~Hale Feinstein}.} \bibinfo{year}{2018}\natexlab{}.
\newblock \showarticletitle{Gender and Age Preferences of Hotel Room Design}.
\newblock \bibinfo{journal}{\emph{International Journal of Contemporary Hospitality Management}} \bibinfo{volume}{30}, \bibinfo{number}{2} (\bibinfo{date}{Jan.} \bibinfo{year}{2018}), \bibinfo{pages}{874--899}.
\newblock
\showISSN{0959-6119}
\urldef\tempurl%
\url{https://doi.org/10.1108/IJCHM-08-2016-0450}
\showDOI{\tempurl}


\bibitem[{Bourgeois-Bougrine} et~al\mbox{.}(2020)]%
        {Bourgeois-Bougrine2020a}
\bibfield{author}{\bibinfo{person}{Samira {Bourgeois-Bougrine}}, \bibinfo{person}{Peter Richard}, \bibinfo{person}{Jean-Marie Burkhardt}, \bibinfo{person}{Benjamin Frantz}, {and} \bibinfo{person}{Todd Lubart}.} \bibinfo{year}{2020}\natexlab{}.
\newblock \showarticletitle{The {{Expression}} of {{Users}}' {{Creative Potential}} in {{Virtual}} and {{Real Environments}}: {{An Exploratory Study}}}.
\newblock \bibinfo{journal}{\emph{Creativity Research Journal}} \bibinfo{volume}{32}, \bibinfo{number}{1} (\bibinfo{date}{Jan.} \bibinfo{year}{2020}), \bibinfo{pages}{55--65}.
\newblock
\showISSN{1040-0419}
\urldef\tempurl%
\url{https://doi.org/10.1080/10400419.2020.1712162}
\showDOI{\tempurl}


\bibitem[Bredikhina et~al\mbox{.}(2020)]%
        {Bredikhina2020}
\bibfield{author}{\bibinfo{person}{Liudmila Bredikhina}, \bibinfo{person}{Toya Sakaguchi}, {and} \bibinfo{person}{Akihiko Shirai}.} \bibinfo{year}{2020}\natexlab{}.
\newblock \showarticletitle{{{Web3D Distance}} Live Workshop for Children in {{Mozilla Hubs}}}.
\newblock \bibinfo{journal}{\emph{Proceedings - Web3D 2020: 25th ACM Conference on 3D Web Technology}} (\bibinfo{date}{Nov.} \bibinfo{year}{2020}).
\newblock
\urldef\tempurl%
\url{https://doi.org/10.1145/3424616.3424724}
\showDOI{\tempurl}


\bibitem[Brucks and Levav(2022)]%
        {Brucks2022}
\bibfield{author}{\bibinfo{person}{Melanie~S. Brucks} {and} \bibinfo{person}{Jonathan Levav}.} \bibinfo{year}{2022}\natexlab{}.
\newblock \showarticletitle{Virtual Communication Curbs Creative Idea Generation}.
\newblock \bibinfo{journal}{\emph{Nature}} \bibinfo{volume}{605}, \bibinfo{number}{7908} (\bibinfo{date}{May} \bibinfo{year}{2022}), \bibinfo{pages}{108--112}.
\newblock
\showISSN{1476-4687}
\urldef\tempurl%
\url{https://doi.org/10.1038/s41586-022-04643-y}
\showDOI{\tempurl}


\bibitem[Buisine and Guegan(2020)]%
        {Buisine2020}
\bibfield{author}{\bibinfo{person}{St{\'e}phanie Buisine} {and} \bibinfo{person}{J{\'e}r{\^o}me Guegan}.} \bibinfo{year}{2020}\natexlab{}.
\newblock \showarticletitle{Proteus vs. Social Identity Effects on Virtual Brainstorming}.
\newblock \bibinfo{journal}{\emph{Behaviour \& Information Technology}} \bibinfo{volume}{39}, \bibinfo{number}{5} (\bibinfo{date}{May} \bibinfo{year}{2020}), \bibinfo{pages}{594--606}.
\newblock
\showISSN{0144-929X}
\urldef\tempurl%
\url{https://doi.org/10.1080/0144929X.2019.1605408}
\showDOI{\tempurl}


\bibitem[Buisine et~al\mbox{.}(2016)]%
        {Buisine2016}
\bibfield{author}{\bibinfo{person}{St{\'e}phanie Buisine}, \bibinfo{person}{J{\'e}r{\^o}me Guegan}, \bibinfo{person}{Jessy Barr{\'e}}, \bibinfo{person}{Fr{\'e}d{\'e}ric Segonds}, {and} \bibinfo{person}{Am{\'e}ziane Aoussat}.} \bibinfo{year}{2016}\natexlab{}.
\newblock \showarticletitle{Using Avatars to Tailor Ideation Process to Innovation Strategy}.
\newblock \bibinfo{journal}{\emph{Cogn Tech Work}} \bibinfo{volume}{18}, \bibinfo{number}{3} (\bibinfo{date}{Aug.} \bibinfo{year}{2016}), \bibinfo{pages}{583--594}.
\newblock
\showISSN{1435-5566}
\urldef\tempurl%
\url{https://doi.org/10.1007/s10111-016-0378-y}
\showDOI{\tempurl}


\bibitem[Casaneuva(2001)]%
        {Casaneuva2001}
\bibfield{author}{\bibinfo{person}{Juan~S Casaneuva}.} \bibinfo{year}{2001}\natexlab{}.
\newblock \emph{\bibinfo{title}{Presence and Co-Presence in Collaborative Virtual Environments}}.
\newblock \bibinfo{thesistype}{Ph.\,D. Dissertation}. \bibinfo{school}{University of Cape Town}.
\newblock
\urldef\tempurl%
\url{https://open.uct.ac.za/handle/11427/6383}
\showURL{%
\tempurl}


\bibitem[Ceylan et~al\mbox{.}(2008)]%
        {Ceylan2008}
\bibfield{author}{\bibinfo{person}{Canan Ceylan}, \bibinfo{person}{Jan Dul}, {and} \bibinfo{person}{Serpil Aytac}.} \bibinfo{year}{2008}\natexlab{}.
\newblock \showarticletitle{Can the Office Environment Stimulate a Manager's Creativity?}
\newblock \bibinfo{journal}{\emph{Human Factors and Ergonomics in Manufacturing \& Service Industries}} \bibinfo{volume}{18}, \bibinfo{number}{6} (\bibinfo{year}{2008}), \bibinfo{pages}{589--602}.
\newblock
\showISSN{1520-6564}
\urldef\tempurl%
\url{https://doi.org/10.1002/hfm.20128}
\showDOI{\tempurl}


\bibitem[Chilufya and Arvola(2021)]%
        {Chilufya2021}
\bibfield{author}{\bibinfo{person}{Emma~Mainza Chilufya} {and} \bibinfo{person}{Mattias Arvola}.} \bibinfo{year}{2021}\natexlab{}.
\newblock \showarticletitle{Conceptual {{Designing}} of a {{Virtual Receptionist}}: {{Remote Desktop Walkthrough}} and {{Bodystorming}} in {{VR}}}. In \bibinfo{booktitle}{\emph{Proceedings of the 9th {{International Conference}} on {{Human-Agent Interaction}}}} \emph{(\bibinfo{series}{{{HAI}} '21})}. \bibinfo{publisher}{{Association for Computing Machinery}}, \bibinfo{address}{{New York, NY, USA}}, \bibinfo{pages}{112--120}.
\newblock
\showISBNx{978-1-4503-8620-3}
\urldef\tempurl%
\url{https://doi.org/10.1145/3472307.3484171}
\showDOI{\tempurl}


\bibitem[Coburn et~al\mbox{.}(2017)]%
        {Coburn2017}
\bibfield{author}{\bibinfo{person}{Joshua~Q. Coburn}, \bibinfo{person}{Ian Freeman}, {and} \bibinfo{person}{John~L. Salmon}.} \bibinfo{year}{2017}\natexlab{}.
\newblock \showarticletitle{A {{Review}} of the {{Capabilities}} of {{Current Low-Cost Virtual Reality Technology}} and {{Its Potential}} to {{Enhance}} the {{Design Process}}}.
\newblock \bibinfo{journal}{\emph{Journal of Computing and Information Science in Engineering}} \bibinfo{volume}{17}, \bibinfo{number}{3} (\bibinfo{date}{July} \bibinfo{year}{2017}).
\newblock
\showISSN{1530-9827}
\urldef\tempurl%
\url{https://doi.org/10.1115/1.4036921}
\showDOI{\tempurl}


\bibitem[{de Rooij} et~al\mbox{.}(2017)]%
        {Rooij2017}
\bibfield{author}{\bibinfo{person}{Alwin {de Rooij}}, \bibinfo{person}{Sarah {van der Land}}, {and} \bibinfo{person}{Shelly {van Erp}}.} \bibinfo{year}{2017}\natexlab{}.
\newblock \showarticletitle{The {{Creative Proteus Effect}}: {{How Self-Similarity}}, {{Embodiment}}, and {{Priming}} of {{Creative Stereotypes}} with {{Avatars Influences Creative Ideation}}}. In \bibinfo{booktitle}{\emph{Proceedings of the 2017 {{ACM SIGCHI Conference}} on {{Creativity}} and {{Cognition}}}} \emph{(\bibinfo{series}{C\&amp;{{C}} '17})}. \bibinfo{publisher}{{Association for Computing Machinery}}, \bibinfo{address}{{New York, NY, USA}}, \bibinfo{pages}{232--236}.
\newblock
\showISBNx{978-1-4503-4403-6}
\urldef\tempurl%
\url{https://doi.org/10.1145/3059454.3078856}
\showDOI{\tempurl}


\bibitem[Dey et~al\mbox{.}(2017)]%
        {Dey2017}
\bibfield{author}{\bibinfo{person}{Arindam Dey}, \bibinfo{person}{Thammathip Piumsomboon}, \bibinfo{person}{Youngho Lee}, {and} \bibinfo{person}{Mark Billinghurst}.} \bibinfo{year}{2017}\natexlab{}.
\newblock \showarticletitle{Effects of {{Sharing Physiological States}} of {{Players}} in a {{Collaborative Virtual Reality Gameplay}}}. In \bibinfo{booktitle}{\emph{Proceedings of the 2017 {{CHI Conference}} on {{Human Factors}} in {{Computing Systems}}}} \emph{(\bibinfo{series}{{{CHI}} '17})}. \bibinfo{publisher}{{Association for Computing Machinery}}, \bibinfo{address}{{New York, NY, USA}}, \bibinfo{pages}{4045--4056}.
\newblock
\showISBNx{978-1-4503-4655-9}
\urldef\tempurl%
\url{https://doi.org/10.1145/3025453.3026028}
\showDOI{\tempurl}


\bibitem[Dobre et~al\mbox{.}(2022)]%
        {Dobre2022}
\bibfield{author}{\bibinfo{person}{Georgiana~Cristina Dobre}, \bibinfo{person}{Marta Wilczkowiak}, \bibinfo{person}{Marco Gillies}, \bibinfo{person}{Xueni Pan}, {and} \bibinfo{person}{Sean Rintel}.} \bibinfo{year}{2022}\natexlab{}.
\newblock \showarticletitle{Nice Is {{Different}} than {{Good}}: {{Longitudinal Communicative Effects}} of {{Realistic}} and {{Cartoon Avatars}} in {{Real Mixed Reality Work Meetings}}}. In \bibinfo{booktitle}{\emph{Extended {{Abstracts}} of the 2022 {{CHI Conference}} on {{Human Factors}} in {{Computing Systems}}}} \emph{(\bibinfo{series}{{{CHI EA}} '22})}. \bibinfo{publisher}{{Association for Computing Machinery}}, \bibinfo{address}{{New York, NY, USA}}, \bibinfo{pages}{1--7}.
\newblock
\showISBNx{978-1-4503-9156-6}
\urldef\tempurl%
\url{https://doi.org/10.1145/3491101.3519628}
\showDOI{\tempurl}


\bibitem[Ducheneaut et~al\mbox{.}(2009)]%
        {Ducheneaut2009}
\bibfield{author}{\bibinfo{person}{Nicolas Ducheneaut}, \bibinfo{person}{Ming-Hui Wen}, \bibinfo{person}{Nicholas Yee}, {and} \bibinfo{person}{Greg Wadley}.} \bibinfo{year}{2009}\natexlab{}.
\newblock \showarticletitle{Body and {{Mind}}: A {{Study}} of {{Avatar Personalization}} in {{Three Virtual Worlds}}}. In \bibinfo{booktitle}{\emph{Proceedings of the 27th International Conference on {{Human}} Factors in Computing Systems - {{CHI}} 09}}. \bibinfo{publisher}{{ACM Press}}, \bibinfo{address}{{Boston, MA, USA}}, \bibinfo{pages}{1151}.
\newblock
\showISBNx{978-1-60558-246-7}
\urldef\tempurl%
\url{https://doi.org/10.1145/1518701.1518877}
\showDOI{\tempurl}


\bibitem[El~Ali et~al\mbox{.}(2021)]%
        {ElAli2021}
\bibfield{author}{\bibinfo{person}{Abdallah El~Ali}, \bibinfo{person}{Monica {Perusquia-Hernandez}}, \bibinfo{person}{Mariam Hassib}, \bibinfo{person}{Yomna Abdelrahman}, {and} \bibinfo{person}{Joshua Newn}.} \bibinfo{year}{2021}\natexlab{}.
\newblock \showarticletitle{{{MEEC}}: {{Second Workshop}} on {{Momentary Emotion Elicitation}} and {{Capture}}}.
\newblock \bibinfo{journal}{\emph{Conference on Human Factors in Computing Systems - Proceedings}} (\bibinfo{date}{May} \bibinfo{year}{2021}).
\newblock
\urldef\tempurl%
\url{https://doi.org/10.1145/3411763.3441351}
\showDOI{\tempurl}


\bibitem[Eriksson(2021)]%
        {Eriksson2021}
\bibfield{author}{\bibinfo{person}{Thommy Eriksson}.} \bibinfo{year}{2021}\natexlab{}.
\newblock \showarticletitle{Failure and {{Success}} in {{Using Mozilla Hubs}} for {{Online Teaching}} in a {{Movie Production Course}}}.
\newblock \bibinfo{journal}{\emph{2021 7th International Conference of the Immersive Learning Research Network (iLRN)}} (\bibinfo{date}{June} \bibinfo{year}{2021}), \bibinfo{pages}{1--8}.
\newblock
\urldef\tempurl%
\url{https://doi.org/10.23919/ILRN52045.2021.9459321}
\showDOI{\tempurl}


\bibitem[Felnhofer et~al\mbox{.}(2015)]%
        {Felnhofer2015}
\bibfield{author}{\bibinfo{person}{Anna Felnhofer}, \bibinfo{person}{Oswald~D. Kothgassner}, \bibinfo{person}{Mareike Schmidt}, \bibinfo{person}{Anna-Katharina Heinzle}, \bibinfo{person}{Leon Beutl}, \bibinfo{person}{Helmut Hlavacs}, {and} \bibinfo{person}{Ilse {Kryspin-Exner}}.} \bibinfo{year}{2015}\natexlab{}.
\newblock \showarticletitle{Is Virtual Reality Emotionally Arousing? {{Investigating}} Five Emotion Inducing Virtual Park Scenarios}.
\newblock \bibinfo{journal}{\emph{International Journal of Human-Computer Studies}}  \bibinfo{volume}{82} (\bibinfo{date}{Oct.} \bibinfo{year}{2015}), \bibinfo{pages}{48--56}.
\newblock
\showISSN{1071-5819}
\urldef\tempurl%
\url{https://doi.org/10.1016/j.ijhcs.2015.05.004}
\showDOI{\tempurl}


\bibitem[Freeman and Maloney(2021)]%
        {Freeman2021}
\bibfield{author}{\bibinfo{person}{Guo Freeman} {and} \bibinfo{person}{Divine Maloney}.} \bibinfo{year}{2021}\natexlab{}.
\newblock \showarticletitle{Body, {{Avatar}}, and {{Me}}: {{The Presentation}} and {{Perception}} of {{Self}} in {{Social Virtual Reality}}}.
\newblock \bibinfo{journal}{\emph{Proceedings of the ACM on Human-Computer Interaction}} \bibinfo{volume}{4}, \bibinfo{number}{CSCW3} (\bibinfo{year}{2021}), \bibinfo{pages}{239:1--239:27}.
\newblock
\urldef\tempurl%
\url{https://doi.org/10.1145/3432938}
\showDOI{\tempurl}


\bibitem[Freiwald et~al\mbox{.}(2021)]%
        {Freiwald2021}
\bibfield{author}{\bibinfo{person}{Jann~Philipp Freiwald}, \bibinfo{person}{Julius Schenke}, \bibinfo{person}{Nale {Lehmann-Willenbrock}}, {and} \bibinfo{person}{Frank Steinicke}.} \bibinfo{year}{2021}\natexlab{}.
\newblock \showarticletitle{Effects of {{Avatar Appearance}} and {{Locomotion}} on {{Co-Presence}} in {{Virtual Reality Collaborations}}}. In \bibinfo{booktitle}{\emph{Mensch Und {{Computer}} 2021}} \emph{(\bibinfo{series}{{{MuC}} '21})}. \bibinfo{publisher}{{Association for Computing Machinery}}, \bibinfo{address}{{New York, NY, USA}}, \bibinfo{pages}{393--401}.
\newblock
\showISBNx{978-1-4503-8645-6}
\urldef\tempurl%
\url{https://doi.org/10.1145/3473856.3473870}
\showDOI{\tempurl}


\bibitem[Friedman et~al\mbox{.}(2003)]%
        {Friedman2003}
\bibfield{author}{\bibinfo{person}{Ronald~S. Friedman}, \bibinfo{person}{Ayelet Fishbach}, \bibinfo{person}{Jens F{\"o}rster}, {and} \bibinfo{person}{Lioba Werth}.} \bibinfo{year}{2003}\natexlab{}.
\newblock \showarticletitle{Attentional {{Priming Effects}} on {{Creativity}}}.
\newblock \bibinfo{journal}{\emph{Creativity Research Journal}} \bibinfo{volume}{15}, \bibinfo{number}{2-3} (\bibinfo{date}{July} \bibinfo{year}{2003}), \bibinfo{pages}{277--286}.
\newblock
\showISSN{1040-0419, 1532-6934}
\urldef\tempurl%
\url{https://doi.org/10.1080/10400419.2003.9651420}
\showDOI{\tempurl}


\bibitem[Gaver(2012)]%
        {Gaver2012}
\bibfield{author}{\bibinfo{person}{William Gaver}.} \bibinfo{year}{2012}\natexlab{}.
\newblock \showarticletitle{What {{Should We Expect From Research Through Design}}?}
\newblock \bibinfo{journal}{\emph{Proceedings of the SIGCHI Conference on Human Factors in Computing Systems}} (\bibinfo{year}{2012}).
\newblock
\showISBNx{9781450310154}
\urldef\tempurl%
\url{https://doi.org/10.1145/2207676}
\showDOI{\tempurl}


\bibitem[{Gomes de Siqueira} et~al\mbox{.}(2021)]%
        {GomesdeSiqueira2021}
\bibfield{author}{\bibinfo{person}{Alexandre {Gomes de Siqueira}}, \bibinfo{person}{Pedro~Guillermo {Feij{\'o}o-Garc{\'i}a}}, \bibinfo{person}{Jacob Stuart}, {and} \bibinfo{person}{Benjamin Lok}.} \bibinfo{year}{2021}\natexlab{}.
\newblock \showarticletitle{Toward {{Facilitating Team Formation}} and {{Communication Through Avatar Based Interaction}} in {{Desktop-Based Immersive Virtual Environments}}}.
\newblock \bibinfo{journal}{\emph{Frontiers in Virtual Reality}}  \bibinfo{volume}{2} (\bibinfo{year}{2021}).
\newblock
\showISSN{2673-4192}
\urldef\tempurl%
\url{https://www.frontiersin.org/articles/10.3389/frvir.2021.647801}
\showURL{%
\tempurl}


\bibitem[Graessler and Taplick(2019)]%
        {Graessler2019}
\bibfield{author}{\bibinfo{person}{Iris Graessler} {and} \bibinfo{person}{Patrick Taplick}.} \bibinfo{year}{2019}\natexlab{}.
\newblock \showarticletitle{Supporting {{Creativity}} with {{Virtual Reality Technology}}}.
\newblock \bibinfo{journal}{\emph{Proceedings of the Design Society: International Conference on Engineering Design}} \bibinfo{volume}{1}, \bibinfo{number}{1} (\bibinfo{date}{July} \bibinfo{year}{2019}), \bibinfo{pages}{2011--2020}.
\newblock
\showISSN{2220-4342}
\urldef\tempurl%
\url{https://doi.org/10.1017/dsi.2019.207}
\showDOI{\tempurl}


\bibitem[Greeno(1994)]%
        {Greeno1994}
\bibfield{author}{\bibinfo{person}{James~G. Greeno}.} \bibinfo{year}{1994}\natexlab{}.
\newblock \showarticletitle{Gibson's {{Affordances}}}.
\newblock \bibinfo{journal}{\emph{Psychological Review}} \bibinfo{volume}{101}, \bibinfo{number}{2} (\bibinfo{year}{1994}), \bibinfo{pages}{336--342}.
\newblock
\showISSN{0033295X}
\urldef\tempurl%
\url{https://doi.org/10.1037/0033-295X.101.2.336}
\showDOI{\tempurl}


\bibitem[Guan et~al\mbox{.}(2021)]%
        {Guan2021}
\bibfield{author}{\bibinfo{person}{Jue-Qi Guan}, \bibinfo{person}{Liang-Hui Wang}, \bibinfo{person}{Qu Chen}, \bibinfo{person}{Kai Jin}, {and} \bibinfo{person}{Gwo-Jen Hwang}.} \bibinfo{year}{2021}\natexlab{}.
\newblock \showarticletitle{Effects of a Virtual Reality-Based Pottery Making Approach on Junior High School Students' Creativity and Learning Engagement}.
\newblock \bibinfo{journal}{\emph{Interactive Learning Environments}} (\bibinfo{date}{Jan.} \bibinfo{year}{2021}), \bibinfo{pages}{1--17}.
\newblock
\showISSN{1049-4820, 1744-5191}
\urldef\tempurl%
\url{https://doi.org/10.1080/10494820.2021.1871631}
\showDOI{\tempurl}


\bibitem[Guegan et~al\mbox{.}(2021)]%
        {Guegan2021}
\bibfield{author}{\bibinfo{person}{J{\'e}r{\^o}me Guegan}, \bibinfo{person}{Claire Brechet}, {and} \bibinfo{person}{Julien Nelson}.} \bibinfo{year}{2021}\natexlab{}.
\newblock \showarticletitle{Dreamlike and {{Playful Virtual Environments}} to {{Inspire}} {{Children}}'s {{Divergent Thinking}}}.
\newblock \bibinfo{journal}{\emph{Journal of Media Psychology}} \bibinfo{volume}{33}, \bibinfo{number}{1} (\bibinfo{date}{Jan.} \bibinfo{year}{2021}), \bibinfo{pages}{28--38}.
\newblock
\showISSN{1864-1105}
\urldef\tempurl%
\url{https://doi.org/10.1027/1864-1105/a000279}
\showDOI{\tempurl}


\bibitem[Guegan et~al\mbox{.}(2016)]%
        {Guegan2016}
\bibfield{author}{\bibinfo{person}{J{\'e}r{\^o}me Guegan}, \bibinfo{person}{St{\'e}phanie Buisine}, \bibinfo{person}{Fabrice Mantelet}, \bibinfo{person}{Nicolas Maranzana}, {and} \bibinfo{person}{Fr{\'e}d{\'e}ric Segonds}.} \bibinfo{year}{2016}\natexlab{}.
\newblock \showarticletitle{Avatar-Mediated Creativity: {{When}} Embodying Inventors Makes Engineers More Creative}.
\newblock \bibinfo{journal}{\emph{Computers in Human Behavior}}  \bibinfo{volume}{61} (\bibinfo{date}{Aug.} \bibinfo{year}{2016}), \bibinfo{pages}{165--175}.
\newblock
\showISSN{0747-5632}
\urldef\tempurl%
\url{https://doi.org/10.1016/j.chb.2016.03.024}
\showDOI{\tempurl}


\bibitem[Guegan et~al\mbox{.}(2017a)]%
        {Guegan2017}
\bibfield{author}{\bibinfo{person}{J{\'e}r{\^o}me Guegan}, \bibinfo{person}{Julien Nelson}, {and} \bibinfo{person}{Todd Lubart}.} \bibinfo{year}{2017}\natexlab{a}.
\newblock \showarticletitle{The {{Relationship Between Contextual Cues}} in {{Virtual Environments}} and {{Creative Processes}}}.
\newblock \bibinfo{journal}{\emph{Cyberpsychology, Behavior, and Social Networking}} \bibinfo{volume}{20}, \bibinfo{number}{3} (\bibinfo{date}{March} \bibinfo{year}{2017}), \bibinfo{pages}{202--206}.
\newblock
\urldef\tempurl%
\url{https://doi.org/10.1089/cyber.2016.0503}
\showDOI{\tempurl}


\bibitem[Guegan et~al\mbox{.}(2017b)]%
        {Guegan2017a}
\bibfield{author}{\bibinfo{person}{J{\'e}r{\^o}me Guegan}, \bibinfo{person}{Fr{\'e}d{\'e}ric Segonds}, \bibinfo{person}{Jessy Barr{\'e}}, \bibinfo{person}{Nicolas Maranzana}, \bibinfo{person}{Fabrice Mantelet}, {and} \bibinfo{person}{St{\'e}phanie Buisine}.} \bibinfo{year}{2017}\natexlab{b}.
\newblock \showarticletitle{Social Identity Cues to Improve Creativity and Identification in Face-to-Face and Virtual Groups}.
\newblock \bibinfo{journal}{\emph{Computers in Human Behavior}}  \bibinfo{volume}{77} (\bibinfo{date}{Dec.} \bibinfo{year}{2017}), \bibinfo{pages}{140--147}.
\newblock
\showISSN{0747-5632}
\urldef\tempurl%
\url{https://doi.org/10.1016/j.chb.2017.08.043}
\showDOI{\tempurl}


\bibitem[Handley et~al\mbox{.}(2022)]%
        {Handley2022}
\bibfield{author}{\bibinfo{person}{Ryan Handley}, \bibinfo{person}{Bert Guerra}, \bibinfo{person}{Rukkmini Goli}, {and} \bibinfo{person}{Douglas Zytko}.} \bibinfo{year}{2022}\natexlab{}.
\newblock \showarticletitle{Designing {{Social VR}}: {{A Collection}} of {{Design Choices Across Commercial}} and {{Research Applications}}}.
\newblock \bibinfo{journal}{\emph{arXiv}}  \bibinfo{volume}{4} (\bibinfo{date}{Jan.} \bibinfo{year}{2022}), \bibinfo{pages}{1--14}.
\newblock
\urldef\tempurl%
\url{https://doi.org/10.48550/arXiv.2201.02253}
\showDOI{\tempurl}


\bibitem[Heeter(1992)]%
        {Heeter1992}
\bibfield{author}{\bibinfo{person}{Carrie Heeter}.} \bibinfo{year}{1992}\natexlab{}.
\newblock \showarticletitle{Being {{There}}: {{The Subjective Experience}} of {{Presence}}}.
\newblock \bibinfo{journal}{\emph{Presence: Teleoperators and Virtual Environments}} \bibinfo{volume}{1}, \bibinfo{number}{2} (\bibinfo{date}{May} \bibinfo{year}{1992}), \bibinfo{pages}{262--271}.
\newblock
\urldef\tempurl%
\url{https://doi.org/10.1162/pres.1992.1.2.262}
\showDOI{\tempurl}


\bibitem[Heidicker et~al\mbox{.}(2017)]%
        {Heidicker2017}
\bibfield{author}{\bibinfo{person}{Paul Heidicker}, \bibinfo{person}{Eike Langbehn}, {and} \bibinfo{person}{Frank Steinicke}.} \bibinfo{year}{2017}\natexlab{}.
\newblock \showarticletitle{Influence of Avatar Appearance on Presence in Social {{VR}}}.
\newblock \bibinfo{journal}{\emph{IEEE Computer Society}} (\bibinfo{date}{Jan.} \bibinfo{year}{2017}), \bibinfo{pages}{233--234}.
\newblock
\showISBNx{978-1-5090-6716-9}
\urldef\tempurl%
\url{https://doi.org/10.1109/3DUI.2017.7893357}
\showDOI{\tempurl}


\bibitem[Heydarian et~al\mbox{.}(2015)]%
        {Heydarian2015}
\bibfield{author}{\bibinfo{person}{Arsalan Heydarian}, \bibinfo{person}{Evangelos Pantazis}, \bibinfo{person}{Joao~P. Carneiro}, \bibinfo{person}{David Gerber}, {and} \bibinfo{person}{{Burcin Becerik-Gerber}}.} \bibinfo{year}{2015}\natexlab{}.
\newblock \showarticletitle{Towards Understanding End-User Lighting Preferences in Office Spaces by Using Immersive Virtual Environments}.
\newblock In \bibinfo{booktitle}{\emph{Computing in Civil Engineering 2015}}. \bibinfo{pages}{475--482}.
\newblock
\urldef\tempurl%
\url{https://doi.org/10.1061/9780784479247.059}
\showDOI{\tempurl}
\showeprint{https://ascelibrary.org/doi/pdf/10.1061/9780784479247.059}


\bibitem[Hunter et~al\mbox{.}(2007)]%
        {Hunter2007}
\bibfield{author}{\bibinfo{person}{Samuel~T. Hunter}, \bibinfo{person}{Katrina~E. Bedell}, {and} \bibinfo{person}{Michael~D. Mumford}.} \bibinfo{year}{2007}\natexlab{}.
\newblock \showarticletitle{Climate for {{Creativity}}: {{A Quantitative Review}}}.
\newblock \bibinfo{journal}{\emph{Creativity Research Journal}} \bibinfo{volume}{19}, \bibinfo{number}{1} (\bibinfo{date}{May} \bibinfo{year}{2007}), \bibinfo{pages}{69--90}.
\newblock
\showISSN{1040-0419, 1532-6934}
\urldef\tempurl%
\url{https://doi.org/10.1080/10400410709336883}
\showDOI{\tempurl}


\bibitem[Ichino et~al\mbox{.}(2022)]%
        {Ichino2022}
\bibfield{author}{\bibinfo{person}{Junko Ichino}, \bibinfo{person}{Masahiro Ide}, \bibinfo{person}{Hitomi Yokoyama}, \bibinfo{person}{Hirotoshi Asano}, \bibinfo{person}{Hideo Miyachi}, {and} \bibinfo{person}{Daisuke Okabe}.} \bibinfo{year}{2022}\natexlab{}.
\newblock \showarticletitle{"{{I}}'ve Talked without Intending to": {{Self-disclosure}} and {{Reciprocity}} via {{Embodied Avatar}}}.
\newblock \bibinfo{journal}{\emph{Proc. ACM Hum.-Comput. Interact.}} \bibinfo{volume}{6}, \bibinfo{number}{CSCW2} (\bibinfo{date}{Nov.} \bibinfo{year}{2022}), \bibinfo{pages}{482:1--482:23}.
\newblock
\urldef\tempurl%
\url{https://doi.org/10.1145/3555583}
\showDOI{\tempurl}


\bibitem[Inkpen and Sedlins(2011)]%
        {Inkpen2011}
\bibfield{author}{\bibinfo{person}{Kori~M. Inkpen} {and} \bibinfo{person}{Mara Sedlins}.} \bibinfo{year}{2011}\natexlab{}.
\newblock \showarticletitle{Me and My Avatar: Exploring Users' Comfort with Avatars for Workplace Communication}. In \bibinfo{booktitle}{\emph{Proceedings of the {{ACM}} 2011 Conference on {{Computer}} Supported Cooperative Work}} \emph{(\bibinfo{series}{{{CSCW}} '11})}. \bibinfo{publisher}{{Association for Computing Machinery}}, \bibinfo{address}{{New York, NY, USA}}, \bibinfo{pages}{383--386}.
\newblock
\showISBNx{978-1-4503-0556-3}
\urldef\tempurl%
\url{https://doi.org/10.1145/1958824.1958883}
\showDOI{\tempurl}


\bibitem[Isbister(2006)]%
        {Isbister2006}
\bibfield{author}{\bibinfo{person}{Katherine Isbister}.} \bibinfo{year}{2006}\natexlab{}.
\newblock \bibinfo{booktitle}{\emph{Better {{Game Characters}} by {{Design}}: {{A Psychological Approach}}}}.
\newblock \bibinfo{publisher}{{Elsevier/Morgan Kaufmann Series in Interactive 3D Technology}}.
\newblock
\showISBNx{1-55860-921-0}
\urldef\tempurl%
\url{https://doi.org/10.1016/B978-1-55860-921-1.50012-0}
\showDOI{\tempurl}


\bibitem[Isbister et~al\mbox{.}(2022)]%
        {Isbister2022}
\bibfield{author}{\bibinfo{person}{Katherine Isbister}, \bibinfo{person}{Joshua {McVeigh-Schultz}}, \bibinfo{person}{Anya Osborne}, {and} \bibinfo{person}{Victor Li}.} \bibinfo{year}{2022}\natexlab{}.
\newblock \showarticletitle{Augmenting {{Social Presence}} in {{VR Meetings}}}.
\newblock \bibinfo{journal}{\emph{CHI'22}} (\bibinfo{year}{2022}), \bibinfo{pages}{1--4}.
\newblock
\urldef\tempurl%
\url{https://drive.google.com/file/d/1hSpEpxN\_d-W04Yjmx7kpSjlDexETkBYd/view}
\showURL{%
\tempurl}


\bibitem[Jo et~al\mbox{.}(2016)]%
        {Jo2016}
\bibfield{author}{\bibinfo{person}{Dongsik Jo}, \bibinfo{person}{Ki-Hong Kim}, {and} \bibinfo{person}{Gerard~Jounghyun Kim}.} \bibinfo{year}{2016}\natexlab{}.
\newblock \showarticletitle{Effects of Avatar and Background Representation Forms to Co-Presence in Mixed Reality ({{MR}}) Tele-Conference Systems}. In \bibinfo{booktitle}{\emph{{{SIGGRAPH ASIA}} 2016 {{Virtual Reality}} Meets {{Physical Reality}}: {{Modelling}} and {{Simulating Virtual Humans}} and {{Environments}}}} \emph{(\bibinfo{series}{{{SA}} '16})}. \bibinfo{publisher}{{Association for Computing Machinery}}, \bibinfo{address}{{New York, NY, USA}}, \bibinfo{pages}{1--4}.
\newblock
\showISBNx{978-1-4503-4548-4}
\urldef\tempurl%
\url{https://doi.org/10.1145/2992138.2992146}
\showDOI{\tempurl}


\bibitem[Jonas et~al\mbox{.}(2019)]%
        {Jonas2019}
\bibfield{author}{\bibinfo{person}{Marcel Jonas}, \bibinfo{person}{Steven Said}, \bibinfo{person}{Daniel Yu}, \bibinfo{person}{Chris Aiello}, \bibinfo{person}{Nicholas Furlo}, {and} \bibinfo{person}{Douglas Zytko}.} \bibinfo{year}{2019}\natexlab{}.
\newblock \showarticletitle{Towards a Taxonomy of Social {{VR}} Application Design}. In \bibinfo{booktitle}{\emph{{{CHI PLAY}} 2019 - {{Extended Abstracts}} of the {{Annual Symposium}} on {{Computer-Human Interaction}} in {{Play}}}}, Vol.~\bibinfo{volume}{19}. \bibinfo{publisher}{{Association for Computing Machinery, Inc}}, \bibinfo{pages}{437--444}.
\newblock
\showISBNx{978-1-4503-6871-1}
\urldef\tempurl%
\url{https://doi.org/10.1145/3341215.3356271}
\showDOI{\tempurl}


\bibitem[Jung et~al\mbox{.}(2013)]%
        {Jung2013}
\bibfield{author}{\bibinfo{person}{Rex~E. Jung}, \bibinfo{person}{Brittany~S. Mead}, \bibinfo{person}{Jessica Carrasco}, {and} \bibinfo{person}{Ranee~A. Flores}.} \bibinfo{year}{2013}\natexlab{}.
\newblock \showarticletitle{The Structure of Creative Cognition in the Human Brain}.
\newblock \bibinfo{journal}{\emph{Front Hum Neurosci}}  \bibinfo{volume}{7} (\bibinfo{year}{2013}), \bibinfo{pages}{330}.
\newblock
\showISSN{1662-5161}
\urldef\tempurl%
\url{https://doi.org/10.3389/fnhum.2013.00330}
\showDOI{\tempurl}


\bibitem[Junuzovic et~al\mbox{.}(2012)]%
        {Junuzovic2012}
\bibfield{author}{\bibinfo{person}{Sasa Junuzovic}, \bibinfo{person}{Kori Inkpen}, \bibinfo{person}{John Tang}, \bibinfo{person}{Mara Sedlins}, {and} \bibinfo{person}{Kristie Fisher}.} \bibinfo{year}{2012}\natexlab{}.
\newblock \showarticletitle{To See or Not to See: {{A}} Study Comparing Four-Way Avatar, Video, and Audio Conferencing for Work}.
\newblock \bibinfo{journal}{\emph{GROUP'12 - Proceedings of the ACM 2012 International Conference on Support Group Work}} (\bibinfo{year}{2012}), \bibinfo{pages}{31--34}.
\newblock
\showISBNx{9781450314862}
\urldef\tempurl%
\url{https://doi.org/10.1145/2389176.2389181}
\showDOI{\tempurl}


\bibitem[Kafai et~al\mbox{.}(2010)]%
        {Kafai2010}
\bibfield{author}{\bibinfo{person}{Yasmin~B. Kafai}, \bibinfo{person}{Deborah~A. Fields}, {and} \bibinfo{person}{Melissa~S. Cook}.} \bibinfo{year}{2010}\natexlab{}.
\newblock \showarticletitle{Your {{Second Selves}}: {{Player-Designed Avatars}}}.
\newblock \bibinfo{journal}{\emph{Games and Culture}} \bibinfo{volume}{5}, \bibinfo{number}{1} (\bibinfo{date}{Jan.} \bibinfo{year}{2010}), \bibinfo{pages}{23--42}.
\newblock
\showISSN{1555-4120}
\urldef\tempurl%
\url{https://doi.org/10.1177/1555412009351260}
\showDOI{\tempurl}


\bibitem[Kay et~al\mbox{.}(2004)]%
        {Kay2004}
\bibfield{author}{\bibinfo{person}{Aaron~C. Kay}, \bibinfo{person}{S.~Christian Wheeler}, \bibinfo{person}{John~A. Bargh}, {and} \bibinfo{person}{Lee Ross}.} \bibinfo{year}{2004}\natexlab{}.
\newblock \showarticletitle{Material Priming: {{The}} Influence of Mundane Physical Objects on Situational Construal and Competitive Behavioral Choice}.
\newblock \bibinfo{journal}{\emph{Organizational Behavior and Human Decision Processes}} \bibinfo{volume}{95}, \bibinfo{number}{1} (\bibinfo{date}{Sept.} \bibinfo{year}{2004}), \bibinfo{pages}{83--96}.
\newblock
\showISSN{0749-5978}
\urldef\tempurl%
\url{https://doi.org/10.1016/j.obhdp.2004.06.003}
\showDOI{\tempurl}


\bibitem[Kendon(1990)]%
        {Kendon1990}
\bibfield{author}{\bibinfo{person}{Adam Kendon}.} \bibinfo{year}{1990}\natexlab{}.
\newblock \bibinfo{booktitle}{\emph{Conducting Interaction: {{Patterns}} of Behavior in Focused Encounters.}}
\newblock \bibinfo{publisher}{{Cambridge University Press}}, \bibinfo{address}{{New York, NY, US}}.
\newblock
\showISBNx{0-521-38036-7 (Hardcover); 0-521-38938-0 (Paperback)}


\bibitem[Kilteni et~al\mbox{.}(2012)]%
        {Kilteni2012}
\bibfield{author}{\bibinfo{person}{Konstantina Kilteni}, \bibinfo{person}{Jean-Marie Normand}, \bibinfo{person}{Maria~V. {Sanchez-Vives}}, {and} \bibinfo{person}{Mel Slater}.} \bibinfo{year}{2012}\natexlab{}.
\newblock \showarticletitle{Extending {{Body Space}} in {{Immersive Virtual Reality}}: {{A Very Long Arm Illusion}}}.
\newblock \bibinfo{journal}{\emph{PLOS ONE}} \bibinfo{volume}{7}, \bibinfo{number}{7} (\bibinfo{date}{July} \bibinfo{year}{2012}), \bibinfo{pages}{e40867}.
\newblock
\showISSN{1932-6203}
\urldef\tempurl%
\url{https://doi.org/10.1371/journal.pone.0040867}
\showDOI{\tempurl}


\bibitem[Knez(1995)]%
        {Knez1995}
\bibfield{author}{\bibinfo{person}{Igor Knez}.} \bibinfo{year}{1995}\natexlab{}.
\newblock \showarticletitle{Effects of Indoor Lighting on Mood and Cognition}.
\newblock \bibinfo{journal}{\emph{Journal of Environmental Psychology}} \bibinfo{volume}{15}, \bibinfo{number}{1} (\bibinfo{date}{March} \bibinfo{year}{1995}), \bibinfo{pages}{39--51}.
\newblock
\showISSN{0272-4944}
\urldef\tempurl%
\url{https://doi.org/10.1016/0272-4944(95)90013-6}
\showDOI{\tempurl}


\bibitem[Kohler et~al\mbox{.}(2009)]%
        {Kohler2009}
\bibfield{author}{\bibinfo{person}{Thomas Kohler}, \bibinfo{person}{Kurt Matzler}, {and} \bibinfo{person}{Johann F{\"u}ller}.} \bibinfo{year}{2009}\natexlab{}.
\newblock \showarticletitle{Avatar-Based Innovation: {{Using}} Virtual Worlds for Real-World Innovation}.
\newblock \bibinfo{journal}{\emph{Technovation}} \bibinfo{volume}{29}, \bibinfo{number}{6} (\bibinfo{date}{June} \bibinfo{year}{2009}), \bibinfo{pages}{395--407}.
\newblock
\showISSN{0166-4972}
\urldef\tempurl%
\url{https://doi.org/10.1016/j.technovation.2008.11.004}
\showDOI{\tempurl}


\bibitem[Latoschik et~al\mbox{.}(2017)]%
        {Latoschik2017}
\bibfield{author}{\bibinfo{person}{Marc~Erich Latoschik}, \bibinfo{person}{Daniel Roth}, \bibinfo{person}{Dominik Gall}, \bibinfo{person}{Jascha Achenbach}, \bibinfo{person}{Thomas Waltemate}, {and} \bibinfo{person}{Mario Botsch}.} \bibinfo{year}{2017}\natexlab{}.
\newblock \showarticletitle{The {{Effect}} of {{Avatar Realism}} in {{Immersive Social Virtual Realities}}}.
\newblock  \bibinfo{volume}{17}, \bibinfo{number}{10} (\bibinfo{year}{2017}).
\newblock
\urldef\tempurl%
\url{https://doi.org/10.1145/3139131.3139156}
\showDOI{\tempurl}


\bibitem[Le et~al\mbox{.}(2017)]%
        {Le2017}
\bibfield{author}{\bibinfo{person}{Khanh~Duy Le}, \bibinfo{person}{Morten Fjeld}, \bibinfo{person}{Ali Alavi}, {and} \bibinfo{person}{Andreas Kunz}.} \bibinfo{year}{2017}\natexlab{}.
\newblock \showarticletitle{Immersive Environment for Distributed Creative Collaboration}.
\newblock \bibinfo{journal}{\emph{Proceedings of the ACM Symposium on Virtual Reality Software and Technology, VRST}}  \bibinfo{volume}{Part F1319} (\bibinfo{date}{Nov.} \bibinfo{year}{2017}).
\newblock
\urldef\tempurl%
\url{https://doi.org/10.1145/3139131.3139163}
\showDOI{\tempurl}


\bibitem[Lee et~al\mbox{.}(2019)]%
        {Lee2019a}
\bibfield{author}{\bibinfo{person}{Jee~Hyun Lee}, \bibinfo{person}{Eun~Kyoung Yang}, {and} \bibinfo{person}{Zhong~Yuan Sun}.} \bibinfo{year}{2019}\natexlab{}.
\newblock \showarticletitle{Design {{Cognitive Actions Stimulating Creativity}} in the {{VR Design Environment}}}. In \bibinfo{booktitle}{\emph{Proceedings of the 2019 on {{Creativity}} and {{Cognition}}}} \emph{(\bibinfo{series}{C\&amp;{{C}} '19})}. \bibinfo{publisher}{{Association for Computing Machinery}}, \bibinfo{address}{{New York, NY, USA}}, \bibinfo{pages}{604--611}.
\newblock
\showISBNx{978-1-4503-5917-7}
\urldef\tempurl%
\url{https://doi.org/10.1145/3325480.3326575}
\showDOI{\tempurl}


\bibitem[Li et~al\mbox{.}(2020)]%
        {Li2020}
\bibfield{author}{\bibinfo{person}{Jie Li}, \bibinfo{person}{Vinoba Vinayagamoorthy}, \bibinfo{person}{Raz Schwartz}, \bibinfo{person}{Wijnand Ijsselsteijn}, \bibinfo{person}{David~A. Shamma}, {and} \bibinfo{person}{Pablo Cesar}.} \bibinfo{year}{2020}\natexlab{}.
\newblock \showarticletitle{Social {{VR}}: {{A New}} Medium for Remote Communication and Collaboration}. In \bibinfo{booktitle}{\emph{Conference on {{Human Factors}} in {{Computing Systems}} - {{Proceedings}}}}. \bibinfo{publisher}{{Association for Computing Machinery}}.
\newblock
\showISBNx{978-1-4503-6819-3}
\urldef\tempurl%
\url{https://doi.org/10.1145/3334480.3375160}
\showDOI{\tempurl}


\bibitem[Li et~al\mbox{.}(2022)]%
        {Li2022}
\bibfield{author}{\bibinfo{person}{Jialang~Victor Li}, \bibinfo{person}{Max Kreminski}, \bibinfo{person}{Sean~M Fernandes}, \bibinfo{person}{Anya Osborne}, \bibinfo{person}{Joshua {McVeigh-Schultz}}, {and} \bibinfo{person}{Katherine Isbister}.} \bibinfo{year}{2022}\natexlab{}.
\newblock \showarticletitle{Conversation {{Balance}}: {{A Shared VR Visualization}} to {{Support Turn-taking}} in {{Meetings}}}. In \bibinfo{booktitle}{\emph{{{CHI Conference}} on {{Human Factors}} in {{Computing Systems Extended Abstracts}}}} \emph{(\bibinfo{series}{{{CHI EA}} '22})}. \bibinfo{publisher}{{Association for Computing Machinery}}, \bibinfo{address}{{New York, NY, USA}}, \bibinfo{pages}{1--4}.
\newblock
\showISBNx{978-1-4503-9156-6}
\urldef\tempurl%
\url{https://doi.org/10.1145/3491101.3519879}
\showDOI{\tempurl}


\bibitem[Lombard and Ditton(1997)]%
        {Lombard1997}
\bibfield{author}{\bibinfo{person}{Matthew Lombard} {and} \bibinfo{person}{Theresa Ditton}.} \bibinfo{year}{1997}\natexlab{}.
\newblock \showarticletitle{At the {{Heart}} of {{It All}}: {{The Concept}} of {{Presence}}}.
\newblock \bibinfo{journal}{\emph{Journal of Computer-Mediated Communication}} \bibinfo{volume}{3}, \bibinfo{number}{2} (\bibinfo{date}{Sept.} \bibinfo{year}{1997}), \bibinfo{pages}{JCMC321}.
\newblock
\showISSN{1083-6101}
\urldef\tempurl%
\url{https://doi.org/10.1111/j.1083-6101.1997.tb00072.x}
\showDOI{\tempurl}


\bibitem[Maloney et~al\mbox{.}(2020a)]%
        {Maloney2020b}
\bibfield{author}{\bibinfo{person}{Divine Maloney}, \bibinfo{person}{Guo Freeman}, {and} \bibinfo{person}{Donghee Yvette~Wohn}.} \bibinfo{year}{2020}\natexlab{a}.
\newblock \showarticletitle{"{{Talking}} without {{A Voice}}": {{Understanding Non-Verbal Communication}} in {{Social Virtual Reality}}}.
\newblock \bibinfo{journal}{\emph{Proc. ACM Hum.-Comput. Interact.}} (\bibinfo{year}{2020}), \bibinfo{pages}{25}.
\newblock
\urldef\tempurl%
\url{https://doi.org/10.1145/3415246}
\showDOI{\tempurl}


\bibitem[Maloney et~al\mbox{.}(2020b)]%
        {Maloney2020c}
\bibfield{author}{\bibinfo{person}{Divine Maloney}, \bibinfo{person}{Samaneh Zamanifard}, {and} \bibinfo{person}{Guo Freeman}.} \bibinfo{year}{2020}\natexlab{b}.
\newblock \showarticletitle{Anonymity vs. {{Familiarity}}: {{Self-Disclosure}} and {{Privacy}} in {{Social Virtual Reality}}}. In \bibinfo{booktitle}{\emph{26th {{ACM Symposium}} on {{Virtual Reality Software}} and {{Technology}}}} \emph{(\bibinfo{series}{{{VRST}} '20})}. \bibinfo{publisher}{{Association for Computing Machinery}}, \bibinfo{address}{{New York, NY, USA}}, \bibinfo{pages}{1--9}.
\newblock
\showISBNx{978-1-4503-7619-8}
\urldef\tempurl%
\url{https://doi.org/10.1145/3385956.3418967}
\showDOI{\tempurl}


\bibitem[M{\'a}rquez~Segura et~al\mbox{.}(2016)]%
        {Segura2016}
\bibfield{author}{\bibinfo{person}{Elena M{\'a}rquez~Segura}, \bibinfo{person}{Laia Turmo~Vidal}, \bibinfo{person}{Asreen Rostami}, {and} \bibinfo{person}{Annika Waern}.} \bibinfo{year}{2016}\natexlab{}.
\newblock \showarticletitle{Embodied {{Sketching}}}. In \bibinfo{booktitle}{\emph{Proceedings of the 2016 {{CHI Conference}} on {{Human Factors}} in {{Computing Systems}}}} \emph{(\bibinfo{series}{{{CHI}} '16})}. \bibinfo{publisher}{{Association for Computing Machinery}}, \bibinfo{address}{{New York, NY, USA}}, \bibinfo{pages}{6014--6027}.
\newblock
\showISBNx{978-1-4503-3362-7}
\urldef\tempurl%
\url{https://doi.org/10.1145/2858036.2858486}
\showDOI{\tempurl}


\bibitem[Marshall et~al\mbox{.}(2011)]%
        {Marshall2011}
\bibfield{author}{\bibinfo{person}{Paul Marshall}, \bibinfo{person}{Yvonne Rogers}, {and} \bibinfo{person}{Nadia Pantidi}.} \bibinfo{year}{2011}\natexlab{}.
\newblock \showarticletitle{Using {{F-formations}} to {{Analyse Spatial Patterns}} of {{Interaction}} in {{Physical Environments}}}.
\newblock \bibinfo{journal}{\emph{Proceedings of the ACM 2011 conference on Computer supported cooperative work - CSCW '11}} (\bibinfo{year}{2011}).
\newblock
\showISBNx{9781450305563}
\urldef\tempurl%
\url{https://doi.org/10.1145/1958824}
\showDOI{\tempurl}


\bibitem[{McVeigh-Schultz} and Isbister(2021a)]%
        {McVeigh-Schultz2021}
\bibfield{author}{\bibinfo{person}{Joshua {McVeigh-Schultz}} {and} \bibinfo{person}{Katherine Isbister}.} \bibinfo{year}{2021}\natexlab{a}.
\newblock \showarticletitle{A ``beyond Being There'' for {{VR}} Meetings: Envisioning the Future of Remote Work}.
\newblock \bibinfo{journal}{\emph{Human\textendash Computer Interaction}} (\bibinfo{date}{Dec.} \bibinfo{year}{2021}), \bibinfo{pages}{1--21}.
\newblock
\showISSN{0737-0024}
\urldef\tempurl%
\url{https://doi.org/10.1080/07370024.2021.1994860}
\showDOI{\tempurl}


\bibitem[{McVeigh-Schultz} and Isbister(2021b)]%
        {McVeigh-Schultz2021a}
\bibfield{author}{\bibinfo{person}{Joshua {McVeigh-Schultz}} {and} \bibinfo{person}{Katherine Isbister}.} \bibinfo{year}{2021}\natexlab{b}.
\newblock \showarticletitle{The {{Case}} for {{Weird Social}} in {{VR}}/{{XR}}}.
\newblock \bibinfo{journal}{\emph{Conference on Human Factors in Computing Systems - Proceedings}} (\bibinfo{date}{May} \bibinfo{year}{2021}).
\newblock
\urldef\tempurl%
\url{https://doi.org/10.1145/3411763.3450377}
\showDOI{\tempurl}


\bibitem[{McVeigh-Schultz} et~al\mbox{.}(2019)]%
        {McVeigh-Schultz2019}
\bibfield{author}{\bibinfo{person}{Joshua {McVeigh-Schultz}}, \bibinfo{person}{Anya Osborne}, {and} \bibinfo{person}{Katherine Isbister}.} \bibinfo{year}{2019}\natexlab{}.
\newblock \showarticletitle{Shaping Pro-Social Interaction in {{VR}} an Emerging Design Framework}. In \bibinfo{booktitle}{\emph{Conference on {{Human Factors}} in {{Computing Systems}} - {{Proceedings}}}} \emph{(\bibinfo{series}{{{CHI}} '19})}. \bibinfo{publisher}{{Association for Computing Machinery}}, \bibinfo{address}{{Glasgow, Scotland UK}}, \bibinfo{pages}{1--12}.
\newblock
\showISBNx{978-1-4503-5970-2}
\urldef\tempurl%
\url{https://doi.org/10.1145/3290605.3300794}
\showDOI{\tempurl}


\bibitem[{McVeigh-Schultz} et~al\mbox{.}(2021)]%
        {McVeigh-Schultz2021b}
\bibfield{author}{\bibinfo{person}{Joshua {McVeigh-Schultz}}, \bibinfo{person}{Anya Osborne}, \bibinfo{person}{Max Kreminski}, \bibinfo{person}{Sean Fernandes}, \bibinfo{person}{Sabrina Fielder}, \bibinfo{person}{Victor Li}, {and} \bibinfo{person}{Katherine Isbister}.} \bibinfo{year}{2021}\natexlab{}.
\newblock \showarticletitle{Social {{Superpowers}} in {{Social VR}}}.
\newblock \bibinfo{journal}{\emph{Social VR: A New Medium for Remote Communication \& Collaboration 2021 Workshop in 2021 ACM CHI Virtual Conference on Human Factors in Computing Systems}} (\bibinfo{year}{2021}), \bibinfo{pages}{1--5}.
\newblock
\urldef\tempurl%
\url{https://442e4efe-5603-4db5-bb2c-b1a81a4eb29a.filesusr.com/ugd/3ad93e\_8cf5a82bf8d746a3a7141289f4db3db5.pdf}
\showURL{%
\tempurl}


\bibitem[Mednick(1962)]%
        {Mednick1962}
\bibfield{author}{\bibinfo{person}{Sarnoff Mednick}.} \bibinfo{year}{1962}\natexlab{}.
\newblock \showarticletitle{The Associative Basis of the Creative Process}.
\newblock \bibinfo{journal}{\emph{Psychological Review}}  \bibinfo{volume}{69} (\bibinfo{year}{1962}), \bibinfo{pages}{220--232}.
\newblock
\showISSN{1939-1471}
\urldef\tempurl%
\url{https://doi.org/10.1037/h0048850}
\showDOI{\tempurl}


\bibitem[Mori et~al\mbox{.}(2012)]%
        {Mori2012}
\bibfield{author}{\bibinfo{person}{Masahiro Mori}, \bibinfo{person}{Karl~F. MacDorman}, {and} \bibinfo{person}{Norri Kageki}.} \bibinfo{year}{2012}\natexlab{}.
\newblock \showarticletitle{The {{Uncanny Valley}} [{{From}} the {{Field}}]}.
\newblock \bibinfo{journal}{\emph{IEEE Robotics \& Automation Magazine}} \bibinfo{volume}{19}, \bibinfo{number}{2} (\bibinfo{date}{June} \bibinfo{year}{2012}), \bibinfo{pages}{98--100}.
\newblock
\showISSN{1558-223X}
\urldef\tempurl%
\url{https://doi.org/10.1109/MRA.2012.2192811}
\showDOI{\tempurl}


\bibitem[Moustafa and Steed(2018)]%
        {Moustafa2018}
\bibfield{author}{\bibinfo{person}{Fares Moustafa} {and} \bibinfo{person}{Anthony Steed}.} \bibinfo{year}{2018}\natexlab{}.
\newblock \showarticletitle{A {{Longitudinal Study}} of {{Small Group Interaction}} in {{Social Virtual Reality}}}.
\newblock \bibinfo{journal}{\emph{ACM}}  \bibinfo{volume}{10} (\bibinfo{year}{2018}), \bibinfo{pages}{1--10}.
\newblock
\urldef\tempurl%
\url{https://doi.org/10.1145/3281505}
\showDOI{\tempurl}


\bibitem[Nelson and Guegan(2019)]%
        {Nelson2019}
\bibfield{author}{\bibinfo{person}{Julien Nelson} {and} \bibinfo{person}{J{\'e}r{\^o}me Guegan}.} \bibinfo{year}{2019}\natexlab{}.
\newblock \showarticletitle{``{{I}}'d like to Be under the Sea'': {{Contextual}} Cues in Virtual Environments Influence the Orientation of Idea Generation}.
\newblock \bibinfo{journal}{\emph{Computers in Human Behavior}}  \bibinfo{volume}{90} (\bibinfo{date}{Jan.} \bibinfo{year}{2019}), \bibinfo{pages}{93--102}.
\newblock
\showISSN{0747-5632}
\urldef\tempurl%
\url{https://doi.org/10.1016/j.chb.2018.08.001}
\showDOI{\tempurl}


\bibitem[Nijstad and Stroebe(2006)]%
        {Nijstad2006}
\bibfield{author}{\bibinfo{person}{Bernard~A. Nijstad} {and} \bibinfo{person}{Wolfgang Stroebe}.} \bibinfo{year}{2006}\natexlab{}.
\newblock \showarticletitle{How the {{Group Affects}} the {{Mind}}: {{A Cognitive Model}} of {{Idea Generation}} in {{Groups}}}.
\newblock \bibinfo{journal}{\emph{Pers Soc Psychol Rev}} \bibinfo{volume}{10}, \bibinfo{number}{3} (\bibinfo{date}{Aug.} \bibinfo{year}{2006}), \bibinfo{pages}{186--213}.
\newblock
\showISSN{1088-8683}
\urldef\tempurl%
\url{https://doi.org/10.1207/s15327957pspr1003_1}
\showDOI{\tempurl}


\bibitem[Olaosebikan et~al\mbox{.}(2022)]%
        {Olaosebikan2022}
\bibfield{author}{\bibinfo{person}{Monsurat Olaosebikan}, \bibinfo{person}{Claudia Aranda~Barrios}, \bibinfo{person}{Blessing Kolawole}, \bibinfo{person}{Lenore Cowen}, {and} \bibinfo{person}{Orit Shaer}.} \bibinfo{year}{2022}\natexlab{}.
\newblock \showarticletitle{Identifying {{Cognitive}} and {{Creative Support Needs}} for {{Remote Scientific Collaboration}} Using {{VR}}: {{Practices}}, {{Affordances}}, and {{Design Implications}}}. In \bibinfo{booktitle}{\emph{Creativity and {{Cognition}}}} \emph{(\bibinfo{series}{C\&amp;{{C}} '22})}. \bibinfo{publisher}{{Association for Computing Machinery}}, \bibinfo{address}{{New York, NY, USA}}, \bibinfo{pages}{97--110}.
\newblock
\showISBNx{978-1-4503-9327-0}
\urldef\tempurl%
\url{https://doi.org/10.1145/3527927.3532797}
\showDOI{\tempurl}


\bibitem[Olin et~al\mbox{.}(2020)]%
        {Olin2020}
\bibfield{author}{\bibinfo{person}{Patrick~Aggergaard Olin}, \bibinfo{person}{Ahmad~Mohammad Issa}, \bibinfo{person}{Tiare Feuchtner}, {and} \bibinfo{person}{Kaj Gr{\o}nb{\ae}k}.} \bibinfo{year}{2020}\natexlab{}.
\newblock \showarticletitle{Designing for {{Heterogeneous Cross-Device Collaboration}} and {{Social Interaction}} in {{Virtual Reality}}}. In \bibinfo{booktitle}{\emph{32nd {{Australian Conference}} on {{Human-Computer Interaction}}}} \emph{(\bibinfo{series}{{{OzCHI}} '20})}. \bibinfo{publisher}{{Association for Computing Machinery}}, \bibinfo{address}{{Sydney, NSW, Australia}}, \bibinfo{pages}{112--127}.
\newblock
\showISBNx{978-1-4503-8975-4}
\urldef\tempurl%
\url{https://doi.org/10.1145/3441000.3441070}
\showDOI{\tempurl}


\bibitem[Osborne(2024)]%
        {Osborne2024}
\bibfield{author}{\bibinfo{person}{Anya Osborne}.} \bibinfo{year}{2024}\natexlab{}.
\newblock \emph{\bibinfo{title}{Design of {{Social Affordances}} for {{Meetings}} in {{Social Virtual Reality}}}}.
\newblock \bibinfo{thesistype}{Ph.\,D. Dissertation}. \bibinfo{school}{UC Santa Cruz}.
\newblock
\urldef\tempurl%
\url{https://escholarship.org/uc/item/8h51c6s0}
\showURL{%
\tempurl}


\bibitem[Osborne et~al\mbox{.}(2023)]%
        {Osborne2023}
\bibfield{author}{\bibinfo{person}{Anya Osborne}, \bibinfo{person}{Sabrina Fielder}, \bibinfo{person}{Joshua {Mcveigh-Schultz}}, \bibinfo{person}{Timothy Lang}, \bibinfo{person}{Max Kreminski}, \bibinfo{person}{George Butler}, \bibinfo{person}{Jialang~Victor Li}, \bibinfo{person}{Diana~R. Sanchez}, {and} \bibinfo{person}{Katherine Isbister}.} \bibinfo{year}{2023}\natexlab{}.
\newblock \showarticletitle{Being {{Social}} in {{VR Meetings}}: {{A Landscape Analysis}} of {{Current Tools}}}. In \bibinfo{booktitle}{\emph{Proceedings of the 2023 {{ACM Designing Interactive Systems Conference}}}} \emph{(\bibinfo{series}{{{DIS}} '23})}. \bibinfo{publisher}{{Association for Computing Machinery}}, \bibinfo{address}{{New York, NY, USA}}, \bibinfo{pages}{1789--1809}.
\newblock
\showISBNx{978-1-4503-9893-0}
\urldef\tempurl%
\url{https://doi.org/10.1145/3563657.3595959}
\showDOI{\tempurl}


\bibitem[Osborne et~al\mbox{.}(2019)]%
        {Osborne2019}
\bibfield{author}{\bibinfo{person}{Anya Osborne}, \bibinfo{person}{Joshua {Mcveigh-Schultz}}, {and} \bibinfo{person}{Katherine Isbister}.} \bibinfo{year}{2019}\natexlab{}.
\newblock \showarticletitle{Understanding {{Emerging Design Practices}} for {{Avatar Systems}} in the {{Commercial Social VR Ecology}}}. In \bibinfo{booktitle}{\emph{Proceedings of the 2019 on {{Designing Interactive Systems Conference}}}}. \bibinfo{publisher}{{ACM}}, \bibinfo{address}{{New York, NY, USA}}.
\newblock
\showISBNx{978-1-4503-5850-7}
\urldef\tempurl%
\url{https://doi.org/10.1145/3322276.3322352}
\showDOI{\tempurl}


\bibitem[Parks(1994)]%
        {Parks1994}
\bibfield{author}{\bibinfo{person}{M.~R. Parks}.} \bibinfo{year}{1994}\natexlab{}.
\newblock \showarticletitle{Communicative Competence and Interpersonal Control}.
\newblock In \bibinfo{booktitle}{\emph{Handbook of Interpersonal Communication} (\bibinfo{edition}{2} ed.)}, \bibfield{editor}{\bibinfo{person}{G.~R. Miller}} (Ed.). \bibinfo{publisher}{{SAGE}}, \bibinfo{address}{{Thousand Oaks, CA}}, \bibinfo{pages}{589--620}.
\newblock


\bibitem[Paulus and Yang(2000)]%
        {Paulus2000}
\bibfield{author}{\bibinfo{person}{Paul~B. Paulus} {and} \bibinfo{person}{Huei~Chuan Yang}.} \bibinfo{year}{2000}\natexlab{}.
\newblock \showarticletitle{Idea {{Generation}} in {{Groups}}: {{A Basis}} for {{Creativity}} in {{Organizations}}}.
\newblock \bibinfo{journal}{\emph{Organizational Behavior and Human Decision Processes}} \bibinfo{volume}{82}, \bibinfo{number}{1} (\bibinfo{date}{May} \bibinfo{year}{2000}), \bibinfo{pages}{76--87}.
\newblock
\showISSN{0749-5978}
\urldef\tempurl%
\url{https://doi.org/10.1006/OBHD.2000.2888}
\showDOI{\tempurl}


\bibitem[Pe{\~n}a and Blackburn(2013)]%
        {Peña2013}
\bibfield{author}{\bibinfo{person}{Jorge Pe{\~n}a} {and} \bibinfo{person}{Kate Blackburn}.} \bibinfo{year}{2013}\natexlab{}.
\newblock \showarticletitle{The {{Priming Effects}} of {{Virtual Environments}} on {{Interpersonal Perceptions}} and {{Behaviors}}}.
\newblock \bibinfo{journal}{\emph{Journal of Communication}} \bibinfo{volume}{63}, \bibinfo{number}{4} (\bibinfo{date}{Aug.} \bibinfo{year}{2013}), \bibinfo{pages}{703--720}.
\newblock
\showISSN{0021-9916}
\urldef\tempurl%
\url{https://doi.org/10.1111/jcom.12043}
\showDOI{\tempurl}


\bibitem[Posner(2011)]%
        {Posner2011}
\bibfield{author}{\bibinfo{person}{Michael~I. Posner}.} \bibinfo{year}{2011}\natexlab{}.
\newblock \bibinfo{booktitle}{\emph{Cognitive {{Neuroscience}} of {{Attention}}, {{Second Edition}}}}.
\newblock \bibinfo{publisher}{{Guilford Press}}.
\newblock
\showISBNx{978-1-60918-987-7}


\bibitem[Praetorius and G{\"o}rlich(2020)]%
        {Praetorius2020}
\bibfield{author}{\bibinfo{person}{Anna~Samira Praetorius} {and} \bibinfo{person}{Daniel G{\"o}rlich}.} \bibinfo{year}{2020}\natexlab{}.
\newblock \showarticletitle{How {{Avatars Influence User Behavior}}: {{A Review}} on the {{Proteus Effect}} in {{Virtual Environments}} and {{Video Games}}}. In \bibinfo{booktitle}{\emph{Proceedings of the 15th {{International Conference}} on the {{Foundations}} of {{Digital Games}}}} \emph{(\bibinfo{series}{{{FDG}} '20})}. \bibinfo{publisher}{{Association for Computing Machinery}}, \bibinfo{address}{{New York, NY, USA}}, \bibinfo{pages}{1--9}.
\newblock
\showISBNx{978-1-4503-8807-8}
\urldef\tempurl%
\url{https://doi.org/10.1145/3402942.3403019}
\showDOI{\tempurl}


\bibitem[Praetorius et~al\mbox{.}(2021)]%
        {Praetorius2021}
\bibfield{author}{\bibinfo{person}{Anna~Samira Praetorius}, \bibinfo{person}{Lara Krautmacher}, \bibinfo{person}{Gabriela Tullius}, {and} \bibinfo{person}{Crist{\'o}bal Curio}.} \bibinfo{year}{2021}\natexlab{}.
\newblock \showarticletitle{User-{{Avatar Relationships}} in {{Various Contexts}}: {{Does Context Influence}} a {{Users}}' {{Perception}} and {{Choice}} of an {{Avatar}}?}. In \bibinfo{booktitle}{\emph{Mensch Und {{Computer}} 2021}} \emph{(\bibinfo{series}{{{MuC}} '21})}. \bibinfo{publisher}{{Association for Computing Machinery}}, \bibinfo{address}{{New York, NY, USA}}, \bibinfo{pages}{275--280}.
\newblock
\showISBNx{978-1-4503-8645-6}
\urldef\tempurl%
\url{https://doi.org/10.1145/3473856.3474007}
\showDOI{\tempurl}


\bibitem[Ratcliffe et~al\mbox{.}(2021)]%
        {Ratcliffe2021}
\bibfield{author}{\bibinfo{person}{Jack Ratcliffe}, \bibinfo{person}{Francesco Soave}, \bibinfo{person}{Melynda Hoover}, \bibinfo{person}{Francisco~Raul Ortega}, \bibinfo{person}{Nick {Bryan-Kinns}}, \bibinfo{person}{Laurissa Tokarchuk}, {and} \bibinfo{person}{Ildar Farkhatdinov}.} \bibinfo{year}{2021}\natexlab{}.
\newblock \showarticletitle{Remote {{XR Studies}}: {{Exploring Three Key Challenges}} of {{Remote XR Experimentation}}}.
\newblock \bibinfo{journal}{\emph{Conference on Human Factors in Computing Systems - Proceedings}} (\bibinfo{date}{May} \bibinfo{year}{2021}), \bibinfo{pages}{4}.
\newblock
\urldef\tempurl%
\url{https://doi.org/10.1145/3411763.3442472}
\showDOI{\tempurl}


\bibitem[Rattner(2019)]%
        {Rattner2019}
\bibfield{author}{\bibinfo{person}{D.M. Rattner}.} \bibinfo{year}{2019}\natexlab{}.
\newblock \bibinfo{booktitle}{\emph{My Creative Space: {{How}} to Design Your Home to Stimulate Ideas and Spark Innovation}}.
\newblock \bibinfo{publisher}{{Skyhorse}}.
\newblock
\showISBNx{978-1-5107-3671-9}
\urldef\tempurl%
\url{https://books.google.com/books?id=3SKwDwAAQBAJ}
\showURL{%
\tempurl}


\bibitem[Rieuf et~al\mbox{.}(2015)]%
        {Rieuf2015}
\bibfield{author}{\bibinfo{person}{Vincent Rieuf}, \bibinfo{person}{Carole Bouchard}, {and} \bibinfo{person}{Am{\'e}ziane Aoussat}.} \bibinfo{year}{2015}\natexlab{}.
\newblock \showarticletitle{Immersive Moodboards, a Comparative Study of Industrial Design Inspiration Material}.
\newblock \bibinfo{journal}{\emph{Journal of Design Research}} \bibinfo{volume}{13}, \bibinfo{number}{1} (\bibinfo{date}{Jan.} \bibinfo{year}{2015}), \bibinfo{pages}{78--106}.
\newblock
\showISSN{1748-3050}
\urldef\tempurl%
\url{https://doi.org/10.1504/JDR.2015.067233}
\showDOI{\tempurl}


\bibitem[Riordan and O'Reilly(2011)]%
        {Riordan2011}
\bibfield{author}{\bibinfo{person}{Niamh~O. Riordan} {and} \bibinfo{person}{Philip O'Reilly}.} \bibinfo{year}{2011}\natexlab{}.
\newblock \showarticletitle{S(t)Imulating {{Creativity}} in {{Decision Making}}}.
\newblock \bibinfo{journal}{\emph{Journal of Decision Systems}} \bibinfo{volume}{20}, \bibinfo{number}{3} (\bibinfo{date}{Jan.} \bibinfo{year}{2011}), \bibinfo{pages}{325--351}.
\newblock
\showISSN{1246-0125}
\urldef\tempurl%
\url{https://doi.org/10.3166/jds.20.325-351}
\showDOI{\tempurl}


\bibitem[Riva et~al\mbox{.}(2007)]%
        {Riva2007}
\bibfield{author}{\bibinfo{person}{Giuseppe Riva}, \bibinfo{person}{Fabrizia Mantovani}, \bibinfo{person}{Claret~Samantha Capideville}, \bibinfo{person}{Alessandra Preziosa}, \bibinfo{person}{Francesca Morganti}, \bibinfo{person}{Daniela Villani}, \bibinfo{person}{Andrea Gaggioli}, \bibinfo{person}{Cristina Botella}, {and} \bibinfo{person}{Mariano Alca{\~n}iz}.} \bibinfo{year}{2007}\natexlab{}.
\newblock \showarticletitle{Affective {{Interactions Using Virtual Reality}}: {{The Link}} between {{Presence}} and {{Emotions}}}.
\newblock \bibinfo{journal}{\emph{CyberPsychology \& Behavior}} \bibinfo{volume}{10}, \bibinfo{number}{1} (\bibinfo{date}{Feb.} \bibinfo{year}{2007}), \bibinfo{pages}{45--56}.
\newblock
\showISSN{1094-9313}
\urldef\tempurl%
\url{https://doi.org/10.1089/cpb.2006.9993}
\showDOI{\tempurl}


\bibitem[Robin McVeigh-Schultz et~al\mbox{.}(2024)]%
        {McVeigh-Schultz2024}
\bibfield{author}{\bibinfo{person}{Joshua Robin McVeigh-Schultz}, \bibinfo{person}{Elena M{\'a}rquez~Segura}, {and} \bibinfo{person}{Katherine Isbister}.} \bibinfo{year}{2024}\natexlab{}.
\newblock \showarticletitle{Embodied prototyping in {VR}: Ideation and bodystorming within a custom {VR} sandbox}. In \bibinfo{booktitle}{\emph{Proceedings of {DRS}}}. \bibinfo{publisher}{Design Research Society}.
\newblock


\bibitem[Roth et~al\mbox{.}(2019)]%
        {Roth2019}
\bibfield{author}{\bibinfo{person}{Daniel Roth}, \bibinfo{person}{Gary Bente}, \bibinfo{person}{Peter Kullmann}, \bibinfo{person}{David Mal}, \bibinfo{person}{Chris~Felix Purps}, \bibinfo{person}{Kai Vogeley}, {and} \bibinfo{person}{Marc~Erich Latoschik}.} \bibinfo{year}{2019}\natexlab{}.
\newblock \showarticletitle{Technologies for Social Augmentations in User-Embodied Virtual Reality}. In \bibinfo{booktitle}{\emph{Proceedings of the {{ACM Symposium}} on {{Virtual Reality Software}} and {{Technology}}, {{VRST}}}}, Vol.~\bibinfo{volume}{5}. \bibinfo{publisher}{{Association for Computing Machinery}}, \bibinfo{address}{{New York, NY, USA}}, \bibinfo{pages}{1--12}.
\newblock
\showISBNx{978-1-4503-7001-1}
\urldef\tempurl%
\url{https://doi.org/10.1145/3359996.3364269}
\showDOI{\tempurl}


\bibitem[Roth et~al\mbox{.}(2018)]%
        {Roth2018}
\bibfield{author}{\bibinfo{person}{Daniel Roth}, \bibinfo{person}{Constantin Klelnbeck}, \bibinfo{person}{Tobias Feigl}, \bibinfo{person}{Christopher Mutschler}, {and} \bibinfo{person}{Marc~Erich Latoschik}.} \bibinfo{year}{2018}\natexlab{}.
\newblock \showarticletitle{Beyond {{Replication}}: {{Augmenting Social Behaviors}} in {{Multi-User Virtual Realities}}}.
\newblock \bibinfo{journal}{\emph{25th IEEE Conference on Virtual Reality and 3D User Interfaces, VR 2018 - Proceedings}}  \bibinfo{volume}{1} (\bibinfo{date}{Aug.} \bibinfo{year}{2018}), \bibinfo{pages}{215--222}.
\newblock
\showISBNx{9781538633656}
\urldef\tempurl%
\url{https://doi.org/10.1109/VR.2018.8447550}
\showDOI{\tempurl}


\bibitem[Rowe et~al\mbox{.}(2007)]%
        {Rowe2007}
\bibfield{author}{\bibinfo{person}{G. Rowe}, \bibinfo{person}{J.~B. Hirsh}, {and} \bibinfo{person}{A.~K. Anderson}.} \bibinfo{year}{2007}\natexlab{}.
\newblock \showarticletitle{Positive Affect Increases the Breadth of Attentional Selection}.
\newblock \bibinfo{journal}{\emph{Proceedings of the National Academy of Sciences}} \bibinfo{volume}{104}, \bibinfo{number}{1} (\bibinfo{date}{Jan.} \bibinfo{year}{2007}), \bibinfo{pages}{383--388}.
\newblock
\urldef\tempurl%
\url{https://doi.org/10.1073/pnas.0605198104}
\showDOI{\tempurl}


\bibitem[Salter and Gann(2003)]%
        {Salter2003}
\bibfield{author}{\bibinfo{person}{Ammon Salter} {and} \bibinfo{person}{David Gann}.} \bibinfo{year}{2003}\natexlab{}.
\newblock \showarticletitle{Sources of Ideas for Innovation in Engineering Design}.
\newblock \bibinfo{journal}{\emph{Research Policy}} \bibinfo{volume}{32}, \bibinfo{number}{8} (\bibinfo{date}{Sept.} \bibinfo{year}{2003}), \bibinfo{pages}{1309--1324}.
\newblock
\showISSN{0048-7333}
\urldef\tempurl%
\url{https://doi.org/10.1016/S0048-7333(02)00119-1}
\showDOI{\tempurl}


\bibitem[Sanchez et~al\mbox{.}(2024)]%
        {Sanchez2024}
\bibfield{author}{\bibinfo{person}{Diana~R Sanchez}, \bibinfo{person}{Joshua McVeigh-Schultz}, \bibinfo{person}{Katherine Isbister}, \bibinfo{person}{Monica Tran}, \bibinfo{person}{Kassidy Martinez}, \bibinfo{person}{Marjan Dost}, \bibinfo{person}{Anya Osborne}, \bibinfo{person}{Daniel Diaz}, \bibinfo{person}{Philip Farillas}, \bibinfo{person}{Timothy Lang}, \bibinfo{person}{Alexandra Leeds}, \bibinfo{person}{George Butler}, {and} \bibinfo{person}{Monique Ferronatto}.} \bibinfo{year}{2024}\natexlab{}.
\newblock \showarticletitle{Virtual Reality pursuit: Using individual predispositions towards {VR} to understand perceptions of a virtualized workplace team experience}.
\newblock \bibinfo{journal}{\emph{Virtual Worlds}} \bibinfo{volume}{3}, \bibinfo{number}{4} (\bibinfo{date}{Oct.} \bibinfo{year}{2024}), \bibinfo{pages}{418--435}.
\newblock


\bibitem[Schroeder et~al\mbox{.}(2001)]%
        {Schroeder2001}
\bibfield{author}{\bibinfo{person}{Ralph Schroeder}, \bibinfo{person}{Anthony Steed}, \bibinfo{person}{Ann~Sofie Axelsson}, \bibinfo{person}{Ilona Heldal}, \bibinfo{person}{{\AA}sa Abelin}, \bibinfo{person}{Josef Widestr{\"o}m}, \bibinfo{person}{Alexander Nilsson}, {and} \bibinfo{person}{Mel Slater}.} \bibinfo{year}{2001}\natexlab{}.
\newblock \showarticletitle{Collaborating in Networked Immersive Spaces: As Good as Being There Together?}
\newblock \bibinfo{journal}{\emph{Computers \& Graphics}} \bibinfo{volume}{25}, \bibinfo{number}{5} (\bibinfo{date}{Oct.} \bibinfo{year}{2001}), \bibinfo{pages}{781--788}.
\newblock
\showISSN{0097-8493}
\urldef\tempurl%
\url{https://doi.org/10.1016/S0097-8493(01)00120-0}
\showDOI{\tempurl}


\bibitem[Schwind(2018)]%
        {Schwind2018}
\bibfield{author}{\bibinfo{person}{Valentin Schwind}.} \bibinfo{year}{2018}\natexlab{}.
\newblock \emph{\bibinfo{title}{Implications of the Uncanny Valley of Avatars and Virtual Characters for Human-Computer Interaction}}.
\newblock \bibinfo{thesistype}{Ph.\,D. Dissertation}. \bibinfo{school}{University of Stuttgart}.
\newblock
\urldef\tempurl%
\url{https://doi.org/10.18419/opus-9936}
\showDOI{\tempurl}


\bibitem[Segura et~al\mbox{.}(2019)]%
        {Segura2019}
\bibfield{author}{\bibinfo{person}{Elena~M{\'a}rquez Segura}, \bibinfo{person}{Laia~Turmo Vidal}, \bibinfo{person}{Luis~Parrilla Bel}, {and} \bibinfo{person}{Annika Waern}.} \bibinfo{year}{2019}\natexlab{}.
\newblock \showarticletitle{Using {{Training Technology Probes}} in {{Bodystorming}} for {{Physical Training}}}. In \bibinfo{booktitle}{\emph{Proceedings of the 6th {{International Conference}} on {{Movement}} and {{Computing}}}} \emph{(\bibinfo{series}{{{MOCO}} '19})}. \bibinfo{publisher}{{Association for Computing Machinery}}, \bibinfo{address}{{New York, NY, USA}}, \bibinfo{pages}{1--8}.
\newblock
\showISBNx{978-1-4503-7654-9}
\urldef\tempurl%
\url{https://doi.org/10.1145/3347122.3347132}
\showDOI{\tempurl}


\bibitem[Seymour et~al\mbox{.}(2021)]%
        {Seymour2021}
\bibfield{author}{\bibinfo{person}{Mike Seymour}, \bibinfo{person}{Lingyao~Ivy Yuan}, \bibinfo{person}{Alan~R. Dennis}, {and} \bibinfo{person}{Kai Riemer}.} \bibinfo{year}{2021}\natexlab{}.
\newblock \showarticletitle{Have {{We Crossed}} the {{Uncanny Valley}}? {{Understanding}} Affinity, Trustworthiness, and Preference for More Realistic Virtual Humans in Immersive Environments.}. In \bibinfo{booktitle}{\emph{{{HICSS}}}} \emph{(\bibinfo{series}{3}, Vol.~\bibinfo{volume}{22})}. \bibinfo{publisher}{{Journal of the Association for Information Systems}}.
\newblock
\urldef\tempurl%
\url{https://doi.org/10.17705/1jais.00674}
\showDOI{\tempurl}


\bibitem[Shami et~al\mbox{.}(2010)]%
        {Shami2010}
\bibfield{author}{\bibinfo{person}{N.~Sadat Shami}, \bibinfo{person}{Li~Te Cheng}, \bibinfo{person}{Steven Rohall}, \bibinfo{person}{Andrew Sempere}, {and} \bibinfo{person}{John Patterson}.} \bibinfo{year}{2010}\natexlab{}.
\newblock \showarticletitle{Avatars {{Meet Meetings}}: {{Design Issues}} in {{Integrating Avatars}} in {{Distributed Corporate Meetings}}}.
\newblock \bibinfo{journal}{\emph{Proceedings of the 16th ACM International Conference on Supporting Group Work, GROUP'10}} (\bibinfo{date}{Nov.} \bibinfo{year}{2010}), \bibinfo{pages}{35--44}.
\newblock
\showISBNx{9781450303873}
\urldef\tempurl%
\url{https://doi.org/10.1145/1880071.1880078}
\showDOI{\tempurl}


\bibitem[Shibata and Suzuki(2004)]%
        {Shibata2004}
\bibfield{author}{\bibinfo{person}{Seiji Shibata} {and} \bibinfo{person}{Naoto Suzuki}.} \bibinfo{year}{2004}\natexlab{}.
\newblock \showarticletitle{Effects of an Indoor Plant on Creative Task Performance and Mood}.
\newblock \bibinfo{journal}{\emph{Scandinavian Journal of Psychology}} \bibinfo{volume}{45}, \bibinfo{number}{5} (\bibinfo{year}{2004}), \bibinfo{pages}{373--381}.
\newblock
\showISSN{1467-9450}
\urldef\tempurl%
\url{https://doi.org/10.1111/j.1467-9450.2004.00419.x}
\showDOI{\tempurl}


\bibitem[Shibata and Suzuki(2002)]%
        {Shibata2002}
\bibfield{author}{\bibinfo{person}{{\relax SEIJI} Shibata} {and} \bibinfo{person}{{\relax NAOTO} Suzuki}.} \bibinfo{year}{2002}\natexlab{}.
\newblock \showarticletitle{Effects of the {{Foliage Plant}} on {{Task Performance}} and {{Mood}}}.
\newblock \bibinfo{journal}{\emph{Journal of Environmental Psychology}} \bibinfo{volume}{22}, \bibinfo{number}{3} (\bibinfo{date}{Sept.} \bibinfo{year}{2002}), \bibinfo{pages}{265--272}.
\newblock
\showISSN{0272-4944}
\urldef\tempurl%
\url{https://doi.org/10.1006/jevp.2002.0232}
\showDOI{\tempurl}


\bibitem[Shin et~al\mbox{.}(2019)]%
        {Shin2019}
\bibfield{author}{\bibinfo{person}{Mincheol Shin}, \bibinfo{person}{Se~Jung Kim}, {and} \bibinfo{person}{Frank Biocca}.} \bibinfo{year}{2019}\natexlab{}.
\newblock \showarticletitle{The Uncanny Valley: {{No}} Need for Any Further Judgments When an Avatar Looks Eerie}.
\newblock \bibinfo{journal}{\emph{Computers in Human Behavior}}  \bibinfo{volume}{94} (\bibinfo{date}{May} \bibinfo{year}{2019}), \bibinfo{pages}{100--109}.
\newblock
\showISSN{0747-5632}
\urldef\tempurl%
\url{https://doi.org/10.1016/j.chb.2019.01.016}
\showDOI{\tempurl}


\bibitem[Sivan et~al\mbox{.}(2014)]%
        {Sivan2014}
\bibfield{author}{\bibinfo{person}{Yesha Sivan}, \bibinfo{person}{David Gefen}, \bibinfo{person}{Maged~Kamel Boulos}, \bibinfo{person}{Pekka Alahuhta}, \bibinfo{person}{Emma Nordb{\"a}ck}, \bibinfo{person}{Anu Sivunen}, {and} \bibinfo{person}{Teemu Surakka}.} \bibinfo{year}{2014}\natexlab{}.
\newblock \showarticletitle{Fostering {{Team Creativity}} in {{Virtual Worlds}}}.
\newblock \bibinfo{journal}{\emph{Journal of Virtual Worlds Research}} \bibinfo{volume}{7}, \bibinfo{number}{32} (\bibinfo{year}{2014}), \bibinfo{pages}{1--24}.
\newblock
\showISSN{1941-8477}
\urldef\tempurl%
\url{https://doi.org/10.4101/jvwr.v7i3.7062}
\showDOI{\tempurl}


\bibitem[Smith and Neff(2018)]%
        {Smith2018}
\bibfield{author}{\bibinfo{person}{Harrison~Jesse Smith} {and} \bibinfo{person}{Michael Neff}.} \bibinfo{year}{2018}\natexlab{}.
\newblock \showarticletitle{Communication {{Behavior}} in {{Embodied Virtual Reality}}}. In \bibinfo{booktitle}{\emph{Proceedings of the 2018 {{CHI Conference}} on {{Human Factors}} in {{Computing Systems}}}} \emph{(\bibinfo{series}{{{CHI}} '18})}. \bibinfo{publisher}{{Association for Computing Machinery}}, \bibinfo{address}{{New York, NY, USA}}, \bibinfo{pages}{1--12}.
\newblock
\showISBNx{978-1-4503-5620-6}
\urldef\tempurl%
\url{https://doi.org/10.1145/3173574.3173863}
\showDOI{\tempurl}


\bibitem[Spitzberg(1988)]%
        {Spitzberg1988}
\bibfield{author}{\bibinfo{person}{Brian~H Spitzberg}.} \bibinfo{year}{1988}\natexlab{}.
\newblock \showarticletitle{Communication Competence: {{Measures}} of Perceived Effectiveness.}
\newblock In \bibinfo{booktitle}{\emph{A Handbook for the Study of Human Communication: {{Methods}} and Instruments for Observing, Measuring, and Assessing Communication Processes.}} \bibinfo{publisher}{{Ablex Publishing}}, \bibinfo{address}{{Westport, CT, US}}, \bibinfo{pages}{67--105}.
\newblock
\showISBNx{0-89391-424-X (Hardcover)}


\bibitem[Sra et~al\mbox{.}(2018)]%
        {Sra2018}
\bibfield{author}{\bibinfo{person}{Misha Sra}, \bibinfo{person}{Ken Perlin}, \bibinfo{person}{Luiz Velho}, \bibinfo{person}{Mark Bolas}, \bibinfo{person}{Ken Perlin}, \bibinfo{person}{Mark Bolas}, \bibinfo{person}{Luiz Velho}, \bibinfo{person}{Mark Bolas}, \bibinfo{person}{Ken Perlin}, {and} \bibinfo{person}{Mark Bolas}.} \bibinfo{year}{2018}\natexlab{}.
\newblock \showarticletitle{Novel {{Interaction Techniques}} for {{Collaboration}} in {{VR}}}. In \bibinfo{booktitle}{\emph{Extended {{Abstracts}} of the 2018 {{CHI Conference}} on {{Human Factors}} in {{Computing Systems}}}}, Vol.~\bibinfo{volume}{2018-April}. \bibinfo{publisher}{{Association for Computing Machinery}}, \bibinfo{pages}{1--8}.
\newblock
\showISBNx{978-1-4503-5621-3}
\urldef\tempurl%
\url{https://doi.org/10.1145/3170427.3170628}
\showDOI{\tempurl}


\bibitem[Steed et~al\mbox{.}(2016)]%
        {Steed2016}
\bibfield{author}{\bibinfo{person}{Anthony Steed}, \bibinfo{person}{Sebastian Frlston}, \bibinfo{person}{Maria~Murcia Lopez}, \bibinfo{person}{Jason Drummond}, \bibinfo{person}{Ye Pan}, {and} \bibinfo{person}{David Swapp}.} \bibinfo{year}{2016}\natexlab{}.
\newblock \showarticletitle{An '{{In}} the {{Wild}}' {{Experiment}} on {{Presence}} and {{Embodiment}} Using {{Consumer Virtual Reality Equipment}}}.
\newblock \bibinfo{journal}{\emph{IEEE Transactions on Visualization and Computer Graphics}} \bibinfo{volume}{22}, \bibinfo{number}{4} (\bibinfo{date}{April} \bibinfo{year}{2016}), \bibinfo{pages}{1406--1414}.
\newblock
\urldef\tempurl%
\url{https://doi.org/10.1109/TVCG.2016.2518135}
\showDOI{\tempurl}


\bibitem[Steinicke et~al\mbox{.}(2020)]%
        {Steinicke2020}
\bibfield{author}{\bibinfo{person}{Frank Steinicke}, \bibinfo{person}{Annika Meinecke}, {and} \bibinfo{person}{Nale {Lehmann-Willenbrock}}.} \bibinfo{year}{2020}\natexlab{}.
\newblock \showarticletitle{A {{First Pilot Study}} to {{Compare Virtual Group Meetings}} Using {{Video Conferences}} and ({{Immersive}}) {{Virtual Reality}}}.
\newblock \bibinfo{journal}{\emph{Symposium on Spatial User Interaction}} (\bibinfo{year}{2020}), \bibinfo{pages}{1--2}.
\newblock
\urldef\tempurl%
\url{https://doi.org/10.1145/3385959}
\showDOI{\tempurl}


\bibitem[Steptoe et~al\mbox{.}(2013)]%
        {Steptoe2013}
\bibfield{author}{\bibinfo{person}{William Steptoe}, \bibinfo{person}{Anthony Steed}, {and} \bibinfo{person}{Mel Slater}.} \bibinfo{year}{2013}\natexlab{}.
\newblock \showarticletitle{Human {{Tails}}: {{Ownership}} and {{Control}} of {{Extended Humanoid Avatars}}}.
\newblock \bibinfo{journal}{\emph{IEEE Transactions on Visualization and Computer Graphics}} \bibinfo{volume}{19}, \bibinfo{number}{4} (\bibinfo{date}{April} \bibinfo{year}{2013}), \bibinfo{pages}{583--590}.
\newblock
\showISSN{1941-0506}
\urldef\tempurl%
\url{https://doi.org/10.1109/TVCG.2013.32}
\showDOI{\tempurl}


\bibitem[Stone(1998)]%
        {Stone1998}
\bibfield{author}{\bibinfo{person}{Nancy~J. Stone}.} \bibinfo{year}{1998}\natexlab{}.
\newblock \showarticletitle{Windows and {{Environmental Cues}} on {{Performance}} and {{Mood}}}.
\newblock \bibinfo{journal}{\emph{Environment and Behavior}} \bibinfo{volume}{30}, \bibinfo{number}{3} (\bibinfo{date}{May} \bibinfo{year}{1998}), \bibinfo{pages}{306--321}.
\newblock
\showISSN{0013-9165}
\urldef\tempurl%
\url{https://doi.org/10.1177/001391659803000303}
\showDOI{\tempurl}


\bibitem[Stone(2003)]%
        {Stone2003}
\bibfield{author}{\bibinfo{person}{Nancy~J. Stone}.} \bibinfo{year}{2003}\natexlab{}.
\newblock \showarticletitle{Environmental View and Color for a Simulated Telemarketing Task}.
\newblock \bibinfo{journal}{\emph{Journal of Environmental Psychology}} \bibinfo{volume}{23}, \bibinfo{number}{1} (\bibinfo{date}{March} \bibinfo{year}{2003}), \bibinfo{pages}{63--78}.
\newblock
\showISSN{0272-4944}
\urldef\tempurl%
\url{https://doi.org/10.1016/S0272-4944(02)00107-X}
\showDOI{\tempurl}


\bibitem[Sun et~al\mbox{.}(2019)]%
        {Sun2019}
\bibfield{author}{\bibinfo{person}{Yilu Sun}, \bibinfo{person}{Omar Shaikh}, {and} \bibinfo{person}{Andrea~Stevenson Won}.} \bibinfo{year}{2019}\natexlab{}.
\newblock \showarticletitle{Nonverbal Synchrony in Virtual Reality}.
\newblock \bibinfo{journal}{\emph{PLOS ONE}} \bibinfo{volume}{14}, \bibinfo{number}{9} (\bibinfo{year}{2019}), \bibinfo{pages}{e0221803}.
\newblock
\showISBNx{1932-6203}
\urldef\tempurl%
\url{https://doi.org/10.1371/journal.pone.0221803}
\showDOI{\tempurl}


\bibitem[Taggar(2002)]%
        {Taggar2002}
\bibfield{author}{\bibinfo{person}{Simon Taggar}.} \bibinfo{year}{2002}\natexlab{}.
\newblock \showarticletitle{Individual {{Creativity}} and {{Group Ability}} to {{Utilize Individual Creative Resources}}: {{A Multilevel Model}}}.
\newblock \bibinfo{journal}{\emph{Academy of Management Journal}} \bibinfo{volume}{45}, \bibinfo{number}{2} (\bibinfo{date}{Nov.} \bibinfo{year}{2002}), \bibinfo{pages}{315--330}.
\newblock
\showISSN{0001-4273}
\urldef\tempurl%
\url{https://doi.org/10.5465/3069349}
\showDOI{\tempurl}


\bibitem[Teevan et~al\mbox{.}(2022)]%
        {Teevan2022}
\bibfield{author}{\bibinfo{person}{Jaime Teevan}, \bibinfo{person}{Nancy Baym}, \bibinfo{person}{Jenna Butler}, \bibinfo{person}{Brent Hecht}, \bibinfo{person}{Sonia Jaffe}, \bibinfo{person}{Kate Nowak}, \bibinfo{person}{Abigail Sellen}, \bibinfo{person}{Longqi Yang}, \bibinfo{person}{Marcus Ash}, \bibinfo{person}{Kagonya Awori}, \bibinfo{person}{Mia Bruch}, \bibinfo{person}{Piali Choudhury}, \bibinfo{person}{Adam Coleman}, \bibinfo{person}{Scott Counts}, \bibinfo{person}{Shiraz Cupala}, \bibinfo{person}{Mary Czerwinski}, \bibinfo{person}{Ed Doran}, \bibinfo{person}{Elizabeth Fetterolf}, \bibinfo{person}{Mar Gonzalez~Franco}, \bibinfo{person}{Kunal Gupta}, \bibinfo{person}{Aaron~L Halfaker}, \bibinfo{person}{Constance Hadley}, \bibinfo{person}{Brian Houck}, \bibinfo{person}{Kori Inkpen}, \bibinfo{person}{Shamsi Iqbal}, \bibinfo{person}{Eric Knudsen}, \bibinfo{person}{Stacey Levine}, \bibinfo{person}{Si{\^a}n Lindley}, \bibinfo{person}{Jennifer Neville}, \bibinfo{person}{Jacki O'Neill}, \bibinfo{person}{Rick
  Pollak}, \bibinfo{person}{Victor Poznanski}, \bibinfo{person}{Sean Rintel}, \bibinfo{person}{Neha~Parikh Shah}, \bibinfo{person}{Siddharth Suri}, \bibinfo{person}{Adam~D. Troy}, {and} \bibinfo{person}{Mengting Wan}.} \bibinfo{year}{2022}\natexlab{}.
\newblock \bibinfo{booktitle}{\emph{Microsoft New Future of Work Report 2022}}.
\newblock \bibinfo{type}{{T}echnical {R}eport} MSR-TR-2022-3. \bibinfo{institution}{{Microsoft}}.
\newblock
\urldef\tempurl%
\url{https://www.microsoft.com/en-us/research/publication/microsoft-new-future-of-work-report-2022/}
\showURL{%
\tempurl}


\bibitem[Toumi et~al\mbox{.}(2021)]%
        {Toumi2021}
\bibfield{author}{\bibinfo{person}{Karima Toumi}, \bibinfo{person}{Fabien Girandola}, {and} \bibinfo{person}{Nathalie Bonnardel}.} \bibinfo{year}{2021}\natexlab{}.
\newblock \showarticletitle{Technologies for {{Supporting Creativity}} in {{Design}}: {{A View}} of {{Physical}} and {{Virtual Environments}} with {{Regard}} to {{Cognitive}} and {{Social Processes}}}.
\newblock \bibinfo{journal}{\emph{Creativity. Theories \textendash{} Research - Applications}} \bibinfo{volume}{8}, \bibinfo{number}{1} (\bibinfo{date}{May} \bibinfo{year}{2021}), \bibinfo{pages}{189--212}.
\newblock
\urldef\tempurl%
\url{https://doi.org/10.2478/ctra-2021-0012}
\showDOI{\tempurl}


\bibitem[Trepte and Reinecke(2010)]%
        {Trepte2010}
\bibfield{author}{\bibinfo{person}{Sabine Trepte} {and} \bibinfo{person}{Leonard Reinecke}.} \bibinfo{year}{2010}\natexlab{}.
\newblock \showarticletitle{Avatar Creation and Video Game Enjoyment: {{Effects}} of Life-Satisfaction, Game Competitiveness, and Identification with the Avatar.}
\newblock \bibinfo{journal}{\emph{Journal of Media Psychology: Theories, Methods, and Applications}} \bibinfo{volume}{22}, \bibinfo{number}{4} (\bibinfo{date}{Dec.} \bibinfo{year}{2010}), \bibinfo{pages}{171}.
\newblock
\showISSN{2151-2388}
\urldef\tempurl%
\url{https://doi.org/10.1027/1864-1105/a000022}
\showDOI{\tempurl}


\bibitem[Vosinakis and Koutsabasis(2013)]%
        {Vosinakis2013}
\bibfield{author}{\bibinfo{person}{Spyros Vosinakis} {and} \bibinfo{person}{Panayiotis Koutsabasis}.} \bibinfo{year}{2013}\natexlab{}.
\newblock \showarticletitle{Interaction Design Studio Learning in Virtual Worlds}.
\newblock \bibinfo{journal}{\emph{Virtual Reality}} \bibinfo{volume}{17}, \bibinfo{number}{1} (\bibinfo{date}{March} \bibinfo{year}{2013}), \bibinfo{pages}{59--75}.
\newblock
\showISSN{1434-9957}
\urldef\tempurl%
\url{https://doi.org/10.1007/s10055-013-0221-1}
\showDOI{\tempurl}


\bibitem[Waltemate et~al\mbox{.}(2018)]%
        {Waltemate2018}
\bibfield{author}{\bibinfo{person}{Thomas Waltemate}, \bibinfo{person}{Dominik Gall}, \bibinfo{person}{Daniel Roth}, \bibinfo{person}{Mario Botsch}, {and} \bibinfo{person}{Marc~Erich Latoschik}.} \bibinfo{year}{2018}\natexlab{}.
\newblock \showarticletitle{The {{Impact}} of {{Avatar Personalization}} and {{Immersion}} on {{Virtual Body Ownership}}, {{Presence}}, and {{Emotional Response}}}.
\newblock \bibinfo{journal}{\emph{IEEE Transactions on Visualization and Computer Graphics}} \bibinfo{volume}{24}, \bibinfo{number}{4} (\bibinfo{date}{April} \bibinfo{year}{2018}), \bibinfo{pages}{1643--1652}.
\newblock
\showISSN{1077-2626}
\urldef\tempurl%
\url{https://doi.org/10.1109/TVCG.2018.2794629}
\showDOI{\tempurl}


\bibitem[Wauck et~al\mbox{.}(2018)]%
        {Wauck2018}
\bibfield{author}{\bibinfo{person}{Helen Wauck}, \bibinfo{person}{Gale Lucas}, \bibinfo{person}{Ari Shapiro}, \bibinfo{person}{Andrew Feng}, \bibinfo{person}{Jill Boberg}, {and} \bibinfo{person}{Jonathan Gratch}.} \bibinfo{year}{2018}\natexlab{}.
\newblock \showarticletitle{Analyzing the {{Effect}} of {{Avatar Self-Similarity}} on {{Men}} and {{Women}} in a {{Search}} and {{Rescue Game}}}. In \bibinfo{booktitle}{\emph{Proceedings of the 2018 {{CHI Conference}} on {{Human Factors}} in {{Computing Systems}}}} \emph{(\bibinfo{series}{{{CHI}} '18})}. \bibinfo{publisher}{{Association for Computing Machinery}}, \bibinfo{address}{{New York, NY, USA}}, \bibinfo{pages}{1--12}.
\newblock
\showISBNx{978-1-4503-5620-6}
\urldef\tempurl%
\url{https://doi.org/10.1145/3173574.3174059}
\showDOI{\tempurl}


\bibitem[West(2002)]%
        {West2002}
\bibfield{author}{\bibinfo{person}{Michael~A. West}.} \bibinfo{year}{2002}\natexlab{}.
\newblock \showarticletitle{Sparkling {{Fountains}} or {{Stagnant Ponds}}: {{An Integrative Model}} of {{Creativity}} and {{Innovation Implementation}} in {{Work Groups}}}.
\newblock \bibinfo{journal}{\emph{Applied Psychology}} \bibinfo{volume}{51}, \bibinfo{number}{3} (\bibinfo{year}{2002}), \bibinfo{pages}{355--387}.
\newblock
\showISSN{1464-0597}
\urldef\tempurl%
\url{https://doi.org/10.1111/1464-0597.00951}
\showDOI{\tempurl}


\bibitem[Williamson et~al\mbox{.}(2021)]%
        {Williamson2021}
\bibfield{author}{\bibinfo{person}{Julie~R Williamson}, \bibinfo{person}{Jie Li~CWI Vinoba~Vinayagamoorthy}, \bibinfo{person}{David~A Shamma Pablo~Cesar}, \bibinfo{person}{Jie Li}, \bibinfo{person}{Vinoba Vinayagamoorthy}, \bibinfo{person}{David~A Shamma}, {and} \bibinfo{person}{Pablo Cesar}.} \bibinfo{year}{2021}\natexlab{}.
\newblock \showarticletitle{Proxemics and {{Social Interactions}} in an {{Instrumented Virtual Reality Workshop}}}.
\newblock \bibinfo{journal}{\emph{Proceedings of the 2021 CHI Conference on Human Factors in Computing Systems}}  \bibinfo{volume}{21} (\bibinfo{year}{2021}).
\newblock
\urldef\tempurl%
\url{https://doi.org/10.1145/3411764}
\showDOI{\tempurl}


\bibitem[Williamson et~al\mbox{.}(2022)]%
        {Williamson2022}
\bibfield{author}{\bibinfo{person}{Julie~R. Williamson}, \bibinfo{person}{Joseph O'Hagan}, \bibinfo{person}{John~Alexis {Guerra-Gomez}}, \bibinfo{person}{John~H Williamson}, \bibinfo{person}{Pablo Cesar}, {and} \bibinfo{person}{David~A. Shamma}.} \bibinfo{year}{2022}\natexlab{}.
\newblock \showarticletitle{Digital {{Proxemics}}: {{Designing Social}} and {{Collaborative Interaction}} in {{Virtual Environments}}}. In \bibinfo{booktitle}{\emph{Proceedings of the 2022 {{CHI Conference}} on {{Human Factors}} in {{Computing Systems}}}} \emph{(\bibinfo{series}{{{CHI}} '22})}. \bibinfo{publisher}{{Association for Computing Machinery}}, \bibinfo{address}{{New York, NY, USA}}, \bibinfo{pages}{1--12}.
\newblock
\showISBNx{978-1-4503-9157-3}
\urldef\tempurl%
\url{https://doi.org/10.1145/3491102.3517594}
\showDOI{\tempurl}


\bibitem[Won et~al\mbox{.}(2015b)]%
        {Won2015}
\bibfield{author}{\bibinfo{person}{Andrea~Stevenson Won}, \bibinfo{person}{Jeremy Bailenson}, \bibinfo{person}{Jimmy Lee}, {and} \bibinfo{person}{Jaron Lanier}.} \bibinfo{year}{2015}\natexlab{b}.
\newblock \showarticletitle{Homuncular {{Flexibility}} in {{Virtual Reality}}}.
\newblock \bibinfo{journal}{\emph{Journal of Computer-Mediated Communication}} \bibinfo{volume}{20}, \bibinfo{number}{3} (\bibinfo{date}{May} \bibinfo{year}{2015}), \bibinfo{pages}{241--259}.
\newblock
\showISSN{10836101}
\urldef\tempurl%
\url{https://doi.org/10.1111/jcc4.12107}
\showDOI{\tempurl}


\bibitem[Won et~al\mbox{.}(2015a)]%
        {Won2015a}
\bibfield{author}{\bibinfo{person}{Andrea~Stevenson Won}, \bibinfo{person}{Jeremy~N. Bailenson}, {and} \bibinfo{person}{Jaron Lanier}.} \bibinfo{year}{2015}\natexlab{a}.
\newblock \showarticletitle{Appearance and {{Task Success}} in {{Novel Avatars}}}.
\newblock \bibinfo{journal}{\emph{Presence: Teleoperators and Virtual Environments}} \bibinfo{volume}{24}, \bibinfo{number}{4} (\bibinfo{date}{Nov.} \bibinfo{year}{2015}), \bibinfo{pages}{335--346}.
\newblock
\urldef\tempurl%
\url{https://doi.org/10.1162/PRES_a_00238}
\showDOI{\tempurl}


\bibitem[Won et~al\mbox{.}(2014)]%
        {Won2014}
\bibfield{author}{\bibinfo{person}{Andrea~Stevenson Won}, \bibinfo{person}{Jeremy~N Bailenson}, \bibinfo{person}{Suzanne~C Stathatos}, {and} \bibinfo{person}{Wenqing Dai}.} \bibinfo{year}{2014}\natexlab{}.
\newblock \showarticletitle{Automatically {{Detected Nonverbal Behavior Predicts Creativity}} in {{Collaborating Dyads}}}.
\newblock \bibinfo{journal}{\emph{Journal of Nonverbal Behavior}} \bibinfo{volume}{38}, \bibinfo{number}{3} (\bibinfo{year}{2014}), \bibinfo{pages}{389--408}.
\newblock
\showISBNx{0191-5886, 1573-3653}
\urldef\tempurl%
\url{https://doi.org/10.1007/s10919-014-0186-0}
\showDOI{\tempurl}


\bibitem[Yang et~al\mbox{.}(2018)]%
        {Yang2018}
\bibfield{author}{\bibinfo{person}{Xiaozhe Yang}, \bibinfo{person}{Lin Lin}, \bibinfo{person}{Pei-Yu Cheng}, \bibinfo{person}{Xue Yang}, \bibinfo{person}{Youqun Ren}, {and} \bibinfo{person}{Yueh-Min Huang}.} \bibinfo{year}{2018}\natexlab{}.
\newblock \showarticletitle{Examining Creativity through a Virtual Reality Support System}.
\newblock \bibinfo{journal}{\emph{Education Tech Research Dev}} \bibinfo{volume}{66}, \bibinfo{number}{5} (\bibinfo{date}{Oct.} \bibinfo{year}{2018}), \bibinfo{pages}{1231--1254}.
\newblock
\showISSN{1556-6501}
\urldef\tempurl%
\url{https://doi.org/10.1007/s11423-018-9604-z}
\showDOI{\tempurl}


\bibitem[Yassien et~al\mbox{.}(2020)]%
        {Yassien2020}
\bibfield{author}{\bibinfo{person}{Amal Yassien}, \bibinfo{person}{Passant Elagroudy}, \bibinfo{person}{Elhassan Makled}, {and} \bibinfo{person}{Slim Abdennadher}.} \bibinfo{year}{2020}\natexlab{}.
\newblock \showarticletitle{A {{Design Space}} for {{Social Presence}} in {{VR}}}. In \bibinfo{booktitle}{\emph{Proceedings of the 11th {{Nordic Conference}} on {{Human-Computer Interaction}}: {{Shaping Experiences}}, {{Shaping Society}}}}, Vol.~\bibinfo{volume}{20}. \bibinfo{publisher}{{Association for Computing Machinery}}, \bibinfo{pages}{1--12}.
\newblock
\showISBNx{978-1-4503-7579-5}
\urldef\tempurl%
\url{https://doi.org/10.1145/3419249.3420112}
\showDOI{\tempurl}


\bibitem[Yee and Bailenson(2007)]%
        {Yee2007}
\bibfield{author}{\bibinfo{person}{Nick Yee} {and} \bibinfo{person}{Jeremy Bailenson}.} \bibinfo{year}{2007}\natexlab{}.
\newblock \showarticletitle{The Proteus Effect: {{The}} Effect of Transformed Self-Representation on Behavior}.
\newblock \bibinfo{journal}{\emph{Human Communication Research}} \bibinfo{volume}{33}, \bibinfo{number}{3} (\bibinfo{date}{July} \bibinfo{year}{2007}), \bibinfo{pages}{271--290}.
\newblock
\showISSN{03603989}
\urldef\tempurl%
\url{https://doi.org/10.1111/j.1468-2958.2007.00299.x}
\showDOI{\tempurl}


\bibitem[Yoon et~al\mbox{.}(2019)]%
        {Yoon2019}
\bibfield{author}{\bibinfo{person}{Boram Yoon}, \bibinfo{person}{Hyung-il Kim}, \bibinfo{person}{Gun~A. Lee}, \bibinfo{person}{Mark Billinghurst}, {and} \bibinfo{person}{Woontack Woo}.} \bibinfo{year}{2019}\natexlab{}.
\newblock \showarticletitle{The {{Effect}} of {{Avatar Appearance}} on {{Social Presence}} in an {{Augmented Reality Remote Collaboration}}}. In \bibinfo{booktitle}{\emph{2019 {{IEEE Conference}} on {{Virtual Reality}} and {{3D User Interfaces}} ({{VR}})}}. \bibinfo{publisher}{{IEEE}}, \bibinfo{address}{{Osaka, Japan}}, \bibinfo{pages}{547--556}.
\newblock
\showISSN{2642-5254}
\urldef\tempurl%
\url{https://doi.org/10.1109/VR.2019.8797719}
\showDOI{\tempurl}


\bibitem[Yoshimura and Borst(2020)]%
        {Yoshimura2020}
\bibfield{author}{\bibinfo{person}{Andrew Yoshimura} {and} \bibinfo{person}{Christoph~W. Borst}.} \bibinfo{year}{2020}\natexlab{}.
\newblock \showarticletitle{Evaluation of {{Headset-based Viewing}} and {{Desktop-based Viewing}} of {{Remote Lectures}} in a {{Social VR Platform}}}.
\newblock \bibinfo{journal}{\emph{Proceedings of the ACM Symposium on Virtual Reality Software and Technology, VRST}} (\bibinfo{date}{Nov.} \bibinfo{year}{2020}).
\newblock
\urldef\tempurl%
\url{https://doi.org/10.1145/3385956.3422124}
\showDOI{\tempurl}


\bibitem[Zamanifard and Freeman(2019)]%
        {Zamanifard2019}
\bibfield{author}{\bibinfo{person}{Samaneh Zamanifard} {and} \bibinfo{person}{Guo Freeman}.} \bibinfo{year}{2019}\natexlab{}.
\newblock \showarticletitle{"{{The Togetherness}} That {{We Crave}}": {{Experiencing Social VR}} in {{Long Distance Relationships}}}.
\newblock \bibinfo{journal}{\emph{Conference Companion Publication of the 2019 on Computer Supported Cooperative Work and Social Computing}} (\bibinfo{year}{2019}).
\newblock
\urldef\tempurl%
\url{https://doi.org/10.1145/3311957}
\showDOI{\tempurl}


\bibitem[Zibrek et~al\mbox{.}(2021)]%
        {Zibrek2021}
\bibfield{author}{\bibinfo{person}{Katja Zibrek}, \bibinfo{person}{Xueni Pan}, {and} \bibinfo{person}{Marko Orel}.} \bibinfo{year}{2021}\natexlab{}.
\newblock \showarticletitle{Editorial: {{Meeting Remotely}}\textemdash{{The Challenges}} of {{Optimal Avatar Interaction}} in {{VR}}}.
\newblock \bibinfo{journal}{\emph{Frontiers in Virtual Reality}}  \bibinfo{volume}{2} (\bibinfo{year}{2021}).
\newblock
\showISSN{2673-4192}
\urldef\tempurl%
\url{https://www.frontiersin.org/articles/10.3389/frvir.2021.773258}
\showURL{%
\tempurl}


\end{thebibliography}

\appendix
\section{Research Materials}
\label{appendix: A}

\subsection{Study Two: Conversational Prompts}
\label{appendix: A.1}

If your work needs an appendix, add it before the and note that in the appendix, sections are lettered, not numbered. This document has two appendices, demonstrating the section
and subsection identification method (Table~\ref{tab:Promts} in Appendix~\ref{appendix: A.1}).

\begin{table}[ht]
\centering
\caption{Conversational Prompts in Study Two.}
\label{tab:Promts}
\resizebox{\textwidth}{!}{%
\begin{tabular}{p{0.48\textwidth} p{0.48\textwidth}}
\hline
\textbf{Water conservation prompt} & \textbf{Energy conservation prompt} \\ \hline \addlinespace[0.5ex]
\textbf{Avoid using products that require a lot of water to produce}. For example, a lot of water is used to raise animals for meat. Therefore, reducing one’s meat consumption is a good way to save water. & \textbf{Minimize your use of hot water}. Heating water requires a lot of energy. For example, washing clothes in cold water instead of hot can save energy on water heating. \\ \addlinespace[0.5ex]

\textbf{Avoid using water entirely when other methods are possible}. For example, don’t use a power washer or hose to clean off your porch when you can sweep instead. & \textbf{Heat water in the most energy-efficient way}. You want to make sure that as much as possible of the energy you use is going directly to heating the water that you want to use, and only that water. For example, when heating water for a hot drink, it is more energy efficient to heat water in an electric kettle instead of a microwave. \\ \addlinespace[0.5ex]
\textbf{Stop leaking water}. For example, fixing a leaky faucet would be a good example of a way to save both water and unnecessary expense. & \textbf{Limit your use of electricity during peak usage hours as much as possible}. For example, using electricity during the late afternoon requires more energy than in the morning. \\ \addlinespace[0.5ex]
\textbf{Try to use water in ways appropriate to the time of day}. For example, it’s often against city regulations to water your lawn when it is hot out because water will evaporate quickly and it’s both wasteful and ineffective. & \textbf{Reuse typically single-use items}. For example, reusing a plastic water bottle saves energy because producing plastic bottles uses a lot of energy in both packaging and transport. \\ \addlinespace[0.5ex]
\textbf{Do not use more water than is necessary for a given task/activity}. For example, turn off the water in the shower, instead of allowing it to run while shampooing. & \textbf{Use economy or eco settings on appliances}. Many modern appliances have such options. For example, many space heaters can be set to turn off once the room reaches a particular temperature. \\ \addlinespace[0.5ex]
\textbf{Reuse water whenever possible}. For example, when showering you can keep a bucket in the shower and use the water that collects to water plants. & \textbf{Turn off appliances when you are not using them or set them to be turned off by a set time}. For example, if you often fall asleep watching TV, you can set your TV to turn off by a certain time so that it isn’t on for hours without you watching it. \\ \addlinespace[0.5ex]
\textbf{Patronize businesses that use water efficiently}. Not only does this save water itself, but it also encourages other businesses to follow eco-friendly practices. For example, if you must take your car to a car wash, then go to one that recycles its wash water. & \textbf{Patronize businesses that use energy efficiently}. This helps to support energy-saving practices and also encourages other businesses to follow eco-friendly practices. For example, choose a business that uses solar energy. \\ \hline
\end{tabular}%
}
\end{table}

\end{document}